# Aging, rejuvenation and memory :
# the example of spin glasses

Eric VINCENT

*Service de Physique de l'Etat Condensé (CNRS URA 2464),*
*DSM/DRECAM/SPEC, CEA Saclay, 91191 Gif sur Yvette Cedex, France*

eric.vincent@cea.fr

This paper is written on the basis of a course given at the summer school "Ageing and the glass transition" in the University of Luxembourg. The results presented here are the joint work of the Saclay group: F. Bert, J.-P. Bouchaud, V. Dupuis, J. Hammann, D. Hérisson, F. Ladieu, M. Ocio, D. Parker, E.V., and others.

Abstract - In this paper, we review the general features of the out-of-equilibrium dynamics of spin glasses. We use this example as a guideline for a brief description of glassy dynamics in other disordered systems like structural and polymer glasses, colloids, gels etc. Starting with the simplest experiments, we discuss the scaling laws used to describe the isothermal aging observed in spin glasses after a quench down to the low temperature phase (these scaling laws are the same as established for polymer glasses). We then discuss the rejuvenation and memory effects observed when a spin glass is submitted to temperature variations during aging, and show some examples of similar phenomena in other glassy systems. The rejuvenation and memory effects and their implications are analyzed from the point of view of both energy landscape pictures and of real space pictures. We highlight the fact that both approaches point out the necessity of hierarchical processes involved in aging. We introduce the concept of a slowly growing and strongly temperature dependent dynamical correlation length, which is discussed at the light of a large panel of experiments.

# 1. What is a spin glass ?

A spin glass is a disordered and frustrated system. From the theorist's point of view, the definition of the spin glass is very simple: it is a set of randomly interacting magnetic moments on a lattice. The total energy is simply the sum over interacting neighbours ($S_i$ ,$S_j$) of all coupling energies $J_{ij}S_iS_j$, where the $\{J_{i,j}\}$ are random variables, gaussian or ±J distributed :

$$H = -\sum_{i,j} J_{i,j} S_i S_j \qquad (1) .$$

The impressive number of publications devoted to the spin-glass problem these last decades (see references in e.g. [1,2,3,4]) is in sharp contrast to the rather simple formulation as described by Eq.(1).

From the experimentalist's point of view, the way to obtain a set of randomly interacting magnetic moment is usually to dilute magnetic ions. The canonical example is that of intermetallic alloys, like for instance Cu:$Mn_{3\%}$, in which 3% of (magnetic) Mn atoms are thrown by random in a (non-magnetic) Cu matrix. The Mn magnetic atoms sit at random positions, therefore are separated by random distances, and the oscillating character of the RKKY interaction with respect to distance makes their coupling energy take a random sign. This class of systems corresponds to the historical discovery of spin glasses, which traces back to the studies of strongly diluted magnetic alloys and the Kondo effect [3].

Later on, spin glasses have been identified within insulating compounds. An example that we have studied in details at our laboratory is the the Indium diluted Chromium thiospinel $CdCr_{2x}In_{2(1-x)}$ $S_4$ , with superexchange magnetic interactions between the Cr ions [5]. For x=1, this compound is a ferromagnet with $T_c$=80K. The nearest neighbour interactions are ferromagnetic and dominant for x=1, but the next-nearest ones are antiferromagnetic. Hence, when some (magnetic) Cr ions are substituted by (non-magnetic) In ions, some ferromagnetic bindings are suppressed, and the effect of other antiferromagnetic interactions is enhanced. The balance that globally favours ferromagnetism for zero or small In-dilution is disturbed, and the ferromagnetic phase is replaced by a spin-glass phase for $x \leq 0.85$.

The phase diagram of the CrIn thiospinel is shown in Fig.1.1.a, together with the magnetic behaviour corresponding to various values of x in Fig.1.1.b [5,6]. As usual, the "FC" curves correspond to a measurement procedure in which the sample is cooled in presence of the measuring field, and the "ZFC" curves are obtained after cooling in zero field, applying the field at the lowest temperature and measuring the magnetization while increasing the temperature step by step. In Fig.1.1.b, the x=1 curve shows a very abrupt increase of the magnetization when approaching $T_c$=80K from above, that is characteristic of the ferromagnetic transition. At lower temperatures, magnetic irreversibilities are observed (splitting of the ZFC and FC curves), which are probably due to defects. In the x=0.95 and x=0.90 curves, the ferromagnetic transition is progressively rounded as the level of dilution increases, and the splitting of the FC and ZFC curves at low temperature indicates the reentrance of a spin-glass phase, that has been characterized in other studies [5,6]. For x=0.85, the ferromagnetic phase has

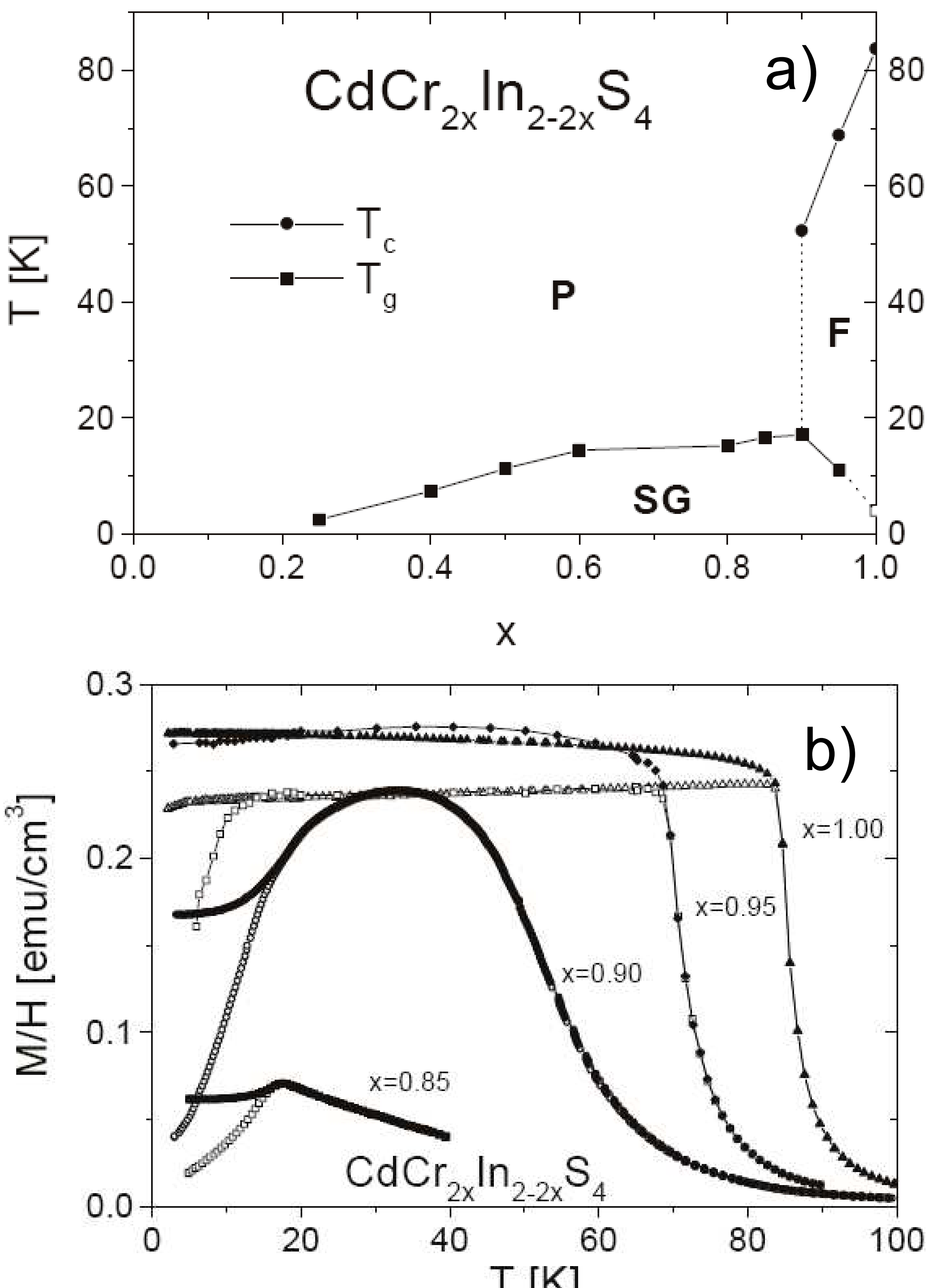

Figure1.1: $CdCr_{2x}In_{2-2x}S_4$ thiospinel compound [5,6]: a) Phase diagram, showing the paramagnetic (P), ferromagnetic (F) and spin glass (SG) phases; b) zero-field cooled (ZFC, open symbols) and field-cooled (FC, filled symbols) magnetizations.

disappeared, and at $T_g$=16.7K the system undergoes a transition from a paramagnetic to a spin-glass phase that presents the same features as observed in intermetallic spin glasses.

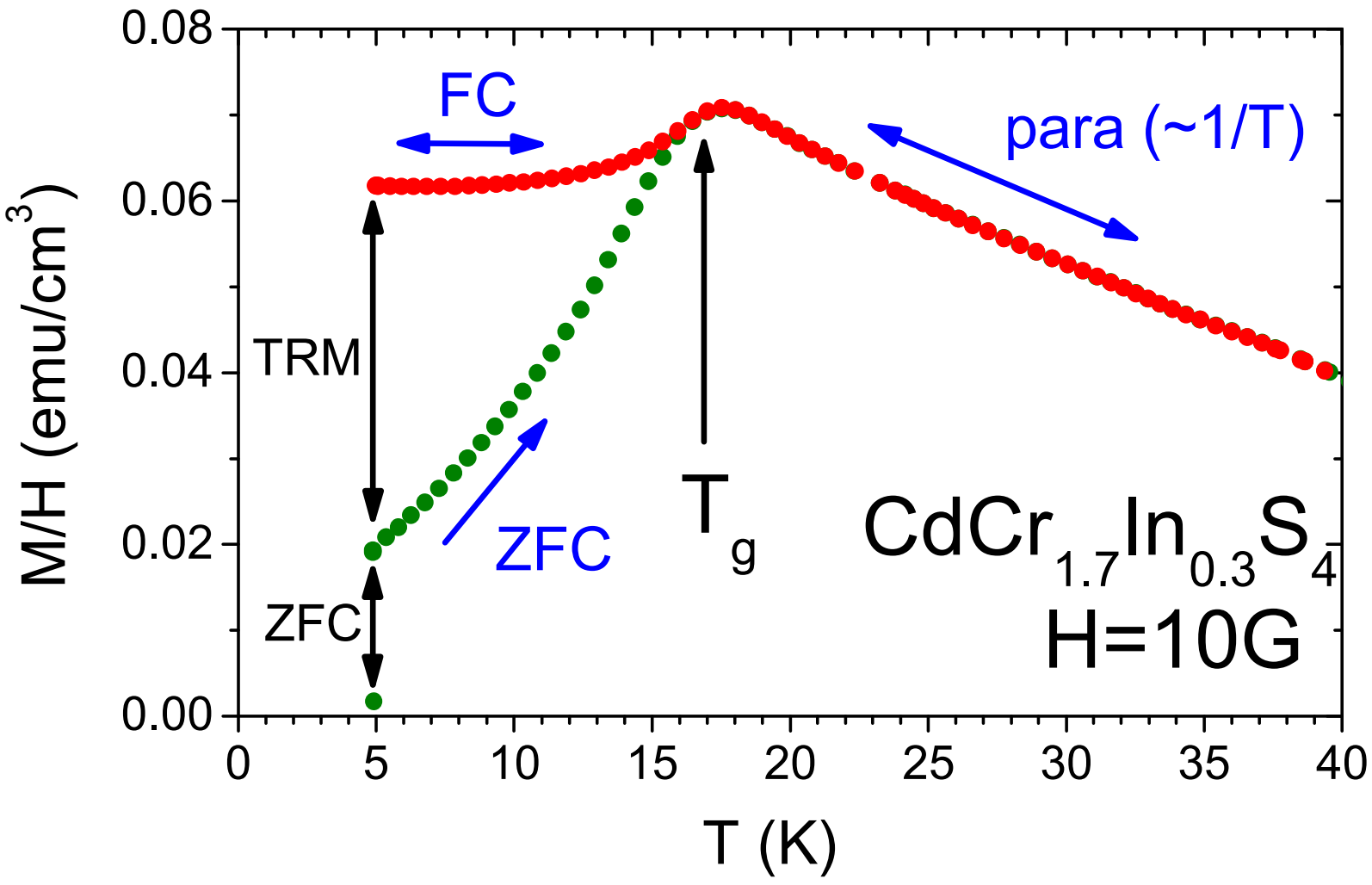

Figure 1.2: Zero-field cooled (ZFC) and field-cooled (FC) magnetization curves of the $CdCr_{1.7}In_{0.3}S_4$ thiospinel spin glass.

Fig.1.2 shows in more details the typical results of a ZFC/FC measurement on a spin glass. It is important to emphasize that a low-temperature splitting of the ZFC/FC curves is not by itself characteristic of a spin glass. This is only the signature of the onset of magnetic irreversibilities, which are not necessarily related to a collective behaviour, as for instance happens with superparamagnetic nanoparticles whose magnetization fluctuations are blocked by the effect of individual anisotropy barriers [7]. However, the (approximate) flatness of the FC curve that is observed here below $T_g$ shows that, when going from the paramagnetic region to low temperatures, the susceptibility increase is rather sharply stopped. This is suggestive of a collective behaviour, and is indeed observed in *concentrated* systems of nanoparticles, where the dipole-dipole interactions are at the origin of a (super-)spin glass-like transition [8,9,10].

While the FC curve can be measured upon decreasing or as well increasing the temperature in presence of the field, because the magnetization value can be considered at equilibrium (in a first approximation, usually within 1%), the ZFC one is fully out of equilibrium. After cooling in zero field and applying the field at some $T<T_g$, the magnetization ZFC(t) relaxes upwards as a function of time. In a symmetric way, starting from a FC state at T, if the field is turned to zero, the “thermo-remanent” magnetization (TRM) relaxes downwards. It has been observed in the early studies of slow dynamics in spin glasses that, for sufficiently low fields, these two “mirror experiments” do yield mirror results: ZFC(t) + TRM(t) = FC (this relation even holds if a slight relaxation of the FC magnetization occurs, FC ≡ FC(t) ) [11].

Another well-studied example of insulating spin-glass is the Sr-diluted Eu sulfur $Eu_xSr_{1-x}$ (e.g. x=0.3) [3], in which the alteration from the EuS ferromagnet to a spin-glass phase occurs in the same way as in the thiospinel. These various examples of spin glasses are helpful for understanding how the situation of randomly interacting moments is realized in “real” spin-glass samples. However, what we want to stress out is that there is a generic spin-glass behaviour which is common to all these systems and independent of the details of the sample chemistry, which the reader will be allowed to forget at least in a first approximation. Metallic as well as insulating spin-glasses show in 3d a well defined phase transition at $T_g$ (attested by the critical behaviour of some quantities), and slow dynamics is observed in the spin-glass phase with the occurrence of such interesting phenomena as aging, rejuvenation and memory effects. Regarding these different aspects, no difference can be traced out between metallic and insulating spin glasses, although the latter are magnetically more concentrated and have shorter range interactions. Certain systematic differences as a function of spin anisotropy have indeed been observed and are explained later in this paper, but, to the best of our present understanding, they are not directly related to their metal/insulator character or to any obvious chemical feature.

Finally, let us note that there is indeed a basic difference between the theoretical spin glass, in which there is a spin at each lattice node, and the experimental spin glass, which is site-diluted. It is not yet clear how far this type of difference may be relevant (for a recent review on the question of universality, see for instance [12]). As will be occasionally evoked along this paper (which is devoted to "experimental" spin glasses), the comparison of "real" (experimental) with "theoretical" spin glasses is not yet totally understood, but significant progresses have been made these last years, as well analytically as numerically [1,13,14,15,16].

# 2. Slow dynamics and aging

A crucial feature of the spin-glass behaviour (and of glassy dynamics in general) is the existence of relaxation processes at all time scales, from the microscopic times ($\sim 10^{-12}$s in spin glasses) to, at least, as long as the experimentalist can wait. The slow relaxation processes are particularly spectacular: in a spin glass, any field change causes a very long-lasting relaxation of the magnetization, and the response to an *ac* field is noticeably delayed. The basic experiments in which glassy dynamics is commonly investigated can be presented in 3 general classes: *dc* response, *ac* response, and spontaneous fluctuations (noise).

## 2.1 DC experiments

The study of the relaxation of the magnetization after a small field change has brought a lot of informations about the glassy features of the spin-glass dynamics. For now we only consider the case of "small fields", that are excitation fields which remain in the limit of linear response, or in other words fields that act as a non-perturbative probe. Usually, this field range (depending on the sample) is limited up to 1 or 10 Oe, a few percents of the field needed to surmount the interactions and recover a paramagnetic state (usually 100-1000 Oe).

Let us first consider the case of the relaxation of the thermo-remanent magnetization (TRM). Magnetization relaxations reveal a "waiting time" dependence of the dynamics that is singled out as "aging"[17,18,19]. In the experimental procedures, the aging time becomes another degree of freedom. As sketched in Fig.2.1, the sample is rapidly cooled in a small field H from above $T_g$ to $T<T_g$, and the sample is kept under field at temperature T during a waiting time $t_w$, after which the field is cut (at $t=0$). Then the relaxation is measured as a function of the observation time t. Fig.2.2 shows the results, which demonstrate the 2 basic features of spin-glass dynamics:

(i) the magnetization relaxation is slow, roughly logarithmic in time (glassy state)

(ii) it strongly depends on the waiting time: the longer $t_w$, the slower the relaxation (aging).

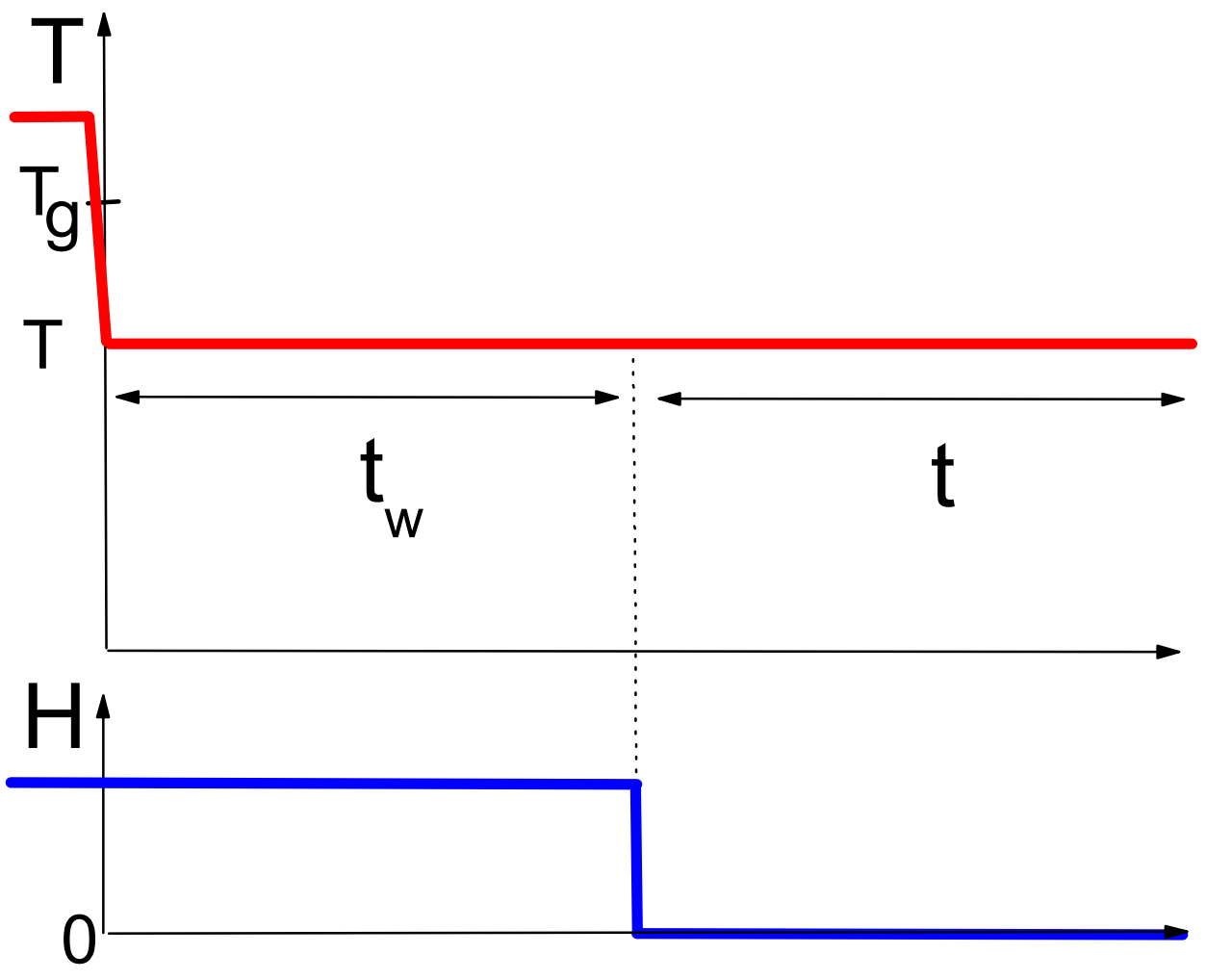

Figure 2.1: Sketch of the TRM measurement procedure.

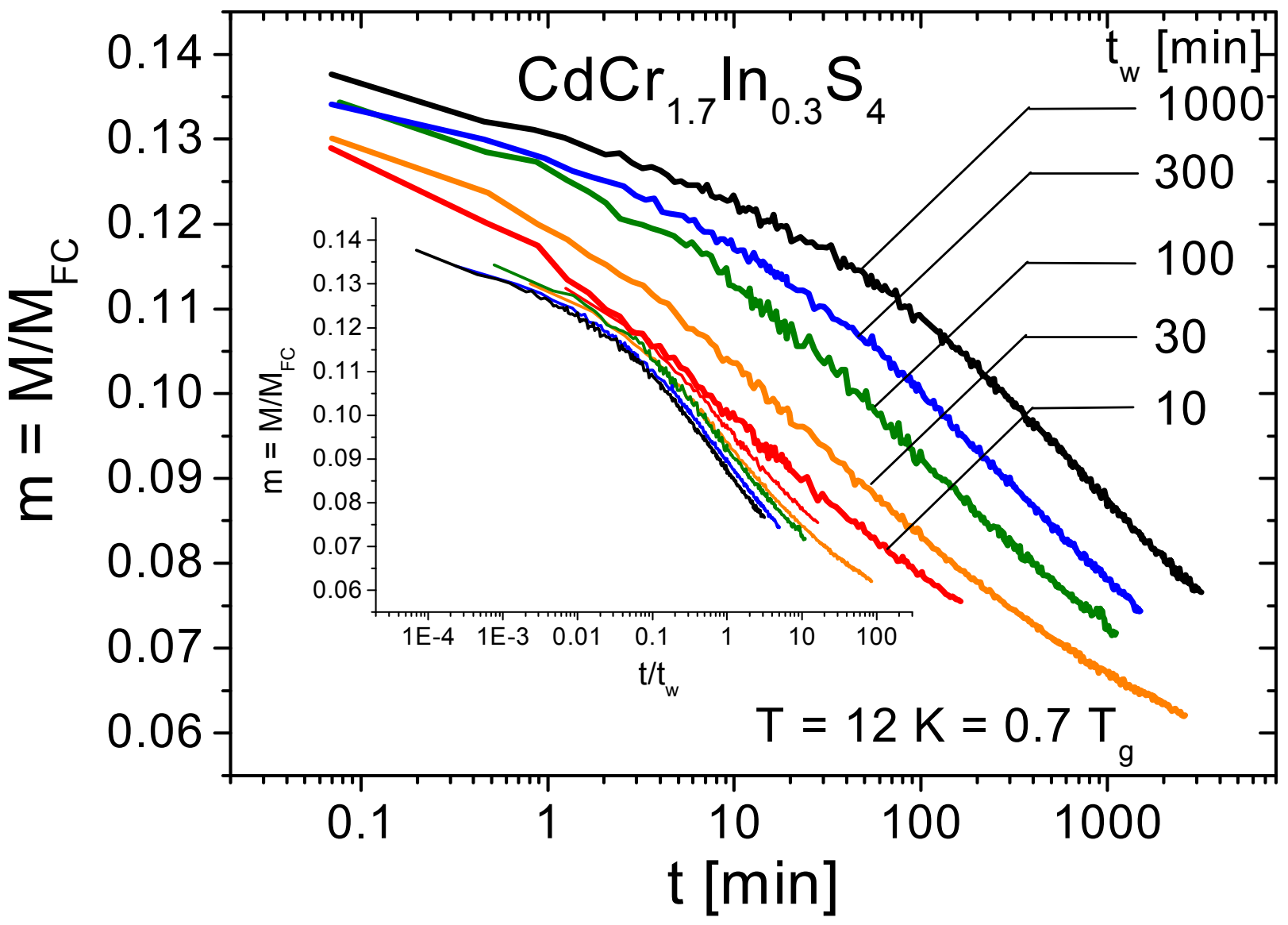

Figure 2.2: Example of TRM relaxations measured for various values of the waiting time $t_w$ (thiospinel spin glass). The inset shows the same curves, plotted as a function of $t/t_w$.

Hence, time translation invariance is lost in the slow dynamics of the spin glass: the relaxation depends on both $t_w$ and $t$, not only on $t$ (non-stationary dynamics). For increasing $t_w$, the response to cutting off the magnetic field becomes slower and slower on two respects: the initial fall-off of the magnetization becomes smaller, and the position of the inflection point of the curves shifts towards longer times. This inflection point approximately occurs at times t of the order of $t_w$ itself. When plotted as a function of $t/t_w$ (inset of Fig.2.2), the curves are gathered together (but they do not superimpose exactly onto each other, with a systematic $t_w$-dependent departure). In a first approximation, we may consider that the curves obey a $t/t_w$ scaling.

The same phenomenon of "aging" has been known for a long time for the mechanical properties of a wide class of materials called "glassy polymers"[20]. When a piece of e.g. PVC is submitted to a mechanical stress, its response (elongation, ...) is logarithmically slow. And the response depends on the time elapsed since the polymer has been quenched below its freezing temperature (Fig.2.2). Like in spin glasses, for increasing aging time the response becomes slower and slower, which was called "physical aging" (as opposed to "chemical aging"). The $t_w$-dependence of the dynamics of glassy polymers has been expressed as a scaling law that can be precisely applied to the case of spin glasses, as is explained below (see also [21]).

Fig.2.3 presents the mirror experiment, in which the sample is cooled in zero field, the field being applied after waiting $t_w$ (ZFC relaxation). The same $t_w$-dependence is observed as in TRM relaxations.

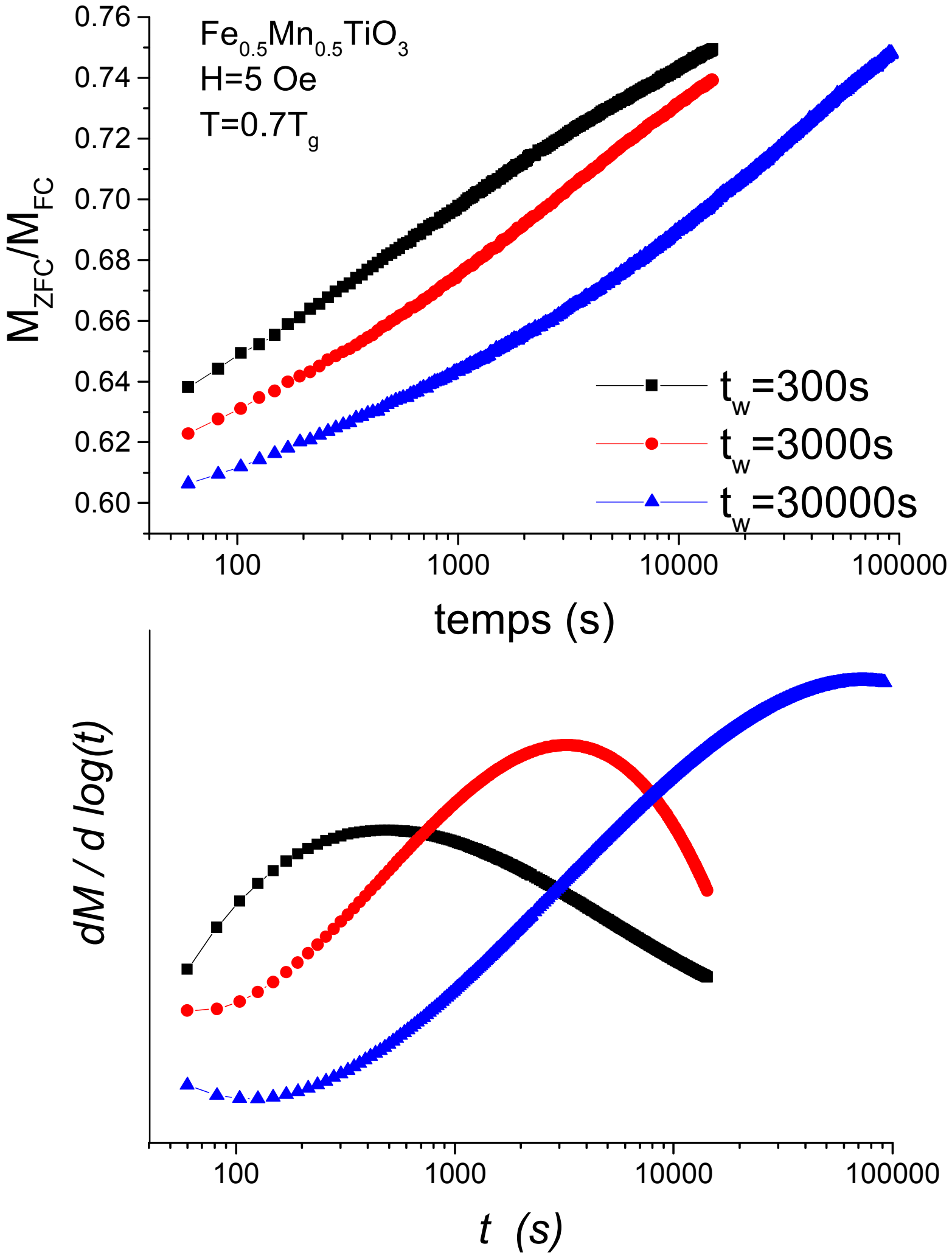

Figure 2.3: ZFC magnetization relaxations of the $Fe_{0.5}Mn_{0.5}TiO_3$ Ising spin glass [68], for 3 values of $t_w$. Top part: magnetization relaxations. Bottom part: logarithmic derivatives *dM/d log t* of the curves from the top part, displaying within a good approximation [22] the distribution of effective response times corresponding to the dynamics of the spin glass after a time of order $t_w$.

Following the suggestion of L. Lundgren et al [22], we also plot (bottom part of Fig.2.3) the logarithmic derivative $dM/d\log t$ of the magnetization M. The curves are bell shaped, with a broad maximum in the region $t \sim t_w$. These curves have an interesting physical interpretation which has been proposed by L. Lundgren and the Uppsala group [22]. The magnetization relaxations are slower than exponential, they can be modelled by a sum of exponential decays $exp(-t/\tau)$, the decay times $\tau$ being distributed as a certain function $g(\tau)$ which is defined in this way as an effective density of relaxation times. Taking the derivative $dM/d\log t$ introduces a $t/\tau\ exp(-t/\tau)$ term in the integrand, which is sharply peaked around $t=\tau$. Approximating this peaked function by a δ-function allows bringing out $g(\tau)$ from the integral over τ and yields

$$dM_{tw}/d\log t \propto g_{tw}(\tau=t). \qquad (2)$$

We have now labelled $M_{tw}$ and $g_{tw}$ by $t_w$ to emphasize that each relaxation curve, taken for a given $t_w$, gives access through its logarithmic time derivative to the density of relaxation times that represents the dynamics of the spin glass at a time of the order of $t_w$ after the quench. Thus, each derivative $dM_{tw}/d\log t$ gives an estimate of the density $g_{tw}(\tau=t)$, and as $t_w$ increases $g_{tw}(\tau)$ shifts towards longer times. This gives a physical picture of the 2 important features listed above:

(i) the effective relaxation times are widely distributed (glassy state)

(ii) this distribution peaks around $\tau=t_w$, which implies that for increasing $t_w$'s the relaxation times become longer (aging, the spin glass becomes "stiffer").

We mentioned above that the departures from a perfect $t/t_w$ scaling are systematic as a function of $t_w$. In addition to the thiospinel example in the inset of Fig.2.1, we show in Fig.2.4 the example of a $Ag{:}Mn_{2.7\%}$ spin glass.

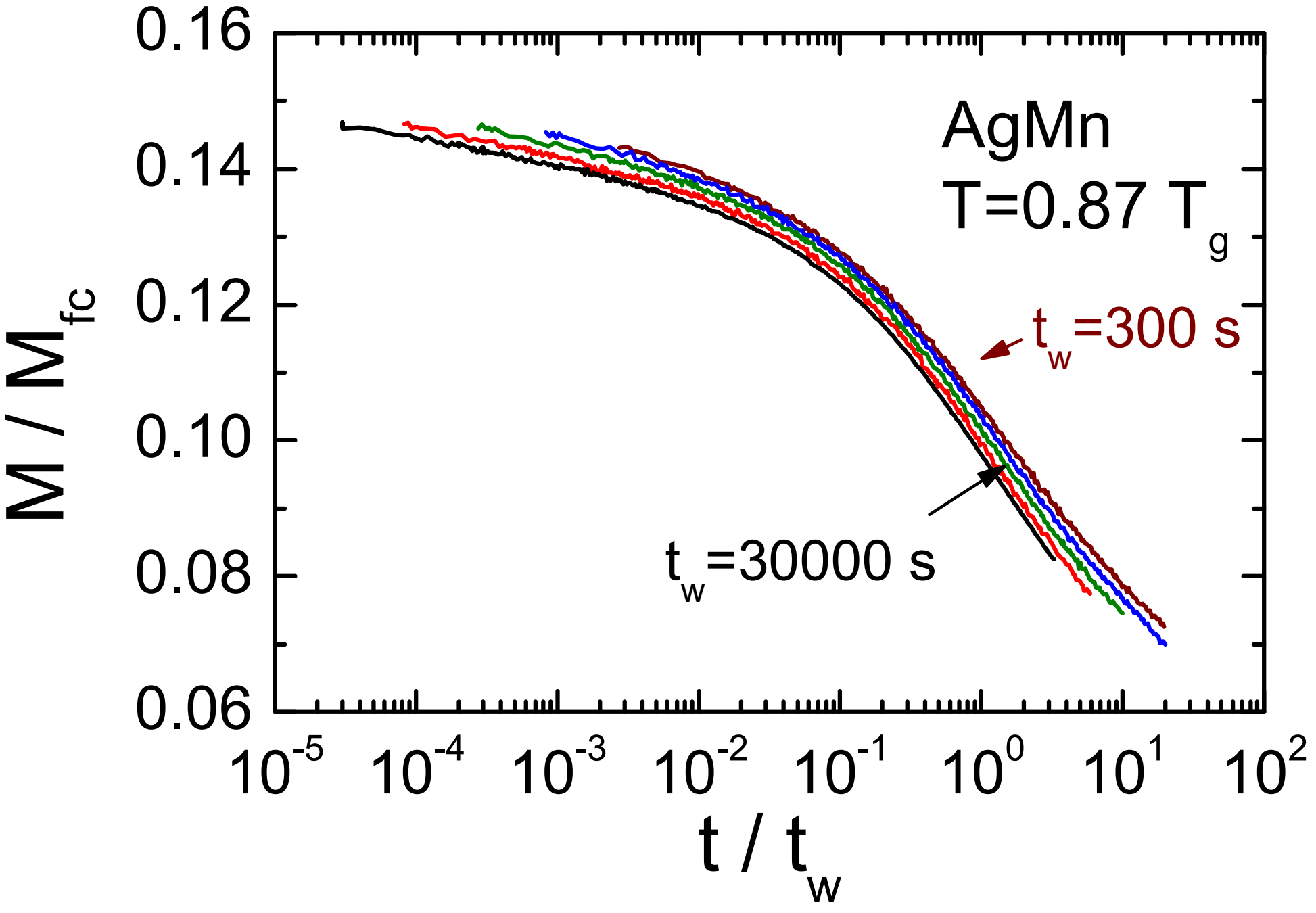

Figure 2.4: TRM relaxation curves of the $Ag{:}Mn_{2.7\%}$ spin glass, plotted as a function of $t/t_w$ (data from [19]). A systematic departure from a $t/t_w$ scaling ("subaging") is observed as a function of $t_w$.

The same trend is observed as in Fig.2.2: as a function of $t/t_w$, the large $t_w$ relaxations decrease faster than the short $t_w$ relaxations. That is, the TRM dependence on $t_w$ is slightly slower than the variation of $t_w$ itself. We call this situation "sub-aging", as opposed to the case of "full aging" that would correspond to full $t/t_w$ scaling.

On a log-scale, the various $t_w$-relaxations are spaced by less than $log\ t_w$ , say by a quantity $\mu\ log\ t_w$ (with $\mu$<1). For increasing $t_w$, the shift of $g_{tw}(\tau)$ towards longer times can therefore be expressed as a shift of the relaxation times that is not exactly $\tau \sim t_w$ but rather $\tau \sim t_w^{\mu}$. But $t/t_w^{\mu}$ itself does not give a full quality scaling of the $t_w$-relaxations. At this point, we have to go one step further than the approximation which consists in defining a density of relaxation times $g_{tw}(\tau)$ at fixed $t_w$ from a given $t_w$-relaxation. Since $g_{tw}(\tau)$ is found to vary with $t_w$, it varies as well during the relaxation itself as a function of time $t$, and the shift of the relaxation times $\tau \sim t_w^{\mu}$ should rather be re-written $\tau \sim (t_w+t)^{\mu}$. This allows the definition of an effective time $\lambda$ [18,19,20], obeying for each individual relaxation process $dm/m$ (of relaxation time $\tau$) to:

$$dm/m = dt/\tau = dt\,/\,(t_w+t)^{\mu} = d\lambda\,/\,t_w^{\mu}. \qquad (3)$$

$\lambda$ defines an artificial time frame in which the spin glass would keep a constant age $t_w$, whereas its age $t_w+t$ constantly increases in the laboratory time frame. Integrating Eq.(3) (setting $\lambda=0$ for $t=0$), $\lambda$ reads

$$\lambda/\,t_w^{\mu} = \{1/(1-\mu)\}\{(t_w+t)^{1-\mu} - t_w^{1-\mu}\} \qquad (4)$$

which reduces to $\lambda \sim t$ for $t<<t_w$.

Then, plotting the relaxation curves of different $t_w$'s as a function of $\lambda/\,t_w^{\mu}$ allows a very precise rescaling onto one unique master curve. This procedure has indeed been first suggested to account for aging in the mechanical properties of polymers. For spin glasses, in more details, the $\lambda/\,t_w^{\mu}$ scaling should be applied to the only aging part of the relaxation, which must be separated from a stationary contribution $\chi_{eq}$ ($\chi_{eq} \sim t^{-\alpha}$, $\alpha$ being a very small exponent, of the order of 0.03-0.1), best evidenced in *ac* experiments (see below) but also present here: $\chi = \chi_{eq} + f(\lambda/\,t_w^{\mu})$ [19]. An example of such a precise rescaling is presented in Fig.2.5.

This rescaling procedure works very well for all known examples of spin glasses. Like in polymers, the exponent $\mu$ is always found lower than one ($\mu \sim 0.8$-0.9, subaging), even if it may sometimes get surprisingly close to 1 (see the example of AgMn in[19], in which $\mu \sim 0.97$ is found, $\mu=1$ remaining excluded by the data with a large range of $t_w$'s explored, from 300 to 30000s). The (simpler) $t/t_w$ scaling with $\mu=1$ can be expected on some rather general grounds [1,23], and the question of the origin of subaging is yet unsolved [24]. It has been proposed that $\mu<1$ arises as an effect of an initial age acquired during the necessarily finite cooling time [25,26,27]. If it is clear that a slower cooling yields a smaller $\mu$, there is no sign in most results (except in the experiment of [25], and for zero cooling time in the numerics of [26]) that $\mu$ could go to 1 for very short cooling times, which always remain long in experiments when compared with microscopic paramagnetic times ($\sim 10^{-12}$s). The dependence of $\mu$ on the amplitude of the magnetic field H has also been carefully checked [28,29].

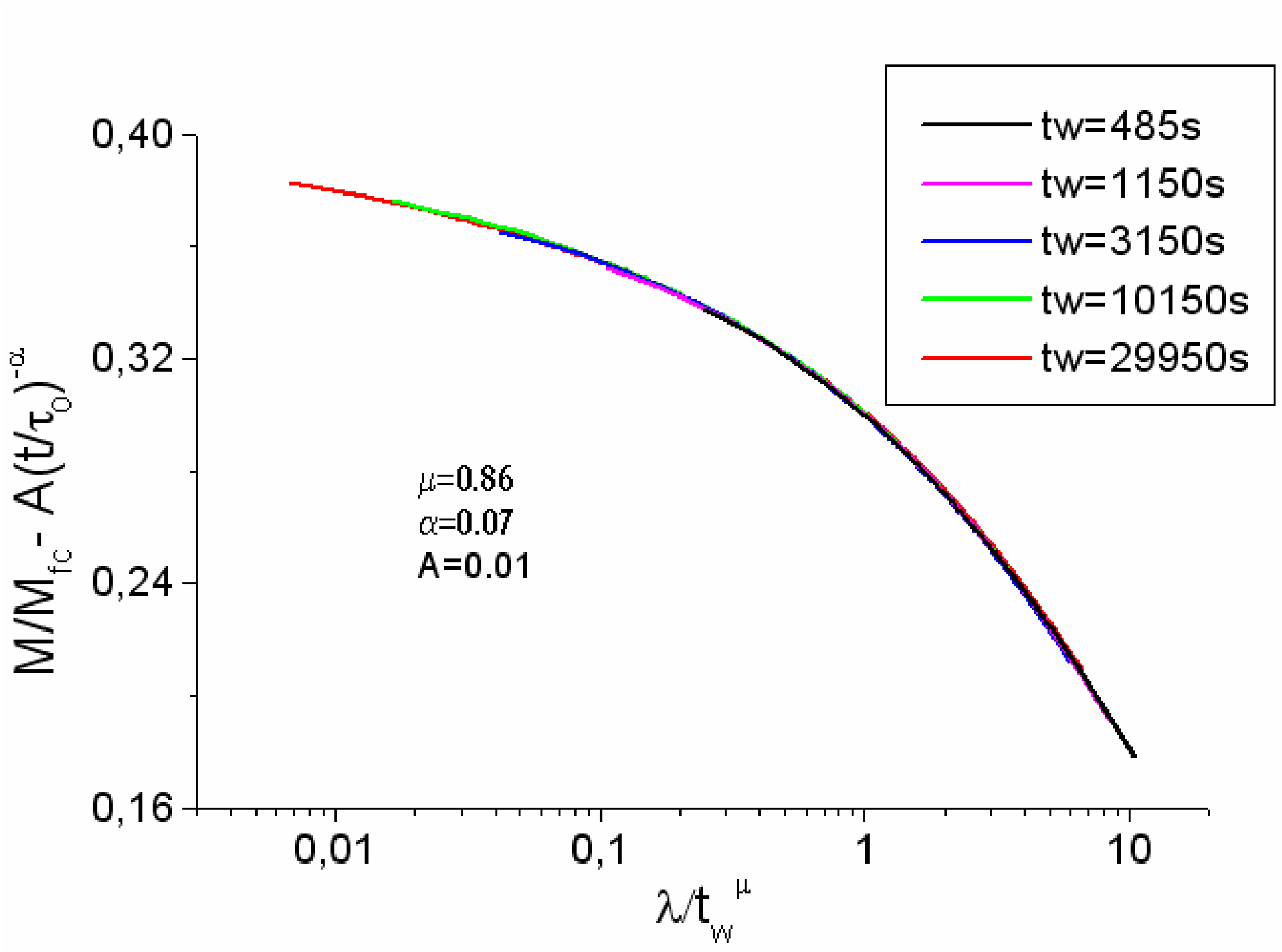

Figure 2.5: Scaling of a set of TRM relaxation curves (thiospinel sample). The aging part of the 5 magnetization curves (obtained by subtracting the stationary part $A(t/\tau_0)^{-\alpha}$ to the total magnetization) shows a fairly good scaling as a function of the reduced variable $\lambda/t_w^{\mu}$.

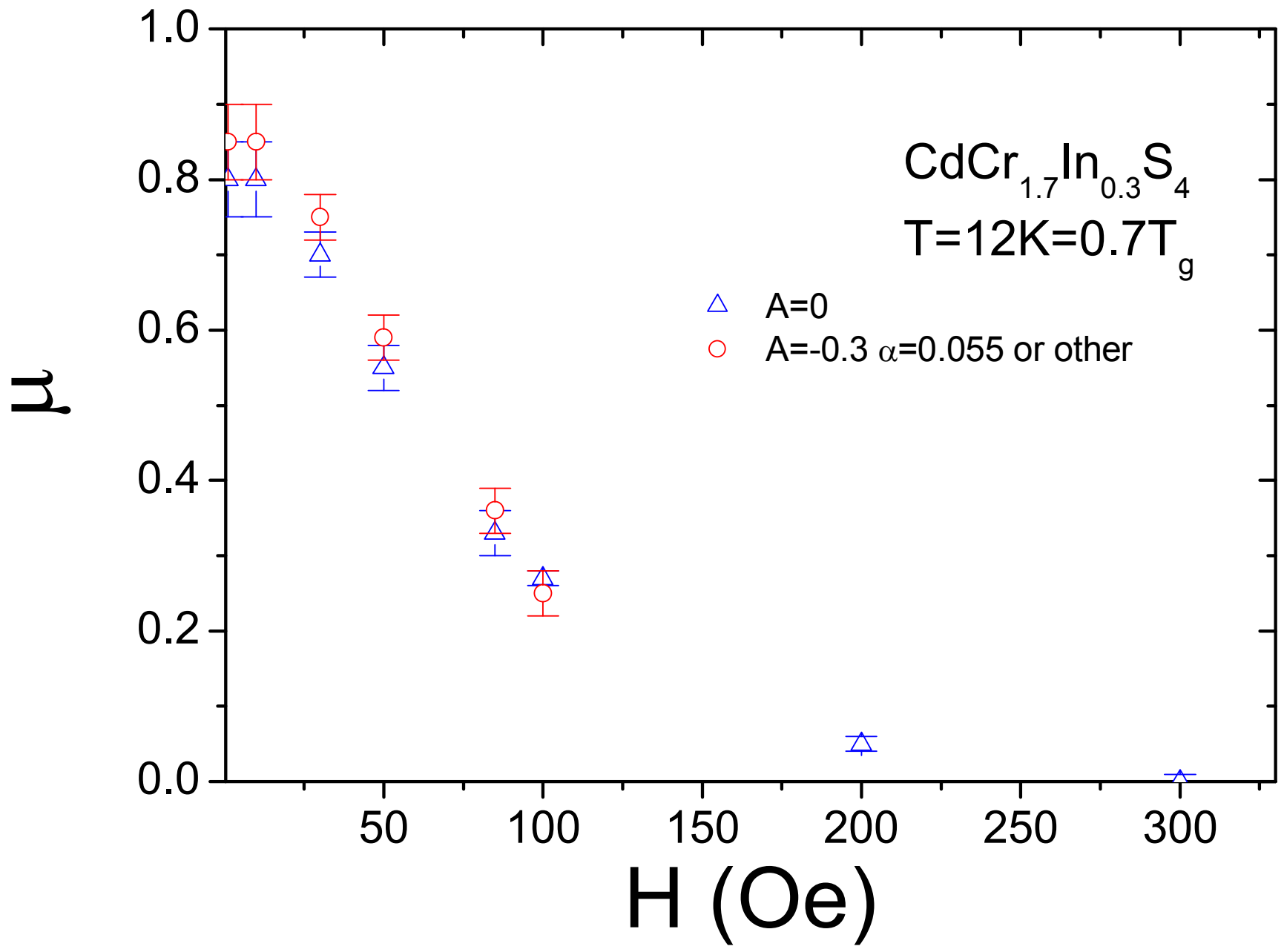

Figure 2.6: Field dependence of the subaging exponent $\mu(H)$ in the thiospinel sample (circles). The triangles show the $\mu(H)$ values obtained when the stationary part of the magnetization is *not* subtracted.

As shown in Fig.2.6, $\mu(H)$ decreases for increasing field, but for vanishing field it seems very unlikely that $\mu$ goes to 1. This region could be precisely explored in experiments by Ocio and Hérisson who took data for fields as low as 0.001 Oe [29]. $\mu$ is found at a plateau value of ~0.85 in the range 10-0.001 Oe (five decades). On the other hand, for increasing fields, $\mu$ eventually goes to zero [28], which means that the field change is enough to erase the effect of previous aging (H>300 Oe in Fig2.6). Let us note that above this value there may still be some slow relaxations (although with no $t_w$-dependence), and that the instantaneous, paramagnetic-like, response to the field is only obtained for still higher fields (600 Oe in the case of Fig.2.6).

Finally, it might well be that $\mu<1$ be related to some finite size effects, as proposed in [30,31], as the result of a saturation of aging in some small parts of the sample (grains?), while larger parts would obey $\mu=1$ for astronomical times. This possible explanation could however not be confirmed experimentally.

An amusing example of subaging has been studied in the rheology of a microgel paste [32], which is of the type used as toothpaste. Here the notion of a freezing temperature is not relevant, but instead the initial state of aging is obtained by applying a strong shear stress, which turns the paste into a fluid, whose viscosity then progressively increases with time (so toothpaste does flow out of the tube when it is pressed, but does not flow from the toothbrush to the ground).

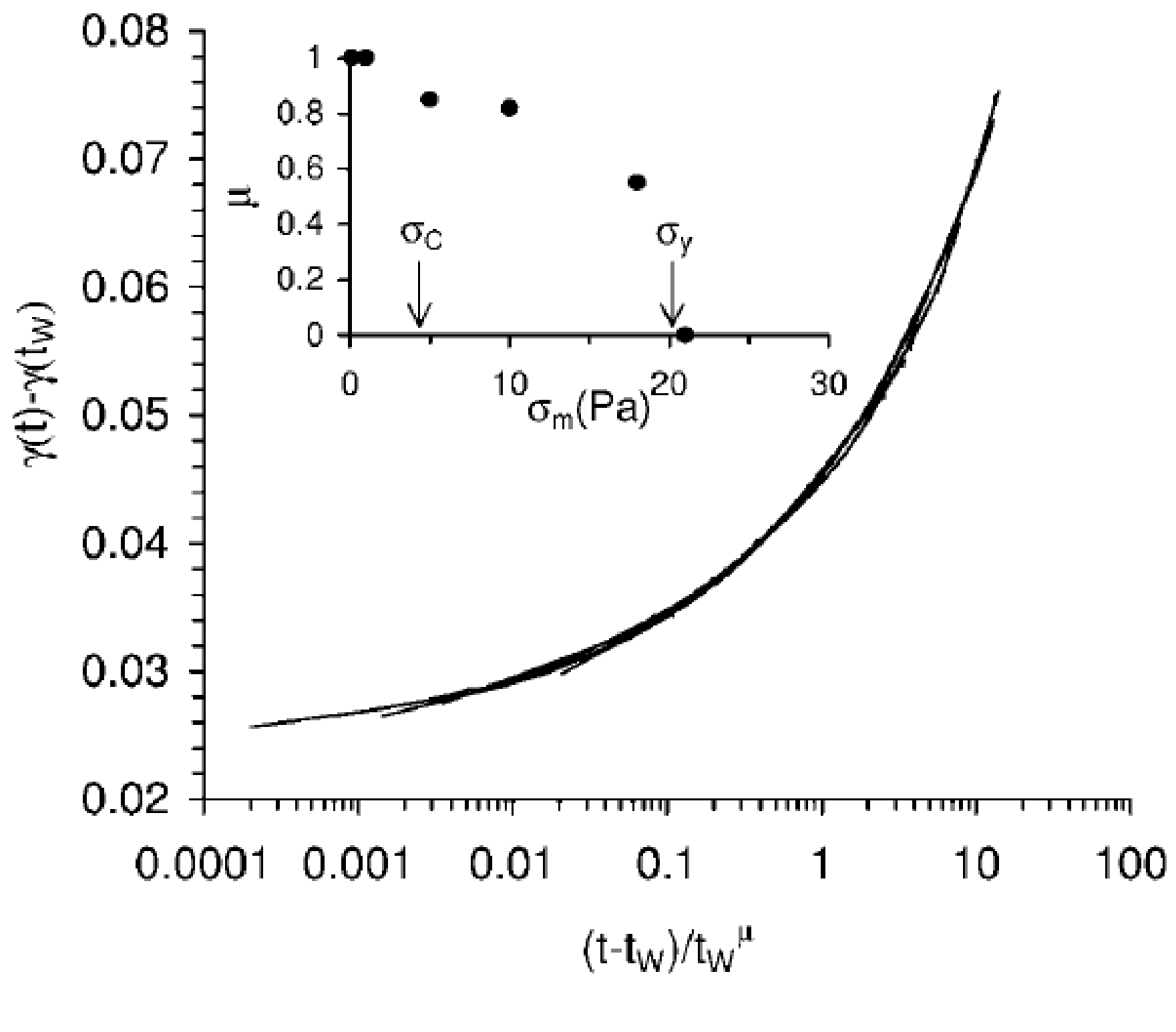

Figure 2.7: From [32], creep curves of a soft deformable microgel (similar to toothpaste), measured at different waiting times ranging from $t_w$ = 15s to 10000s, for a probe stress $\sigma_m$= 10 Pa greater than the yield stress $\sigma_c$, above which the suspension begins to flow. The inset shows the evolution of the subaging exponent $\mu$ as the probe stress increases up to $\sigma_y$, above which aging disappears ($\mu=0$).

At low stresses, the response to a shear excitation is a long-time creep curve, which is slower when the experiment is performed after a longer waiting time. The resulting curves (Fig.2.7) have been scaled together as a function of $t/t_w^{\mu}$ (not far from $\lambda/ t_w^{\mu}$), and $\mu$ is found to decrease as a function of increasing stress, like in glassy polymers, and like in spin glasses as a function of the amplitude of the magnetic field (Fig.2.6). Similar results have been obtained these last years in various examples of colloidal gels [33,34,35].

## 2.2 AC susceptibility

Slow dynamics and aging in the spin-glass phase can also be observed by *ac* susceptibility measurements, in which a small *ac* field (~1 Oe) is applied all along the measurement. Again, the starting point of aging experiments consists in cooling the spin glass from above $T_g$, down to some $T<T_g$ at which the *ac* response is measured as a function of the time elapsing, which is the "age" of the system (equivalent to $t_w+t$ in the *dc* procedures). We find here the same 2 characteristics as observed in *dc* experiments:

(i) the *ac* response is delayed, i.e. the susceptibility has 2 components: an in-phase one $\chi'$, and an out-of-phase one $\chi''$. $\chi''$ is zero above $T_g$ (paramagnetic phase), and rises up as the sample is cooled into the spin-glass phase.

(ii) the susceptibility relaxes down, signing up the occurrence of aging. This relaxation is visible on both $\chi'$ and $\chi''$, but is more important in relative value $\Delta\chi/\chi$ in the out-of-phase component $\chi''$.

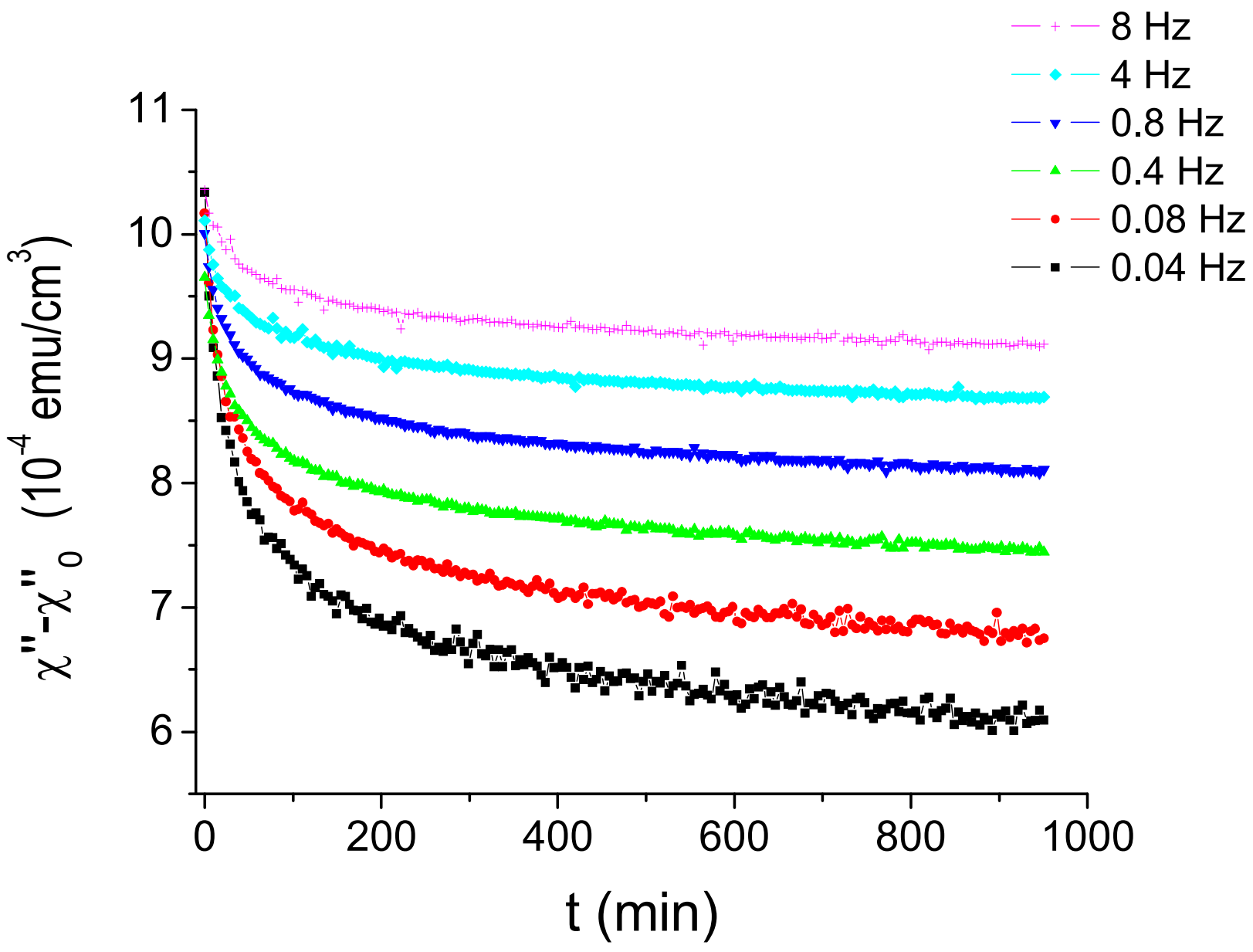

Figure 2.8: Time decay of the out-of-phase susceptibility $\chi''$ of the thiospinel sample after a quench (aging), for different frequencies. The curves have been shifted vertically by an arbitrary amount $\chi''_0$ for the sake of clarity.

Fig.2.8 shows the $\chi''$-relaxation as a function of time for different frequencies $\omega$. A very clear frequency dependence is seen in Fig.2.8: the amplitude (in the fixed experimental time window) of the observed relaxation increases as the frequency $\omega$ decreases. On the other hand, the infinite time limit of $\chi''$ seems very convincingly to be non-zero, pointing out to a finite $\chi''_{eq}$ stationary limit. Once shifted vertically by an arbitrary amount (that should correspond to $\chi''_{eq}$) and plotted as a function of the reduced variable $\omega.t$, the curves can be superposed. Actually, in this *ac* experiment, $1/\omega$ is the typical observation time and plays the same role as $t$ in the *dc* relaxation procedures. The total age of the system is here the time $t$ along which the *ac* relaxation is measured after cooling, equivalent to $t_w+t$ in the *dc* experiment. Hence, the present $\omega.t$ scaling is equivalent to the $t/t_w$ scaling of the *dc* experiments [19]. Strangely enough, there is no sign of subaging ($t_w^{\mu}$ in place of $t_w$) in the scaling behaviour of the *ac* data. Indeed, the superposition of the *ac* curves is not as constraining as that of a series of TRM relaxations over a large range of $t_w$*'s*. But any attempts of an $\omega.t^{\mu}$ scaling of the *ac* data have favoured $\mu\sim1$. One difference with *dc* experiments which may be pointed out is that *ac* measurements are necessarily performed in the $\omega.t\geq1$ regime (sometimes called "quasi-stationary" regime), that corresponds to the limited region of $t/t_w<1$ in *dc* experiments. The possibility of a link between the observation of a subaging behaviour and the time regime explored in the experiments remains open [19].

Similar *ac* procedures are used in the study of structural and polymer glasses. For instance, in [36], the dielectric constant ε of glycerol has been measured following the same procedures as above. The out-of-phase susceptibility ε" shows a strong relaxation as a function of the time following the quench (Fig.2.9).

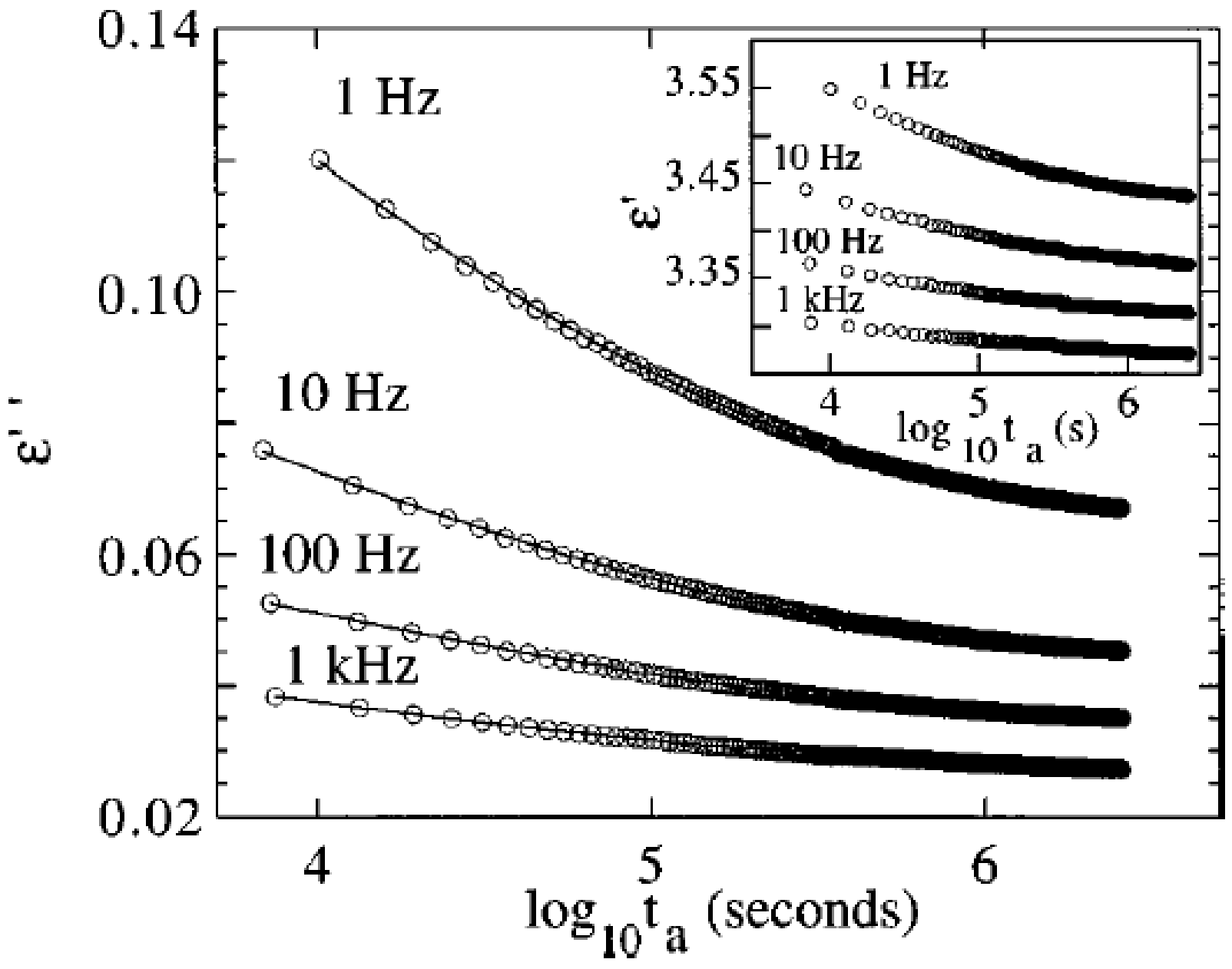

Figure 2.9 : From [36], relaxation of the dielectric constant of glycerol at 178K, as a function of the time following the quench from above $T_g$=190K (aging), for various frequencies. The main part of the figure shows the out-of-phase component *ε"*, the inset shows the in-phase component *ε'*. The amplitude of the relaxation is larger at lower frequencies (same qualitative trend as in spin glasses).

The relaxation has at least the same qualitative frequency dependence as observed in spin glasses: the lower the frequency ω, the larger the relaxation in a given time window. The authors state that no $\omega t$-like scaling is obeyed [36]; however, in the case of this structural glass, one cannot exclude that the influence of the cooling time, probably stronger than in spin glasses, may bring corrections to the effective value $t_{eff}$ of $t$ which could finally yield an $\omega t_{eff}$ scaling.

## 2.3 Noise measurements

The measurement of noise in spin glasses has been a high-level challenge for the experimentalists, because the spontaneous magnetic fluctuations are tiny when compared with the magnetization obtained in response to an external field (in the recent experiment described below, they are equivalent to the response to a field of $\sim 10^{-7}$ Oe). We only recall here the general lines of these remarkable experiments, developed by M. Ocio. The interested reader will refer to his corresponding papers [29,37].

The response to a magnetic field, whose investigation was detailed above, is related to the spontaneous magnetic fluctuations via the Fluctuation-Dissipation relation (FDR), established for ergodic systems at equilibrium. In its integrated form, it relates the relaxation function $\sigma(t',t)$ ($\sigma = m/h$, response at $t$ after cutting off a field $h$ at $t'$, same as the TRM) to the autocorrelation $C$ of the fluctuations of the magnetization $m$, namely $C(t',t)=\langle m(t').m(t)\rangle$:

$$\sigma = C/k_B T\,. \qquad (5)$$

A lot of work has been devoted to extensions of FDR to non-equilibrium situations, for which the aging regime of the spin glass is archetypal [38,39]. A prominent result by Cugliandolo and Kurchan [38] is a modified FD relation which reads

$$\sigma = C.F(C)/k_B T \qquad (6)$$

where $T/F(C)$ takes the meaning of an effective temperature. In this approach, for large $t'$, the obtained correction factor $F(C)$ is a function of the autocorrelation $C$ only, i.e. it does not explicitly depend on $t$ and $t'$ but has a time dependence through the value of $C(t',t)$ only.

This result was one of the strong motivations of the recent noise experiments performed by M. Ocio and D. Hérisson [29]. A decade before, the very first noise measurements were performed by M. Ocio and Ph. Refregier in collaboration with H. Bouchiat and Ph. Monod [37]. In these pioneering experiments, the Fourier transform of the noise could be measured, and compared with the *ac* susceptibility, that was measured in another setup. This early work suffered two limitations: firstly, the comparison between noise and response could only be made up to an unknown calibration factor, and secondly the time regime was limited to the quasi-stationary region $\omega t>1$ (as opposed to the "strongly aging" regime explored in TRM experiments). The results was that the FDR was obeyed as far as could be checked [37].

In the new set of experiments [29], a special setup which allows both types of measurements in the same geometry has been built. For noise measurements, the pickup coil ($3^{rd}$ order gradiometer geometry) which contains the sample is "simply" connected to a *dc* SQUID, and the full signal is recorded as a function of time (not only its Fourier transform). The response function is investigated in the strongly aging regime by means of TRM-relaxation recordings. One bright idea was to use the pickup coil itself as an excitation coil, through which the field is applied by induction of a current in the pickup loop. Thus, the magnetic geometry (rather complex in a gradiometer) is exactly the same for the detection of the magnetization fluctuations as for applying the excitation field, allowing a direct comparison between fluctuations and response.

In order to cancel the self-inductive response to the field variation which triggers the TRM relaxation, a bridge configuration is used for response measurements, in which the main branch involving the sample is balanced by an equivalent one without sample, excited oppositely. The whole experiment is placed in a magnetic shield which lowers the residual magnetic field below $10^{-3}$ Oe. Important care was also taken for eliminating all electromagnetic parasite sources, as well as external low-frequency disturbances such as those accompanying the day-night cycle of the laboratory.

Finally, the measurements were made possible. An absolute calibration was realized with the help of an ultra-pure copper sample, in which the magnetic response and the fluctuations of eddy currents are related through classical (ergodic) FDR. With an ergodic sample like copper, this setup constitutes an absolute thermometer after calibration by only 1 fixed point (one of the unrealized projects of M.Ocio was to develop the use of this method for absolute thermometry).

An example of noise recordings is presented in Fig.2.10.

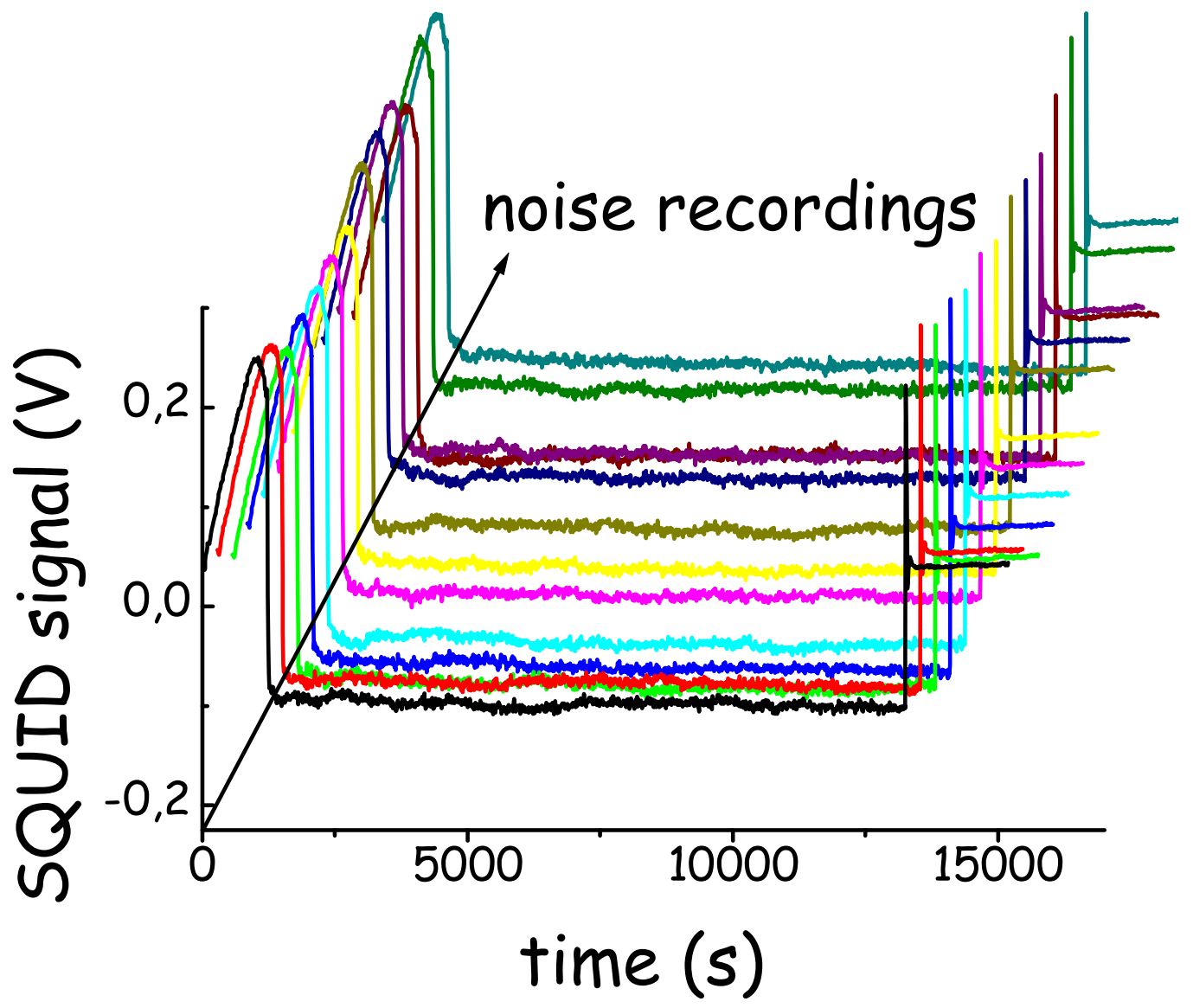

Figure 2.10 : SQUID signal (proportional to the magnetization, with an arbitrary offset voltage) in a series of successive noise recording experiments (data from [29]). Each experiment starts from above $T_g$ ; due to the slight residual field, the magnetization shows a peak when crossing $T_g$.

Each trace shows the SQUID output (proportional to the sample magnetization, with an arbitrary offset) during one experiment, starting from above $T_g$ and cooling. Due to the slight residual field, the trace shows the magnetization peak observed when crossing $T_g$. After cooling, the temperature is stabilized at say $T=0.7T_g$, and the magnetization fluctuations are recorded as a function of time during ~$10^4$ s. After that the sample is re-heated above $T_g$. The experiment is repeated ~300 times. On each of the recorded traces, for any choice of times $(t_w,t)$ the correlation $m(t_w).m(t_w+t)$ can be computed. This value is of course strongly fluctuating from one experiment to the other, but the average over ~300 measurements is taken and after properly subtracting offsets the autocorrelation $C(t_w,t)=< m(t_w).m(t_w+t)>$ is obtained. It is represented in the top part of Fig.2.11 in the same way as usual TRM results, that is as a function of $t$ for various fixed values of $t_w$ .

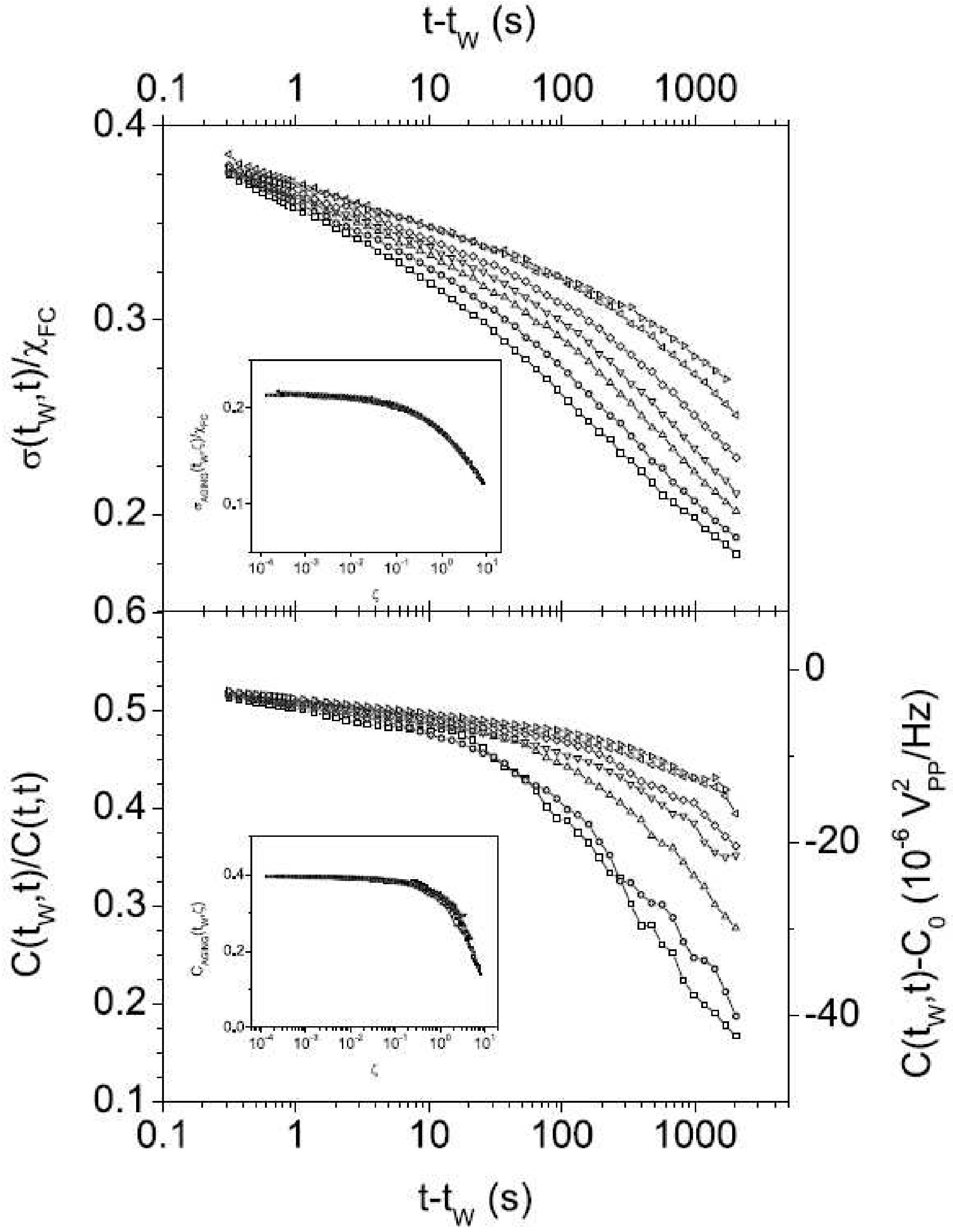

Figure 2.11 : From [29], TRM-relaxation (top) and autocorrelation (bottom) functions, recorded at 13.3K with the thiospinel sample. The different curves correspond, from bottom to top, to $t_w$ = 100, 200, 500, 1 000, 2 000, 5 000 and 10 000 s. The insets shows the respective ageing parts, deduced by the scaling analysis (see text), and plotted as a function of the reduced time variable $\zeta=\lambda/t_w^{\mu}$.

The bottom part of Fig.2.11 shows in the same representation the results obtained from the TRM experiments performed in the same setup, with an excitation field of ~$10^{-3}$ Oe. The two insets show that both noise and response functions obey the same scaling law as a function of the reduced variable $\zeta=\lambda/t_w^{\mu}$ (the same fitting parameters can be used). The comparison between both sets of results is best illustrated in the plot of Fig.2.12, in which the response function $\sigma(t_w,t)$ (or the susceptibility $\chi=1-\sigma$) is plotted as a function of $C(t_w,t)$ for 3 different temperatures $T=0.6$, $0.8$ and $0.9$ $T_g$. See [29] for the details of normalization of $C(t_w,t)$.

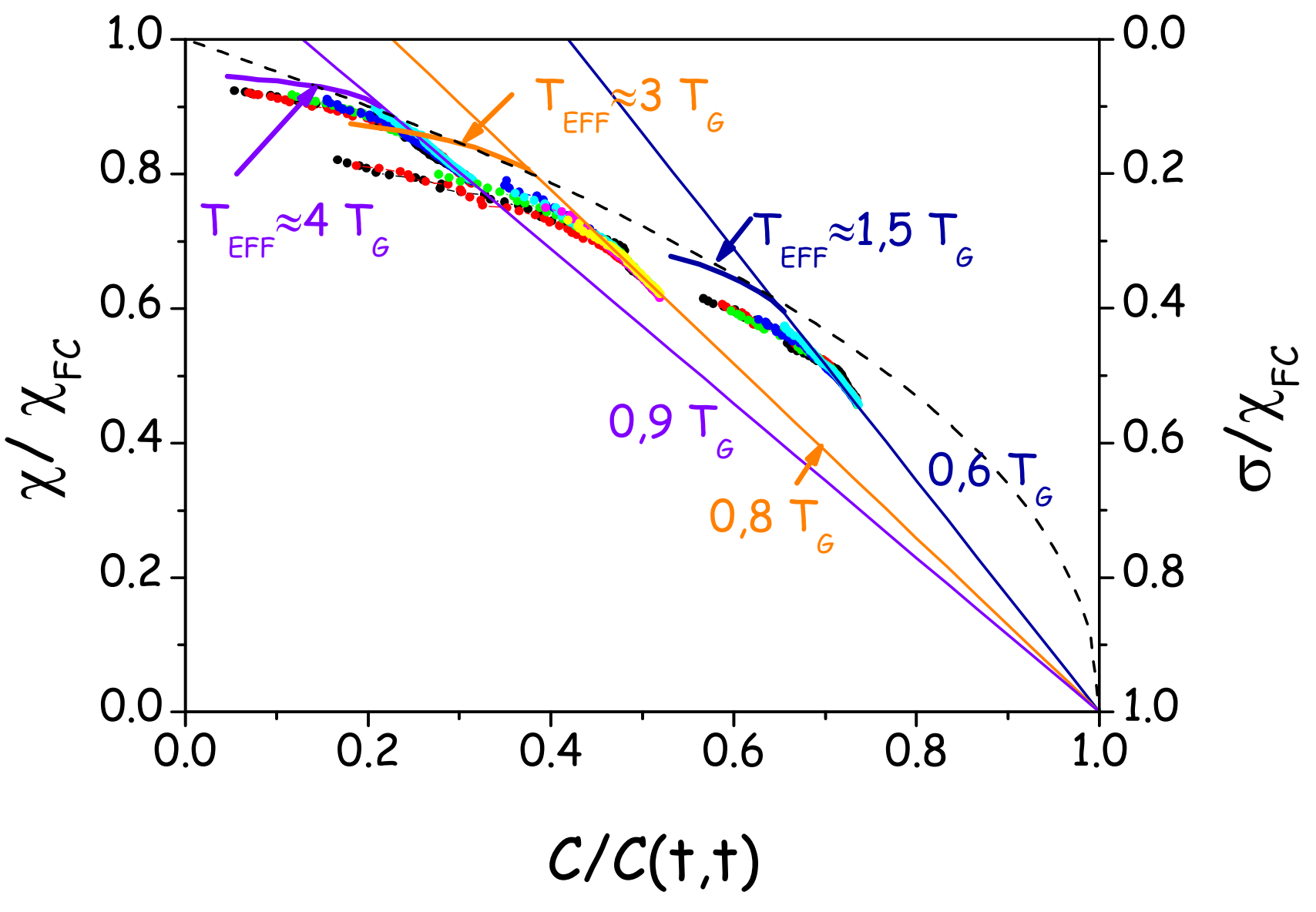

Figure 2.12 : Response function versus autocorrelation, for 3 different temperatures T (data from [29]). The straight lines show the 1/T equilibrium regime. The points are the raw results. The curves are obtained by subtracting the stationary part (equivalent to a long time extrapolation of the data). The dashed line is a $\chi=(1-C)^{0.47}$ fit, in reference to the continuous RSB model [40].

For each of the 3 temperatures, the point cloud is the set of "raw" results obtained for various values of $(t_w,t)$. The straight lines with $1/T$ slope show the expected result when the classical FDR is obeyed with no correction. There is a clear $1/T$ regime for the higher values of $C$, and the results show a crossover towards a weaker slope $1/T_{eff}$ with $T_{eff}>T_g$ as $C$ decreases. These deviations show the first experimental observation in a spin glass of deviations from the normal FDR in the aging regime.

In order to make a more quantitative comparison with the theoretical predictions [38], it is necessary to extrapolate the results in the very long time region. An estimate of this very long time behaviour can tentatively be obtained by extrapolating the existing data to the region where the stationary part of the relaxation $t^{-\alpha}$ has relaxed to zero, i.e. by subtracting to $\sigma$ the stationary part which has been obtained on the basis of a precise rescaling of the curves (as shown in the inset of Fig.2.11). This is shown in Fig.2.12 in solid curves, which are indeed the superposition of the different curves obtained for various $t_w$'s. The different curves are indistinguishable within the present accuracy, which strongly suggests (in the framework of this crude extrapolation) that the correction factor $F(C)$ to the FDR is only a function of $C$, as predicted in [38].

It may be risky to push much further the comparison at this stage, since the extrapolation to long times is problematic, and also there remain some difficulties with the normalization of $C(t_w,t)$ by $C(t,t)$ [29]. One point which is out of doubt is that the data in the aging region do not tend to favour a horizontal slope, as expected in domain growth type models (infinite $T_{eff}$). The observed mean slopes correspond to $T_{eff}(0.6T_g)\sim1.5T_g$, $T_{eff}(0.8T_g)\sim3T_g$, and $T_{eff}(0.9T_g)\sim4T_g$. However, the extrapolated data show some curvature, and do not look like straight lines as would be expected from 1-step RSB type models of spin glasses [1]. In continuous RSB models like the mean-field spin glass [2], a $\chi=(1-C)^{1/2}$ behaviour is predicted [40].The dashed line in Fig.2.12 shows a $\chi=(1-C)^{0.47}$ fit which gives at least a rough account of the results. The next step in this discussion of the first directly comparable noise and response data may arise if a direct experimental determination of $C(t,t)$ is obtained, for example from neutrons scattering data [41].

The autocorrelation function may be more easily accessible in colloidal systems. In the case of colloidal gels, the technique of multispeckle dynamic light scattering allows the direct determination of the dynamical structure factor $f(q,\tau)$, which is the autocorrelation function of the density fluctuations over a time $\tau$ and at a length scale $2\pi/q$. In [34], this autocorrelation has been found to present interesting similarities with the magnetization autocorrelation (and response function) of spin glasses (Fig.2.13).

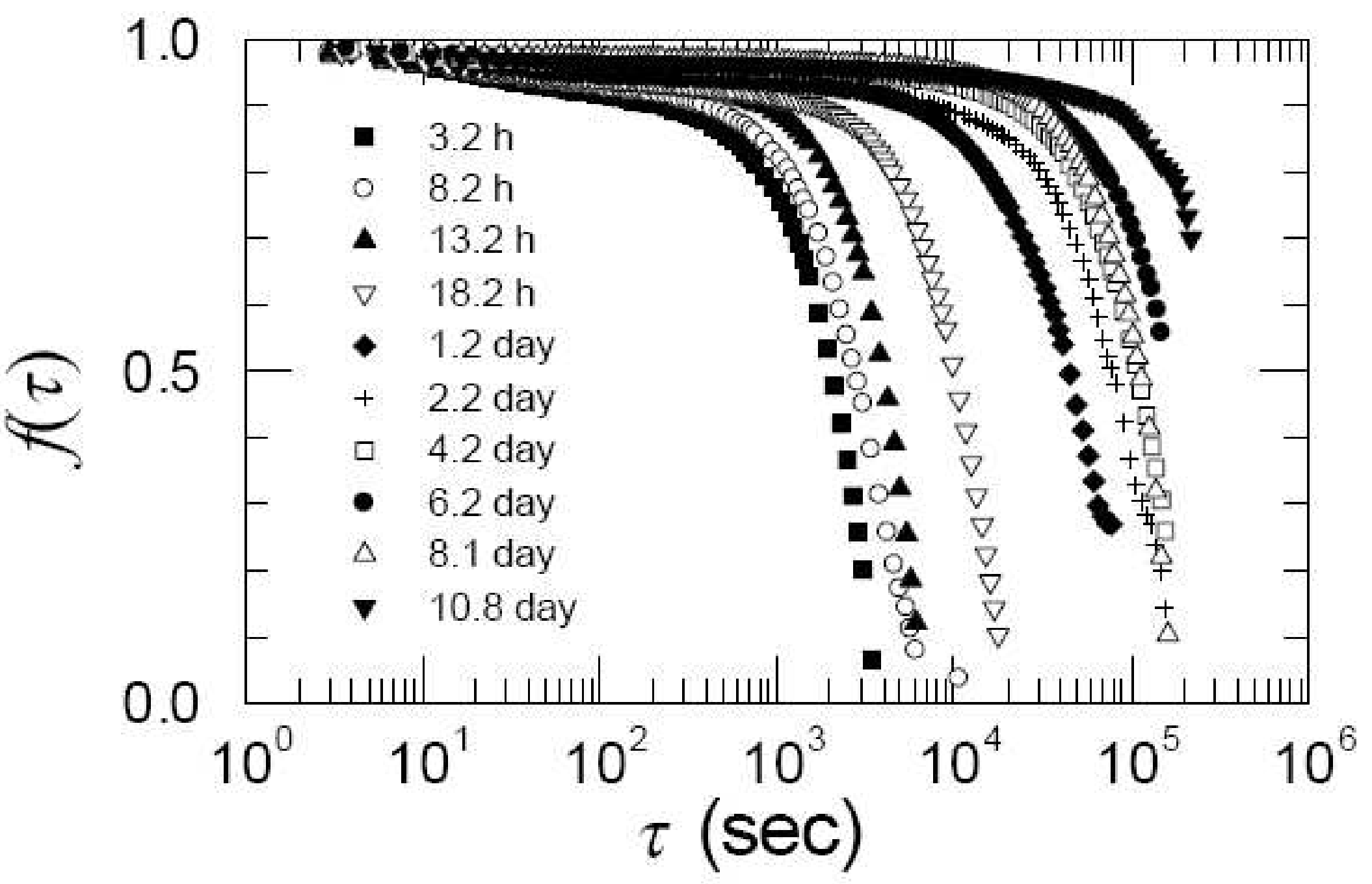

Figure 2.13 : From [34], time evolution of the dynamic structure factor of a gel, measured by multispeckle dynamic light scattering at $q$=6756 cm$^{-1}$. The curves are labelled by the gel age $t_w$.

At fixed $q$ (in the above spin-glass case $q=0$) the time decay of $f(q,\tau)$ has the unusual form $f(q,\tau)\sim exp\{-(\tau/\tau_f)^{1.5}\}$, but like in spin glasses the autocorrelation depends on the waiting time $t_w$ during the gel restructuration through $\tau\sim t_w^{0.9}$ (subaging).

## 2.4 Rejuvenation by a stress

Before turning to the rejuvenation effects which are observed in spin glasses in response to temperature changes, let us mention that a certain kind of rejuvenation effects has been known for a long time in the rheology of glassy materials in response to a mechanical stress [20], and that the equivalent of these phenomena in spin glasses can be traced out in the effect of a (sufficiently strong) variation of the magnetic field [28].

Fig.2.14 shows a typical aging experiment in which the volume relaxation following the quench of an epoxy glass sample is measured [42].

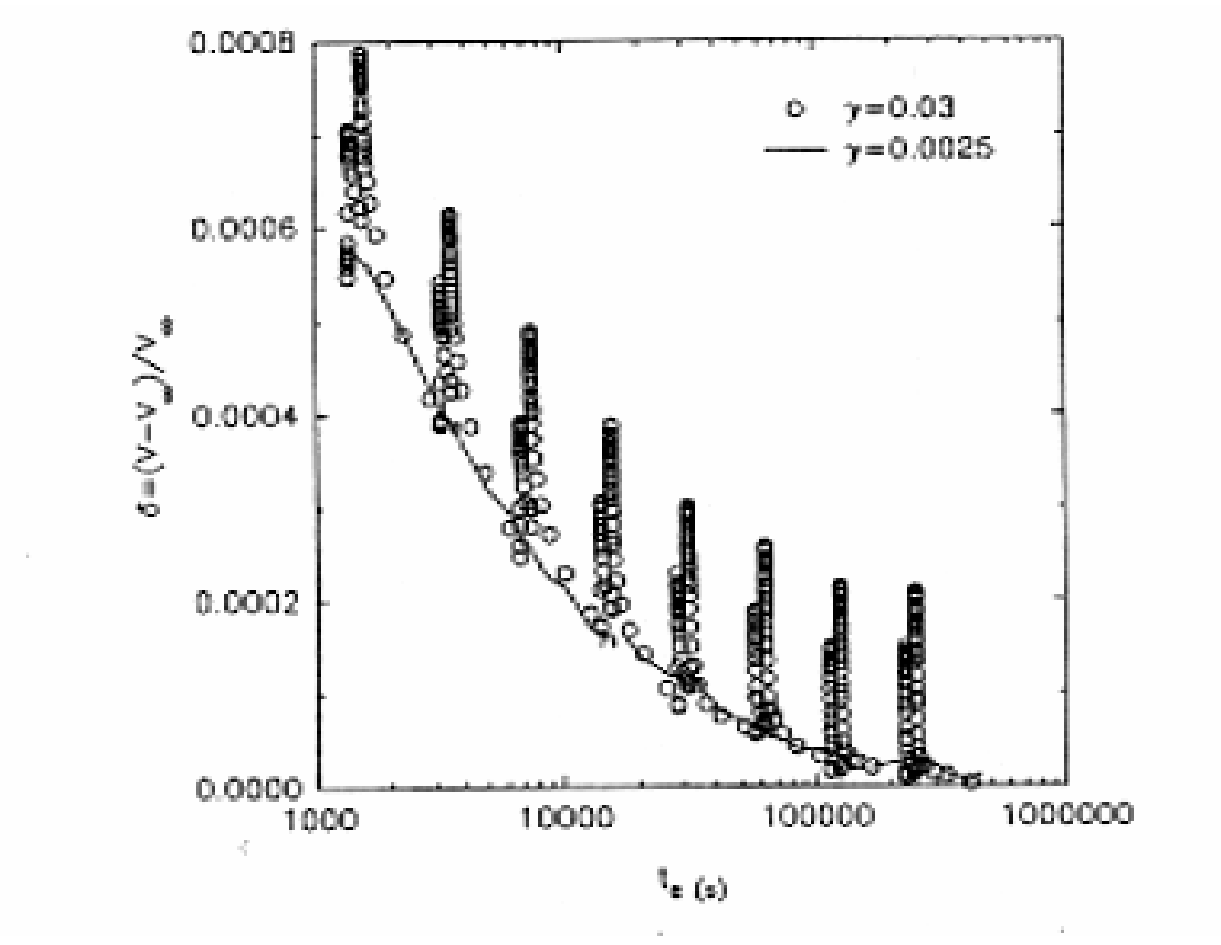

Figure 2.14 : From [42], volume relaxation (in fraction of the long time value) of an epoxy glass sample after a quench. At some times, a stress of amplitude $\gamma$ is applied. For a low stress value (solid line), there is no visible effect. For a higher stress value (circles), rejuvenation can be seen as spikes.

This volume relaxation accompanies the stiffening of the mechanical properties during aging of all structural and polymer glasses. In the experiment of Fig.2.14, at some times a stress of amplitude $\gamma$ is applied. The solid line corresponds to a low $\gamma$ value, for which the stress has no visible effect. But, for a higher $\gamma$ value (open circles), a phenomenon called "rejuvenation" is observed: suddenly the volume increases, and the relaxation is renewed, starting from a value corresponding to a "younger age".

A similar phenomenon can be seen in spin glasses. Fig.2.15 shows an *ac* experiment [28] in which, after 300 min, a *dc* field H=30 Oe is applied (in comparison, the *ac* field, which does not influence aging here, is $H_{ac}=0.3\ Oe$).

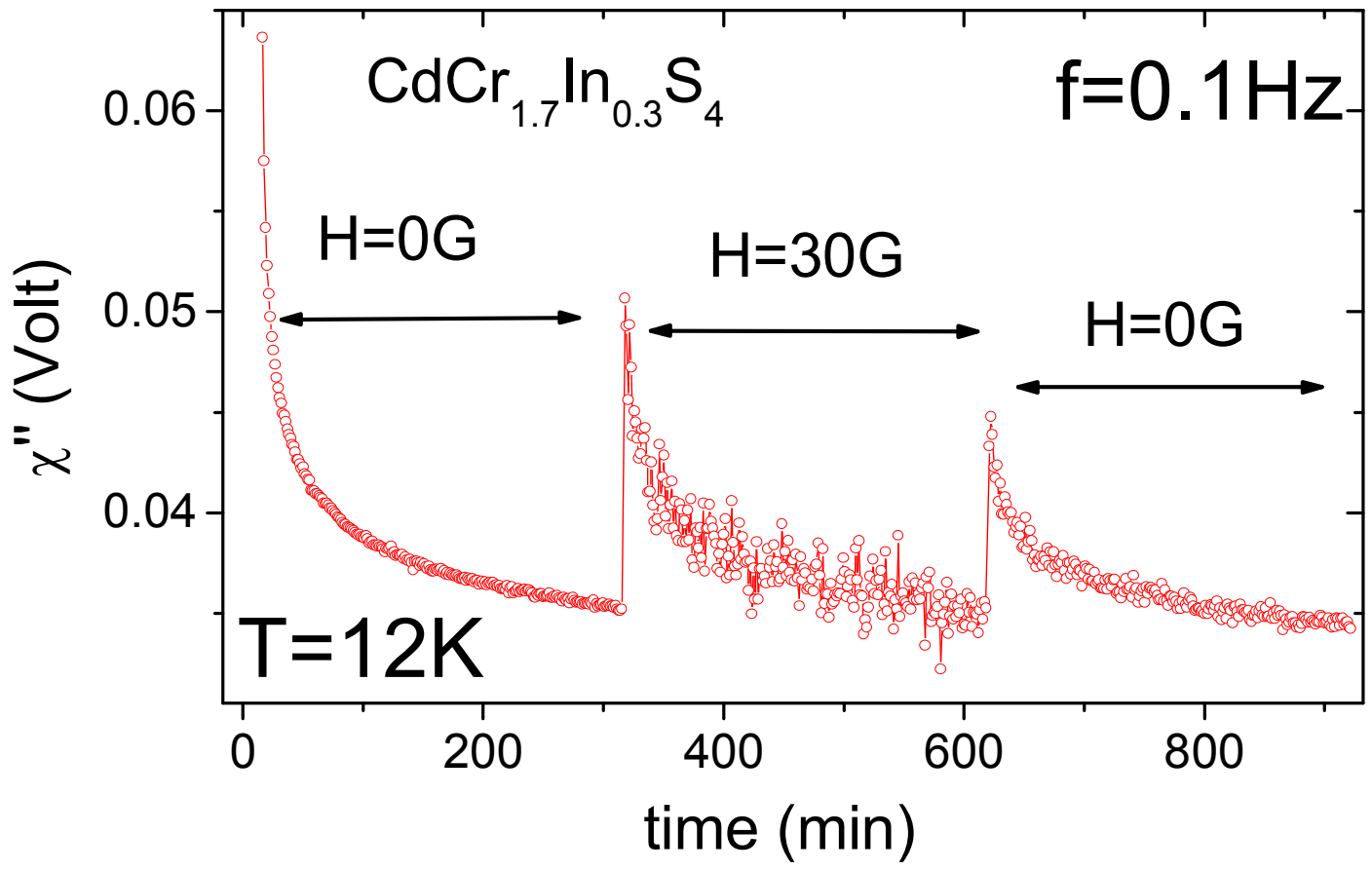

Figure 2.15: AC out-of-phase susceptibility of the thiospinel spin glass after a quench. An additional *dc* field is applied in the middle part of the experiment, inducing rejuvenation (data from [28]).

The slow relaxation of $\chi''$, which is characteristic of aging, shows a sudden drop when the *dc* field is applied, and restarts from a "younger state". When the *dc* field is turned back to zero, a weaker but similar drop is observed. The Zeeman coupling of the spins to the *dc* field in this experiment is strong enough to overcome the more subtle spin rearrangements which progressively occurred during aging as a result of the local minimization of interaction energies. Hence, part of the effect of aging is erased by applying the *dc* field, and aging (partly) restarts from new (rejuvenation effect). The same effect is also visible in the *dc* (TRM) experiments presented above; as shown in Fig.2.6, the $\mu$ exponent of the $t_w$-scaling decreases with the amplitude of the field used for the TRM-relaxation, or in other words, the influence of $t_w$ becomes weaker and weaker as a stronger field perturbation is applied (in the limit $\mu=0$ there is no $t_w$ effect). In the toothpaste experiment (Fig.2.7 [32]), as the shear stress amplitude increases, $\mu$ also decreases.

It is likely that this effect of the magnetic field on a spin glass is the equivalent of the effect of a mechanical stress on a glass, in which the slow rearrangements of atoms (or polymers, or micro-spheres or discs in a colloid) during aging are partly destroyed by applying a shear or elongation stress. The rejuvenation effects as a function of temperature that we present in the next chapter pertain to a different class of phenomena, with the possibility of obtaining almost independent aging evolutions at different temperatures, and memory effects despite rejuvenation.

# 3. Aging, rejuvenation and memory

## 3.1 Cooling rate effects

The state of a glass is strongly influenced by the way it has been cooled. What one usually has in mind is the kind of picture that is shown in Fig.3.1 [43], displaying the evolution of a thermodynamic quantity like the enthalpy or the specific volume as a function of temperature during the cooling process.

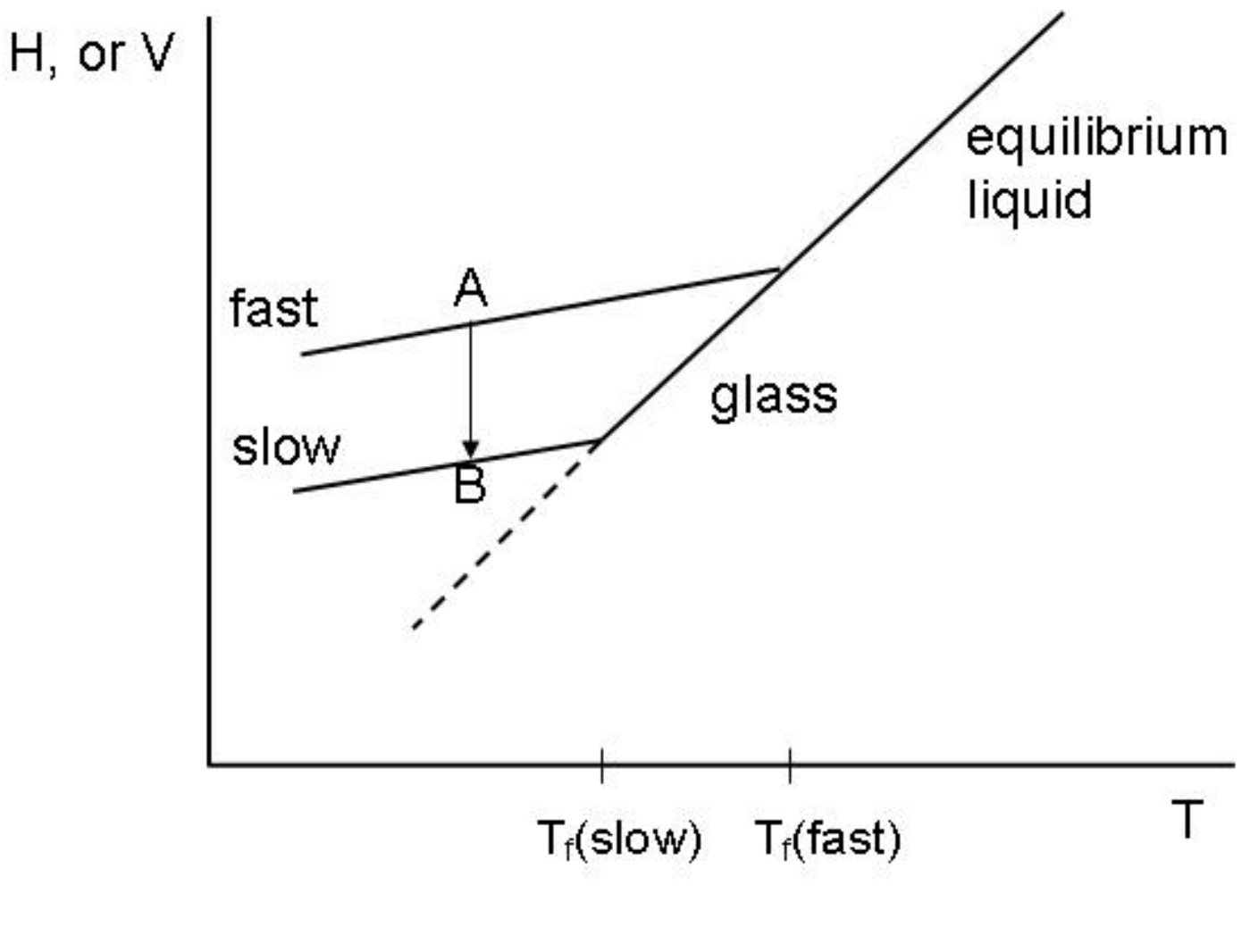

Figure 3.1: Sketch of the typical enthalpy or volume variation with temperature in a glass (freely inspired from e.g. [43]). During cooling, the liquid falls out of equilibrium at a freezing temperature $T_f$ which depends on the cooling rate (“fast”, or “slow”), and becomes a glass. After a fast cooling to point A, aging over very long times will eventually bring the glass to point B, which can be attained much more quickly by a slow cooling.

Above the freezing temperature $T_f$, the glass follows the equilibrium line in the graph of Fig3.1, but when crossing $T_f$ it falls out of equilibrium, reaching a state in which the enthalpy relaxes down slowly (aging). $T_f$ is of course only dynamically defined: for a faster cooling, $T_f$ is higher, and a slower cooling allows the glass to follow the equilibrium line down to lower temperatures. Following the scheme of Fig.3.1, a state B that would be attained after rapidly cooling to A and aging for a long time could more easily be obtained by a slower cooling.

This view of glasses was the starting point of experiments in spin glasses in which we explored how the aging behaviour could be influenced by the temperature history, having in mind that well-suited cooling procedures might bring the spin glass into a strongly aged state which otherwise would require astronomical waiting times to be established [44]. These experiments have brought important surprises. The one presented in Fig.3.2 is representative of the unexpected features which were found in the spin-glass behaviour [45].

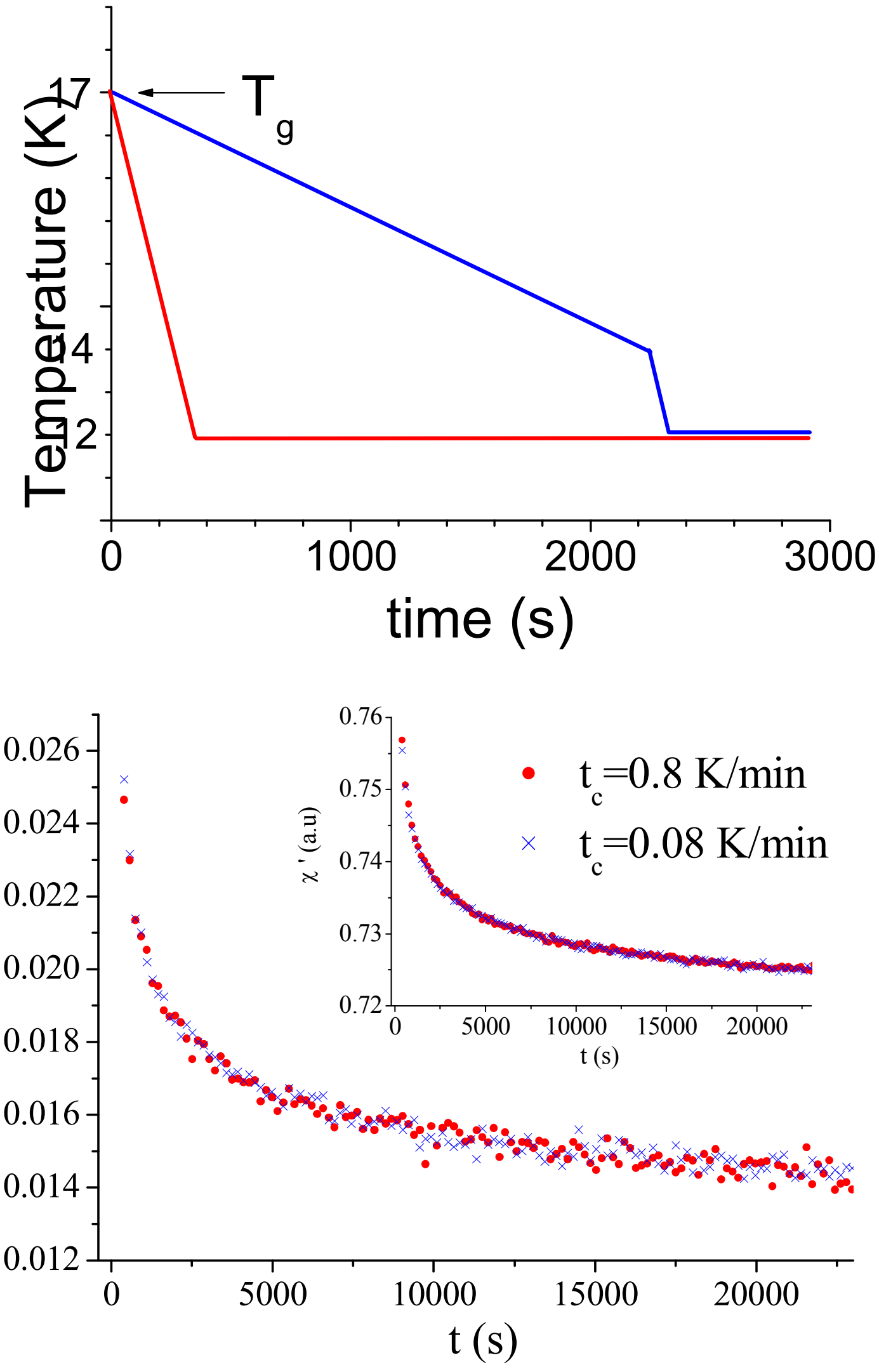

Figure 3.2: Effect of the cooling rate on the relaxation of the *ac* susceptibility in the thiospinel sample (from [45]). Top part: sketch of the procedure, in which a fast and a slow cooling rate are used around $T_g$=16.7K, before measuring at 12K. Bottom part: relaxation of the out-of-phase (main figure) and in-phase (inset) components of the *ac* susceptibility, from the time $t$=0 at which the temperature of 12K has been reached. Full circles: fast cooling. Crosses: slow cooling.

In this experiment, we compare the relaxation of the *ac* susceptibility at $0.7T_g$ after two cooling procedures in which the region of $T_g$ was crossed at cooling rates differing by a factor 10. Both aging relaxations, measured from the time at which the final temperature was reached, are exactly superimposed onto each other, as well for $\chi''$ as for $\chi'$: a slower cooling through $T_g$ does not help bringing the spin glass closer to equilibrium, at least as far as can be seen in this measurement. Note that, in the slow cooling procedure, we used a fast cooling rate in the last Kelvin's; a slower final approach of the landing temperature does indeed influence further aging at this temperature, as shown in the original publication. Our point here is that a slower cooling in the $T_g$ region does not help aging at a lower temperature.

We have studied this apparent "insensitivity" of the spin glass to cooling rate effects in more systematic experiments in which the temperature is changed by steps. Figure 3.3 presents the result of a "negative temperature cycle" experiment [46].

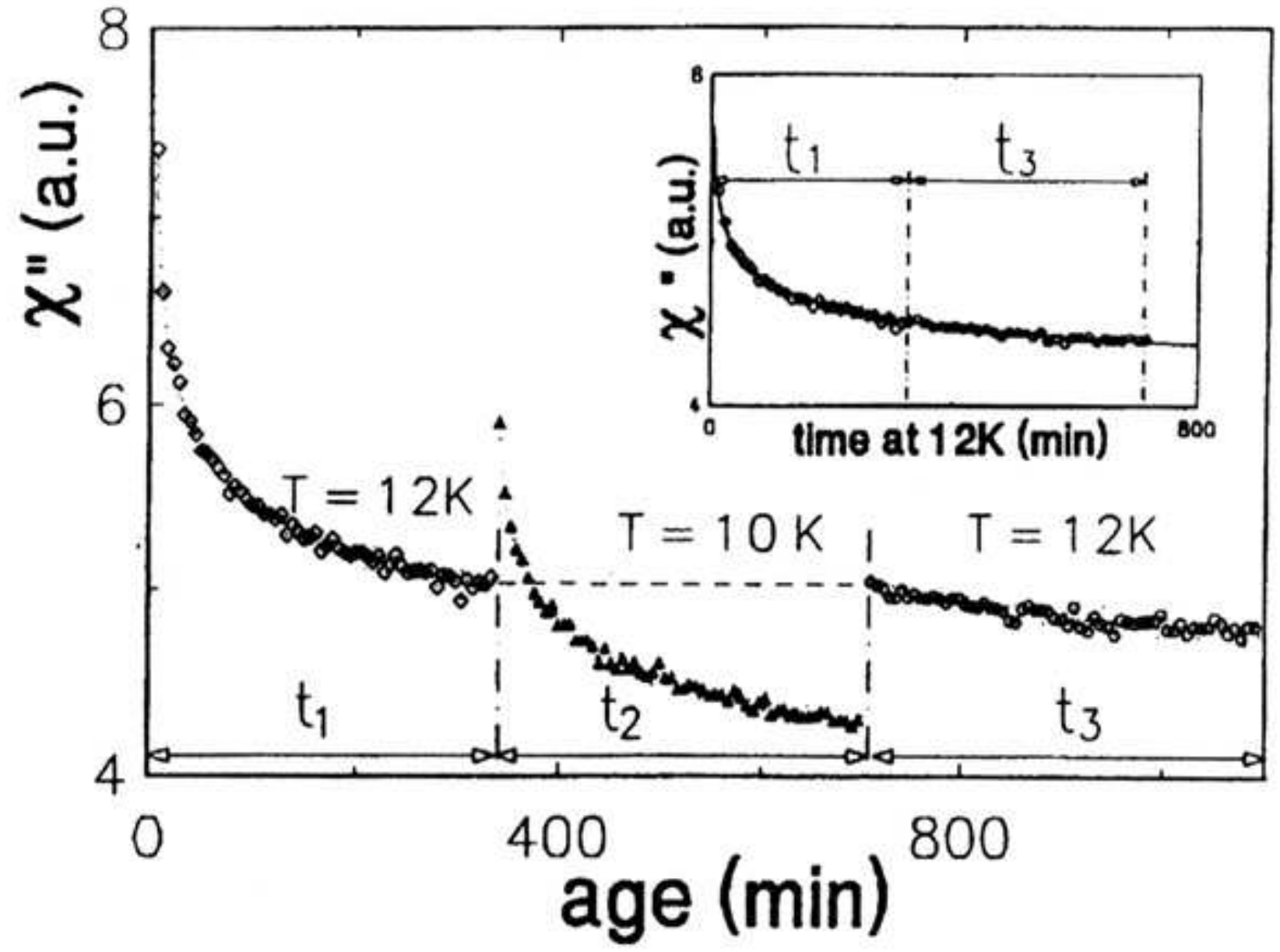

Figure 3.3: Relaxation of the out-of-phase susceptibility $\chi''$ during a negative temperature cycle of amplitude $\Delta T$=2K (frequency 0.01 Hz), showing aging at 12K, rejuvenation at 10K, and memory at 12K (from [46,19]). The inset shows that, despite the rejuvenation at 10K, both parts at 12K are in continuation of each other (memory).

After a normal cooling (typically of ~100s from $1.3T_g$ to $0.7T_g$), the spin glass is kept at constant temperature $T=12K=0.7T_g$ for $t_1=300$ min., during which aging is visible in the strong relaxation of $\chi''$. Then, the temperature is lowered one step further from $T=12$ to $T-\Delta T=10K$. What is observed is not a slowing down of the relaxation, but on the contrary a jump of $\chi''$ and a restart, which we state as a rejuvenation effect upon decreasing the temperature, as if aging was starting anew at $T-\Delta T$. The apparent absence of influence of former aging at $T$ is in agreement with the previous experiment (Fig.3.2) in which "slower cooling does not help".

One may wonder whether this renewed relaxation corresponds to a full rejuvenation of the sample: the answer is no. A first point is that the new relaxation can be identical to the previous one, but only - of course - if $\Delta T$ is sufficiently large, here $\Delta T \geq 2\text{-}3K$. And one should not forget that this identity can only be checked in the very limited time window of the experiments, thus not proving very much concerning the *overall* state of the spin glass. More importantly, the 3rd part of the experiment brings a definitive negative answer.

When the temperature is turned back from $T\text{-}\Delta T=10K$ to $T=12K$, the $\chi''$ relaxation restarts exactly from the point that was attained at the end of the stay at $T$, and goes on in precise continuity with the former one, as if nothing of relevance at $T$ had happened at $T\text{-}\Delta T$. As shown in the inset of Fig.3.3, this can be checked by shifting the 3rd relaxation to the end of the 1st one: they are in continuity, and can be superposed on the reference curve which is obtained in a simple aging at $T$. Hence, during aging at $T\text{-}\Delta T$ and despite the strong associated $\chi''$-relaxation, the spin glass has kept a "memory" of previous aging at $T$, and this memory is retrieved when heating to $T$.

This negative temperature cycle experiment pictures in a spectacular manner the phenomenon of rejuvenation and memory in a spin glass. When examined in more details, however, the situation is not always so simple. Fig.3.4 shows the results of negative temperature cycle experiments performed with various values of $\Delta T$.

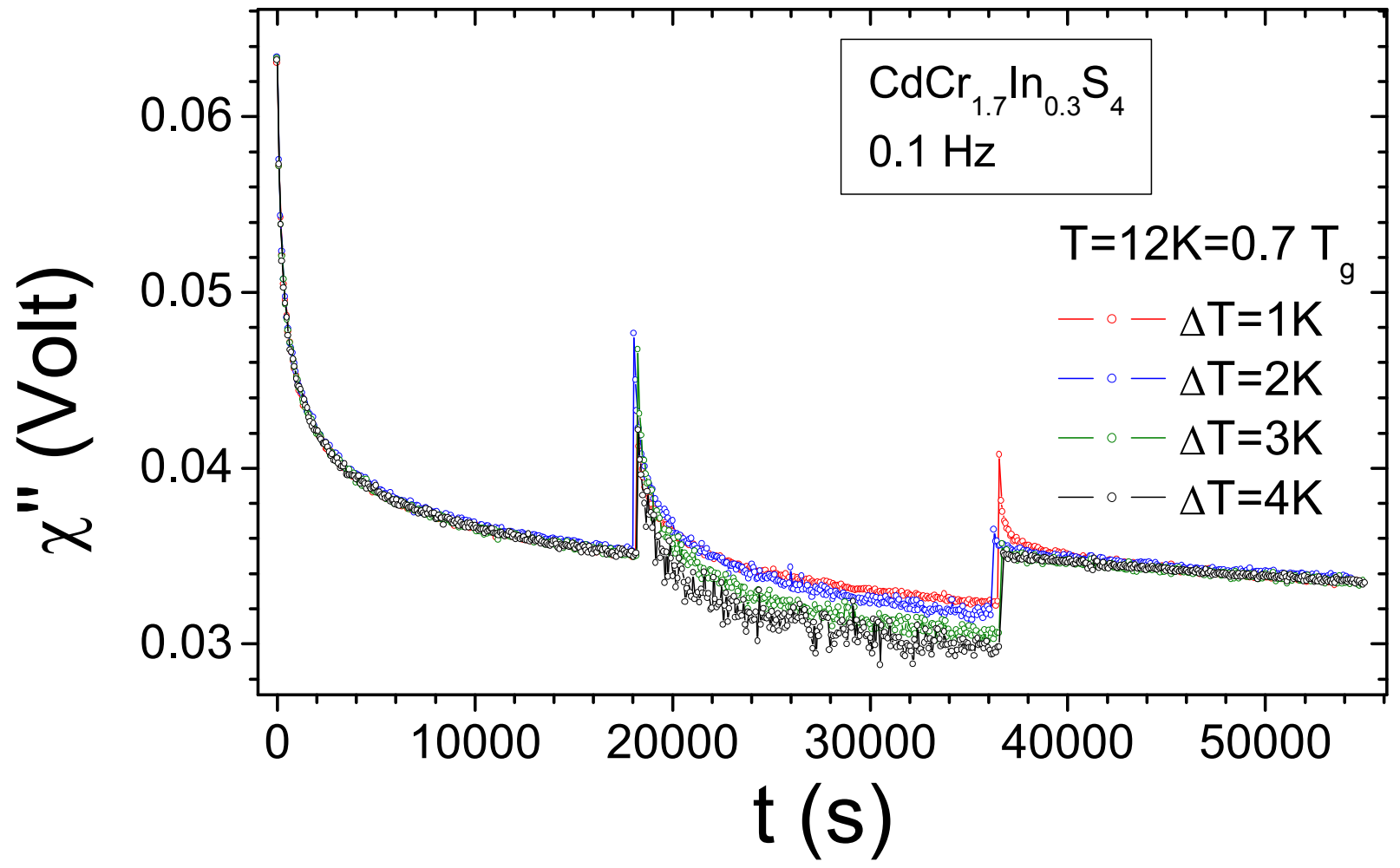

Figure 3.4 : Relaxation of the out-of-phase susceptibility $\chi''$ during negative temperature cycles of different amplitudes (from [48], but see also [46,47] for other examples), ranging from $\Delta T$=1K (upper curve, with the prominent spike) to $\Delta T$=4K (lower curve, no spike and full memory). The frequency is 0.1 Hz.

For $\Delta T=1K$, the beginning of the 3rd part relaxation shows a transient spike, which lasts for ~5000s before the curve merges with those, obtained for higher $\Delta T$'s, that are in continuity with the relaxation at $T$. Thus, for a smaller $\Delta T$ than that corresponding to full memory, there is indeed some contribution at $T$ from aging at $T\text{-}\Delta T$, and this contribution is "incoherent", extending over rather long but finite times (3-5000 s). Note that the data of Fig.3.4 is taken at frequency 0.1Hz, whereas in Fig.3.3 it is taken at 0.01Hz. In Fig.3.4, the points can therefore be taken more rapidly, and a small upturn is visible for $\Delta T=2K$: full memory is only obtained for $\Delta T=3$ and $4K$.

For smaller and smaller values of the temperature interval $\Delta T$, the observed "transient spike" decreases, changes sign (the curve merges with the reference from below), and finally vanishes [46,47]. In this small $\Delta T$ regime, apart from the transient part, aging at $T\text{-}\Delta T$ contributes "coherently" to aging at $T$ as an additional aging time $t_{eff}$, in such a way that the 3rd relaxation must be shifted by $(t_2\text{-}t_{eff})$ to be in continuity with the 1st part. Details on the results in this regime, together with their discussion in terms of a Random Energy Model, can be found in [47].

### 3.2 Memory dip experiments

The ability of the spin glass to keep a memory despite (partial) rejuvenation can be further explored in experiments with multiple temperature steps. The first (double) "memory dip experiments", suggested by P. Nordblad, have been developed in collaboration between the Uppsala and Saclay groups [45]. An example of a "multiple dip experiment" is shown in Fig.3.5 [48,24,49].

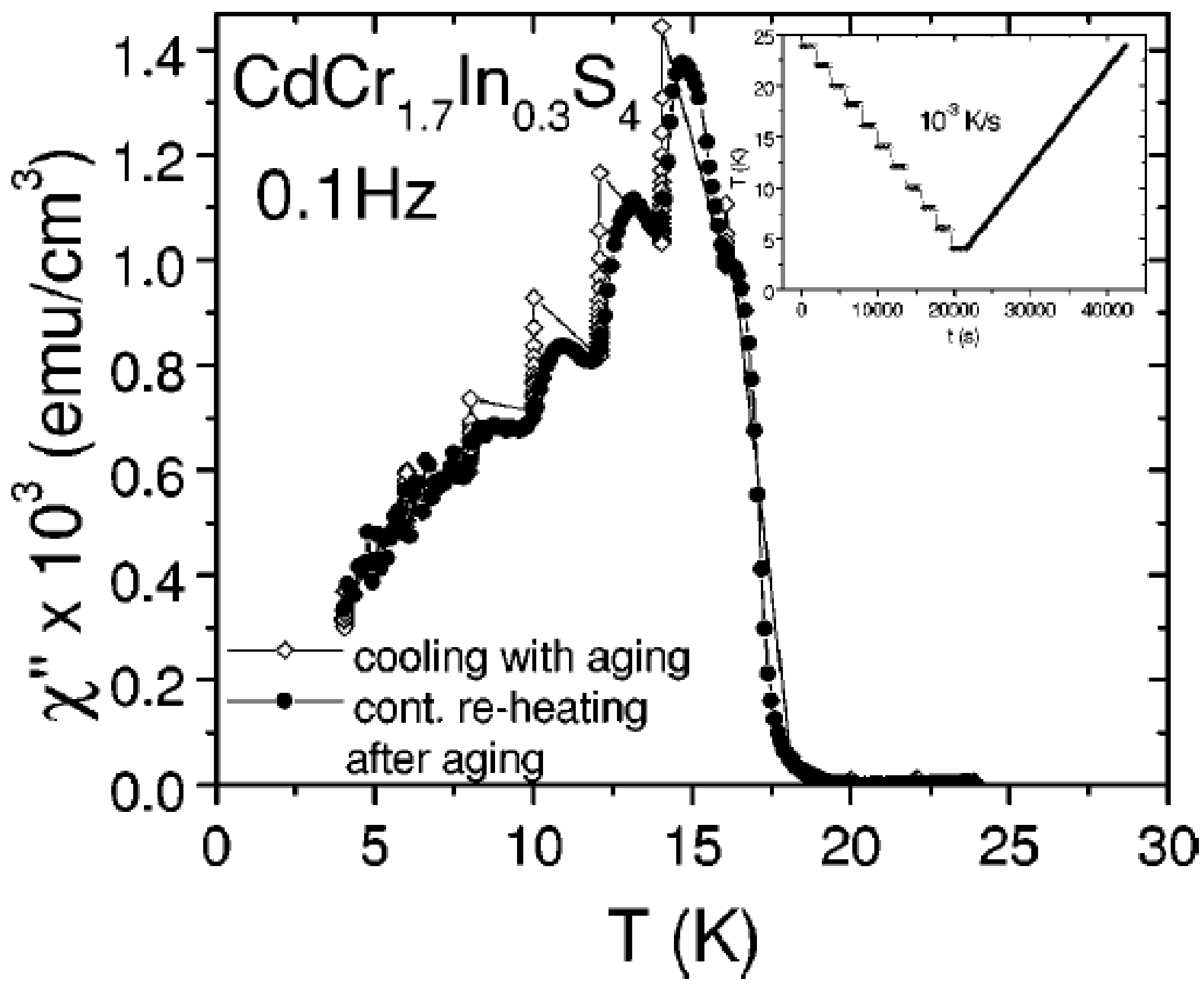

Figure 3.5: An example of multiple rejuvenation and memory steps [48,52,53,24]. The sample was cooled by 2K steps, with an aging of time of 2000 sec at each step (open diamonds). Continuous reheating at 0.001K/s (full circles) shows memory dips at each temperature of aging.

This is an *ac* experiment in which the sample is cooled by 2K steps of duration ~½ hour down to 4K, and then reheated continuously (inset of Fig.3.5). Fig.3.5 shows $\chi''$ as a function of temperature during this procedure, starting from $T>T_g$ where $\chi''=0$ (paramagnetic phase). $\chi''$ rises up when crossing $T_g=16.7K$, and when the cooling is stopped, the relaxation of $\chi''$ due to aging is observed during ½ hour (successive points at the same temperature in the figure). Upon further cooling by another 2K step, the $\chi''$ jump of rejuvenation is seen, and the relaxation due to aging takes place. At each new cooling step, rejuvenation and aging can be seen, and this happens ~6 times in the experiment of Fig.3.5. In the second part of the experiment, the sample is re-heated continuously, at a slow rate (*~0.001K/s*, equal to the average cooling rate) which allows the measurement of $\chi''$. Amazingly, apart from the rather noisy low-*T* region, the memory of each of the aging stages performed during cooling is revealed in shape of "memory dips" in $\chi''(T)$, tracing back the lower value of $\chi''$ which was attained at each of the aging temperatures. Thus, the spin glass is able to keep the simultaneous memory of several (5 or 6!) successive agings performed at lower and lower temperatures. Increasing the temperature afterwards reveals the memories, and meanwhile erases them.

This very asymmetric scheme of rejuvenation upon cooling, topped up by memory effects upon heating, has led the Saclay group to propose a description of these phenomena in terms of a hierarchical organization of the metastable states as a function of temperature, as pictured in Fig.3.6 [44,19].

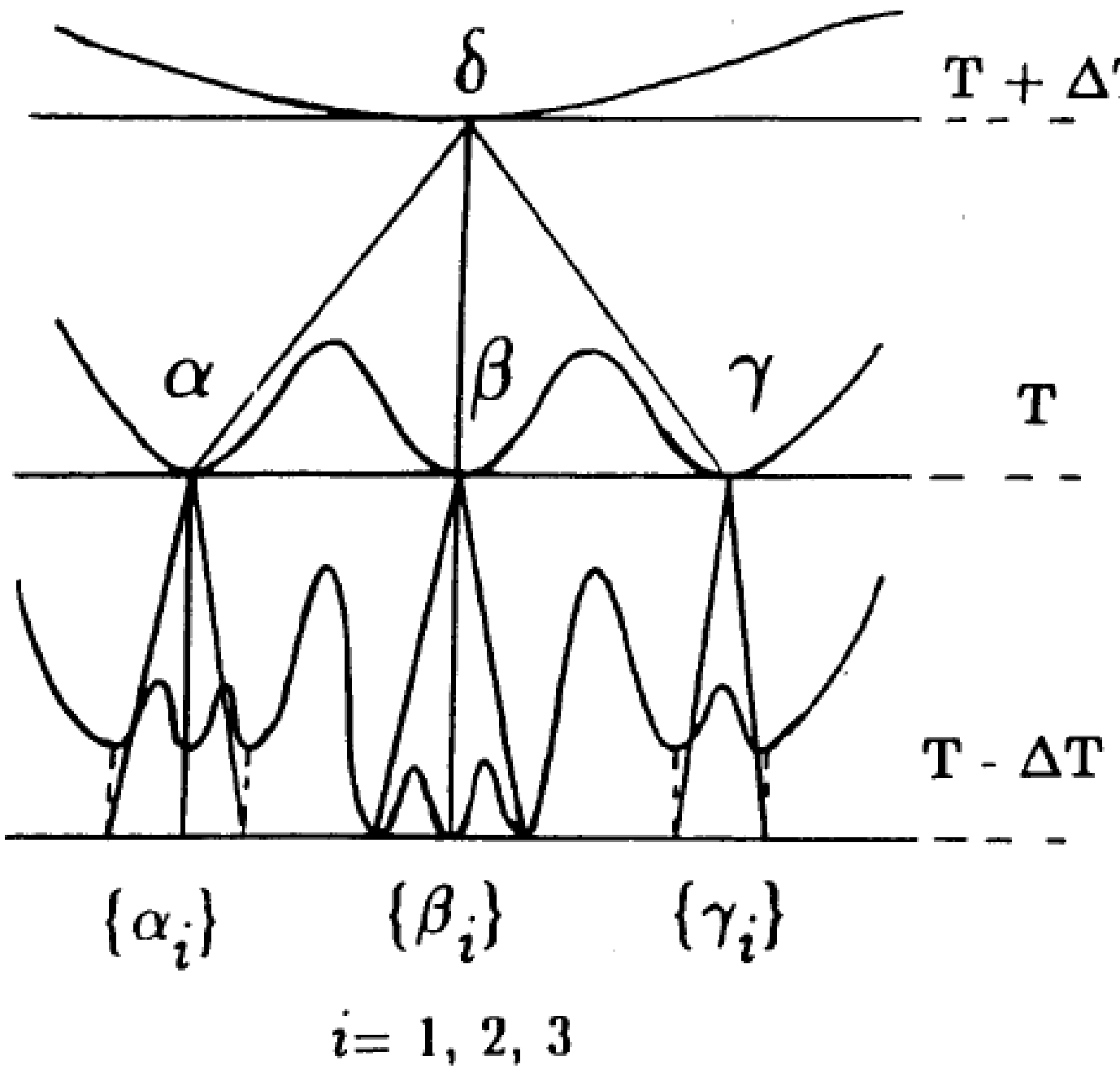

Figure 3.6: Schematic picture of the hierarchical structure of the metastable states as a function of temperature [4,44,19].

This very simple picture sketches the effect of temperature variations in terms of a modification of the free-energy landscape of the metastable states (and not only of a change in the transition rates between them). At fixed temperature T, aging corresponds to the slow exploration by the spin glass of the numerous metastable states. When the temperature is decreased from $T$ to $T\text{-}\Delta T$, the free-energy valleys are considered to subdivide into smaller ones, separated by new barriers. Rejuvenation arises from the transitions that are now needed to equilibrate the population rates of the new sub-valleys: this is a new aging stage. For large enough $\Delta T$ (and on the limited experimental time scale), the transitions can only take place between the sub-valleys, in such a way that the population rates of the main valleys are untouched, keeping the memory of previous aging at $T$. Hence the memory can be retrieved when re-heating and going back to the $T$-landscape. This tree picture, somewhat naïve, is however able to reproduce many features of the experiments when discussed in more details [47]. It has been made quantitative in developments of the Trap Model and the Random Energy Model [50,47]. In the mean-field model of the spin glass with full replica symmetry breaking [2], it has been shown that rejuvenation and memory effects can be expected in the dynamics [51].

Beyond this description of aging and rejuvenation and memory effects in terms of metastable states, it is of course very intriguing to imagine what kind of spin arrangements allow such complex phenomena when the temperature is varied [52,53,54,55]. It is very natural, as proposed in the “droplet model” [56,57], to consider that the spin glass, initially in a random configuration after the quench, slowly builds up from neighbour to neighbour a spin glass local order over larger and larger length scales. Frustration makes the process of minimizing the interaction energy of each spin with its neighbours very slow, making the jump from microscopic times (which are at play in the domain growth of pure ferromagnets, in which $l \sim t^{1/2}$) to macroscopic times corresponding to thermally activated crossing of free-energy barriers. In the droplet model, the spin glass is a kind of “disguised ferromagnet”, having simply two (spin reversal symmetric) ground states, which compete in the slow growth of spin glass ordered domains during aging. Can we see such domains in experiments ? No obvious macroscopic symmetry is expected in spin glass order, therefore no imaging of such domains could be realized until now, in contrast with the case of ferromagnetic domain growth. The only pictures that we have of the growth of a potential spin glass order are obtained from recent numerical simulations. Fig.3.7 shows a nice example given by Berthier and Young in [15], but the reader should not be misled by the apparent simplicity of this ferromagnetic-like picture.

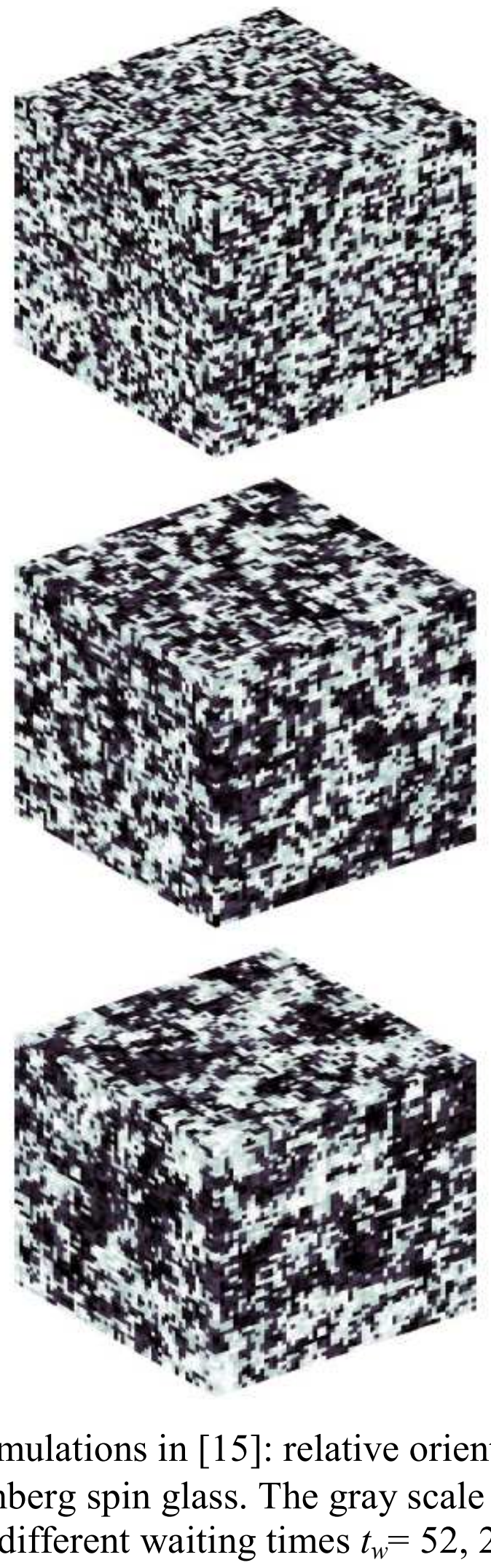

Figure 3.7: From the numerical simulations in [15]: relative orientation $\theta_i$ of the spins $\mathbf{S}_i$ in two copies *(a,b)* of a numerical Heisenberg spin glass. The gray scale stands for $cos\ \theta_i(t_w) = \mathbf{S}_i^a(t_w). \mathbf{S}_i^b(t_w)$. From top to bottom, three different waiting times $t_w$= 52, 27, and 57 797s are represented, showing the slow growth of a local random ordering of the spins.

In Fig.3.7 the grey scale codes the relative orientations of the spins in two copies (replicas) of the system which, starting from different random states, evolve independently by a Monte-Carlo algorithm. The snapshots taken after different waiting times $t_w$ show the growth of uniformly coloured regions. A region with a uniform grey-level colour is a region in which the individual spins have a constant angle from one replica to the other: over this region, seen in independent Monte-Carlo evolutions, the neighbour spins build the same relative angles. This is indeed an image of regions in which the spin evolution is correlated, which are in this sense equivalent to spin glass ordered domains.

If we now come back to the multiple memory experiment in Fig.3.5, thinking of a spin glass order being established on longer and longer length scales during each stage of aging, the observed rejuvenation and memory effects have some implications concerning these dynamic length scales. The restart of dissipative processes when going from $T$ to $T\text{-}\Delta T$ indicates that the spin-spin correlations growing at $T\text{-}\Delta T$ are different from those established at $T$. For thermally activated processes, if correlations extend up to a given length scale $L^*_T$ during aging at $T$, the correlation length $L^*_{T\text{-}\Delta T}$ which is attained at $T\text{-}\Delta T$ during the same time should be smaller, $L^*_{T\text{-}\Delta T} < L^*_T$ . The memory effect imposes here an important constraint: aging up to $L^*_{T\text{-}\Delta T}$ should occur without changing significantly the correlations established at the scale $L^*_T$ , that is, $L^*_{T\text{-}\Delta T} < L^*_T$. In practice, the independence of aging at length scales $L^*_{T\text{-}\Delta T}$ and $L^*_T$ is realized by a strong separation of the related *time* scales $\tau$ : $\tau( L,T\text{-}\Delta T)>>\tau( L,T)$. This necessary separation of the aging length scales with temperature has been coined “temperature-microscope” effect by J.-P. Bouchaud [53]: in an experiment like shown in Fig.3.5, at each stage aging should take place at well-separated length scales $L^*_n<\ldots< L^*_2< L^*_1$, as if the magnification of the microscope was varied by orders of magnitude at each temperature step. This hierarchy of embedded length scales as a function of temperature is the “real space” equivalent of the hierarchy of metastable states in the “phase space” (Fig.3.6).

Do we have examples of systems which present such a hierarchy of reconformation length scales ? This has been proposed for the very generic case of an elastic line in presence of pinning disorder [58,59]. Here, frustration arises from the competition between elastic energy, which tends to make the line straight, and pinning energy, which tends to twist the line to go through all pinning sites. As sketched in Fig.3.8 [59], starting from a random configuration after a quench, the line will progressively “age” by equilibrating slowly (thermally activated dynamics) over larger and larger distances.

At a given temperature $T$ and after some aging, the line can be pictured as a fuzzy ribbon (top of right part in Fig.3.8) which is equilibrated over a length scale $L^*_T$. At smaller length scales, the line continues to fluctuate between configurations which are roughly equivalent at temperature T (thus seen as a fuzzy ribbon). However, when going from $T$ to $T\text{-}\Delta T$, the difference between the equilibrium populations of some of these configurations may become significant, and a new equilibration at shorter length scales $L^*_{T\text{-}\Delta T} < L^*_T$ must take place. These dissipative processes will cause a rejuvenation signal. Meanwhile, processes at length scale $L^*_T$ are frozen at $T\text{-}\Delta T$, and the memory previous aging remains intact despite the rejuvenation processes, which occur at smaller (and well-separated) length scales. This scheme is a good candidate for the mechanism of aging, rejuvenation and memory in spin glasses [53,59,60]. The theory of an elastic line in pinning disorder yields a hierarchy of embedded states and length scales [58].

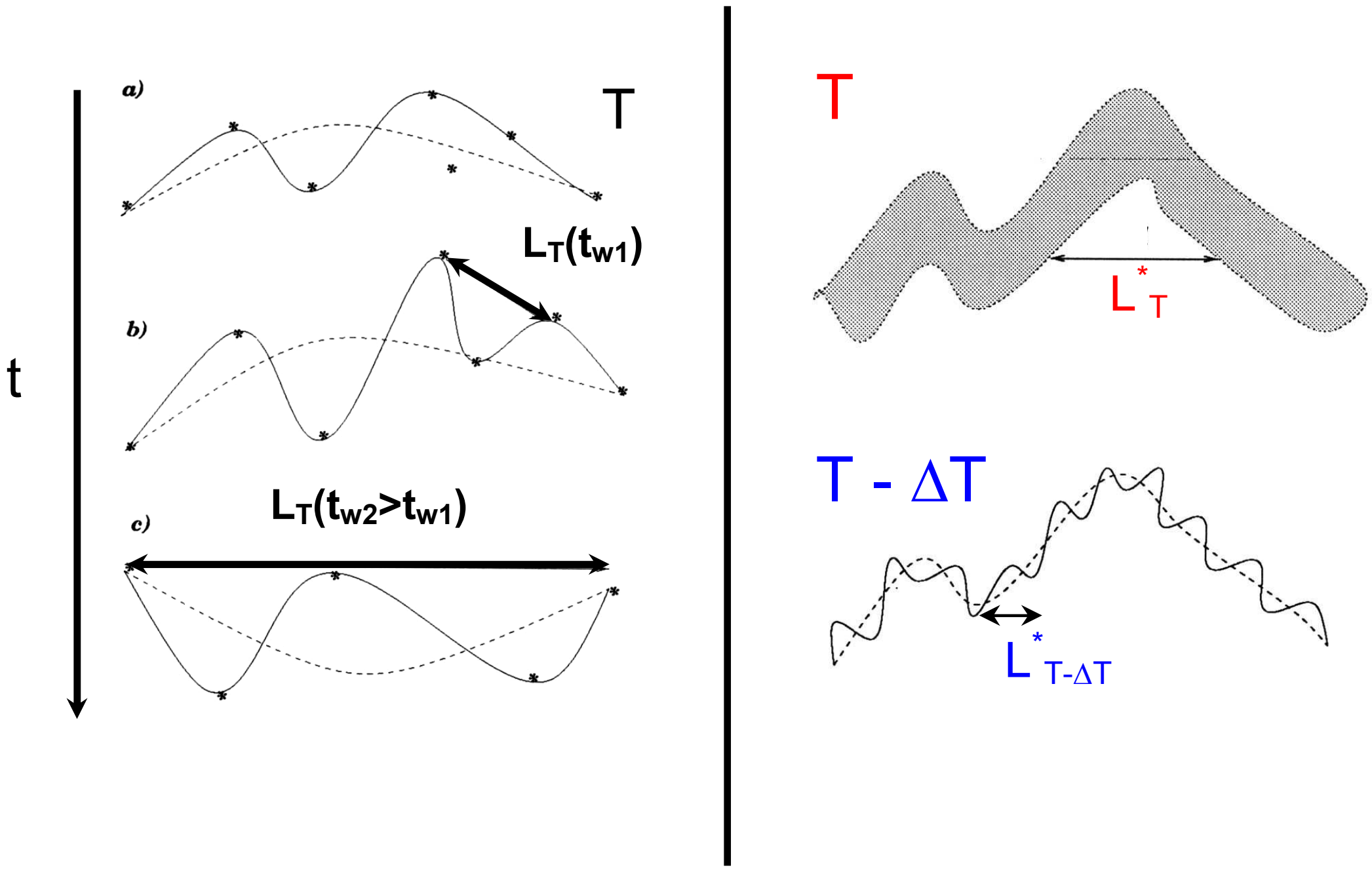

Figure 3.8: From [59], sketch of aging, rejuvenation and memory phenomena in terms of the dynamics of an elastic line in pinning disorder. Left part: at fixed temperature $T$, as time goes on, the line matches the pinning sites over larger and larger distances $L_T$. Right part: as the temperature is lowered from $T$ to $T$-$\Delta T$, rejuvenation processes occur at a smaller length scale $L_{T\text{-}\Delta T}$ , while the memory of reconformations at the larger length scale $L_T$ is preserved.

In the spin glass, it is not yet clear what objects could play the role of pinned elastic lines. Experiments on disordered ferromagnets show that spin glass dynamics can indeed be observed, which is most probably due to the dynamics of the walls [60,6]. Thus, we propose that the observed slow dynamics in spin glasses is explained in terms of wall-like dynamics, but in the present state of the art we cannot identify what are these walls, and what is the nature of the domains which are separated by these walls (see however the "sponge-like" excitations which have been characterized in numerical simulations [14]).

### 3.3 Rejuvenation and memory versus cumulative aging

In the previous section we described a "rejuvenation and memory like" dynamics, implying a hierarchical organization of the metastable states and of the corresponding length scales. This type of dynamics is found in systems which have so many "embedded" degrees of freedom that some of them are available to excitation at any temperature, even independently from each other at sufficiently different temperatures.

In "domain growth like" dynamics, of the type occurring in a ferromagnet, the approach of equilibrium is a one way only evolution through domain growth and wall elimination, in which the size of the domains should always increase. In an ideal ferromagnet, in which no energy barriers impede the domain wall motion, the temperature does not play any role. If we think of activated processes like the pinning of walls on defects, then temperature is relevant, but domain growth should just be accelerated or slowed down by temperature changes. Aging by domain growth processes is "temperature cumulative", in the sense that aging continues additively ("cumulatively") from one temperature to the other. In this type of dynamics, it is not clear how rejuvenation and memory effects may arise. In the droplet theory [56] they are related to "temperature chaos" effects, a scenario introduced in [61,56] which we do not discuss here. Detailed discussions of its possible relevance can be found in [53,54,55].

However, this language should not be misleading, and there is indeed some part of "domain growth" in "rejuvenation and memory" dynamics [21], but in our present understanding what is growing here is an object of the nature of a pinned wall rather than a (compact) domain. For a sufficiently small temperature variation $\Delta T$, no rejuvenation effects are seen in the spin glass: aging continues from $T$ to $T-\Delta T$ (see *ac* experiments in Section 3.1 and [46], or *dc* experiments with negative temperature cycles in [19]). In the hierarchical picture, for small $\Delta T$'s the free-energy landscape is almost identical at $T$ and $T-\Delta T$. In more general words, for small $\Delta T$'s the length scale of the aging processes are almost the same at $T$ and $T-\Delta T$, and aging is cumulative between both temperatures. As soon as $\Delta T$ is large enough, the free-energy landscapes become different, the aging length scales are separated (as is clear in the example of the pinned elastic line), and rejuvenation occurs due to the existence of independent degrees of freedom.
In some spin glass experiments like the one presented in Fig.3.9 [48], this dual aspect of aging dynamics shows up very clearly.

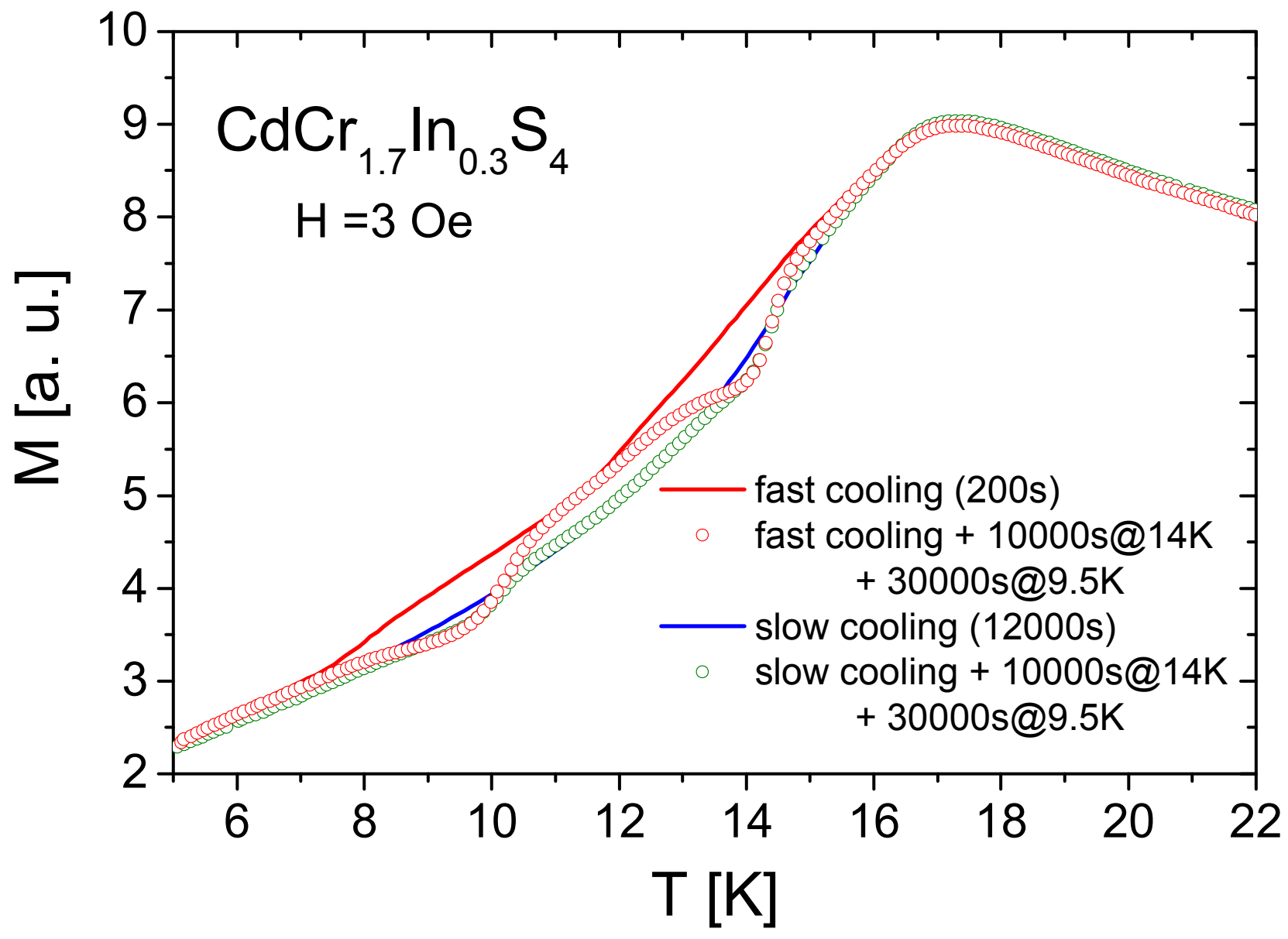

Figure 3.9: Effect of various cooling procedures on the ZFC magnetization of the thiospinel (insulating) spin glass [48]. Comparison of fast and slow coolings, with and without stops.

In this experiment, the sample is zero-field cooled with various thermal histories, and after applying the field at low temperature the magnetization is measured while increasing the temperature continuously at fixed speed (small steps of 0.1K/min). On one hand, we can observe the effect of a slow cooling in comparison with that of a fast cooling: the slow-cooled curve lies below the fast one in the whole temperature range. There is indeed a cooling rate effect in spin glasses, provided that one chooses an appropriate procedure to evidence it. On the other hand, we can evidence memory effects by stopping the cooling at two distinct temperatures and waiting during aging of the spin glass. The magnetization measured during re-heating after this step-cooling procedure shows clear dips at both temperatures at which the sample has been aging (this experimental procedure, very similar to that of the *ac* experiment in Fig.3.5, has been proposed by the Uppsala group [62]). In a third experiment, we can mix both effects, by slowly cooling the sample and interrupting the slow cooling by long waiting times at constant temperature. The resulting magnetization curve is lower than those obtained after faster cooling (temperature cumulative aging), and shows memory dips on top of this lower curve.

In a similar experiment, performed with another spin glass ($Au:Fe_{8\%}$ from [63], in Fig.3.10, metallic sample instead of the insulator of Fig.3.9), we have also plotted (bottom part of Fig.3.10) the difference between the curves obtained after a specific cooling history and the reference one obtained after a fast cooling. Fast oscillations (memory dips) show up on top of a wide bump (cumulative aging).

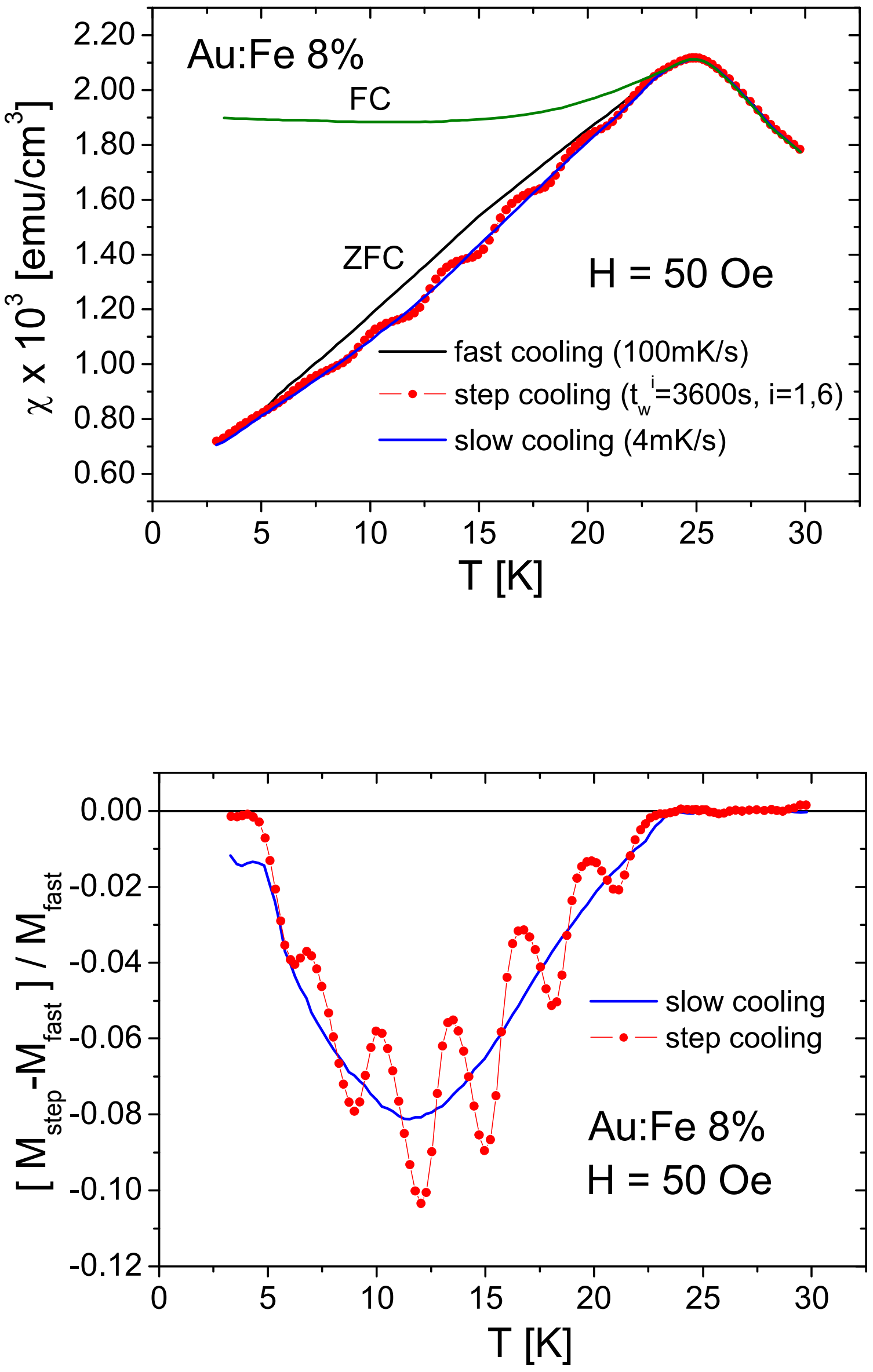

Figure 3.10: Effect of various cooling procedures on the ZFC magnetization of the $Au:Fe_{8\%}$ spin glass (top part). Comparison of fast and slow coolings, with and without stops. Bottom part: difference with the magnetization obtained after fast cooling. From [48].

Thus, the spin glass should not be considered as exempt of cooling rate effects, but rather as being able to show rejuvenation and memory effects in addition to cooling rate effects. How can we now compare the spin glass with "normal" glasses, which are considered to be dominated by cooling rate effects ? [21] New experiments have been designed to search for rejuvenation and memory effects in such systems. And these effects have been found, as is shown in the experiment of Fig.3.11 by the ENS Lyon group [64].

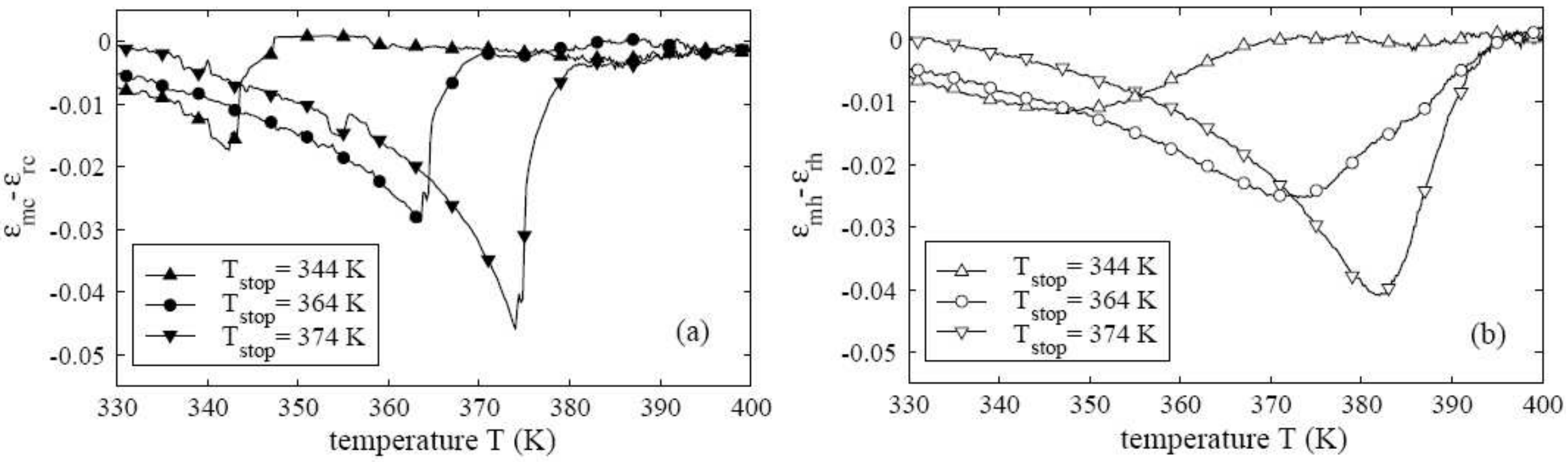

Figure 3.11 : Memory effect in the aging of the PMMA polymer glass, from [64]. The plots show the difference in the dielectric constant between an experiment with a normal cooling and another one with a stop of 10 hours at $T_{stop}$, for 3 values of $T_{stop}$. Left part: cooling. Right part: re-heating, showing a dip at a temperature corresponding to $T_{stop}$.

This experiment uses the same procedure (and the same representation of the results) as in Fig.3.10, but the cooling is only interrupted by one stop at one given temperature. Upon re-heating, the dielectric constant of the PMMA polymer indeed shows a dip, centred at a temperature slightly higher than that of the stop, and the comparison of 3 experiments with stops at 3 different temperatures shows very clearly that the position of the dip follows the temperature of the stop (Fig.3.11). The range of temperatures in which the aging effects are important in PMMA is much narrower than in spin glasses, and the width of the dip may appear to be larger because it spreads over the whole explored temperature range. However, the temperature dependence of the dip position is very clearly evidenced, signing up the occurrence of aging processes which are strongly temperature specific, as is the case in spin glasses.

An even more dramatic example of rejuvenation and memory effects in a structural glass (Fig.3.12) has been obtained in a study of the mechanical response of gelatine by a group of the food company Firmenich SA (Switzerland) [65]. Gelatine is a complex protein made of folded helices, and it has indeed many degrees of freedom related to helix unfolding in the vicinity of room temperature. This experiment is an *ac* measurement of the elastic modulus *G'*, and is again comparable with the *ac* experiment of Fig.3.5. During aging at fixed temperature, *G'* relaxes upwards (aging, the gelatine stiffens), and upon further cooling some rejuvenation can be seen. When re-heating, *G'* shows a dip at the aging temperature, and the authors could even realize a double memory experiment in which two memory dips can be distinguished (Fig.3.12).

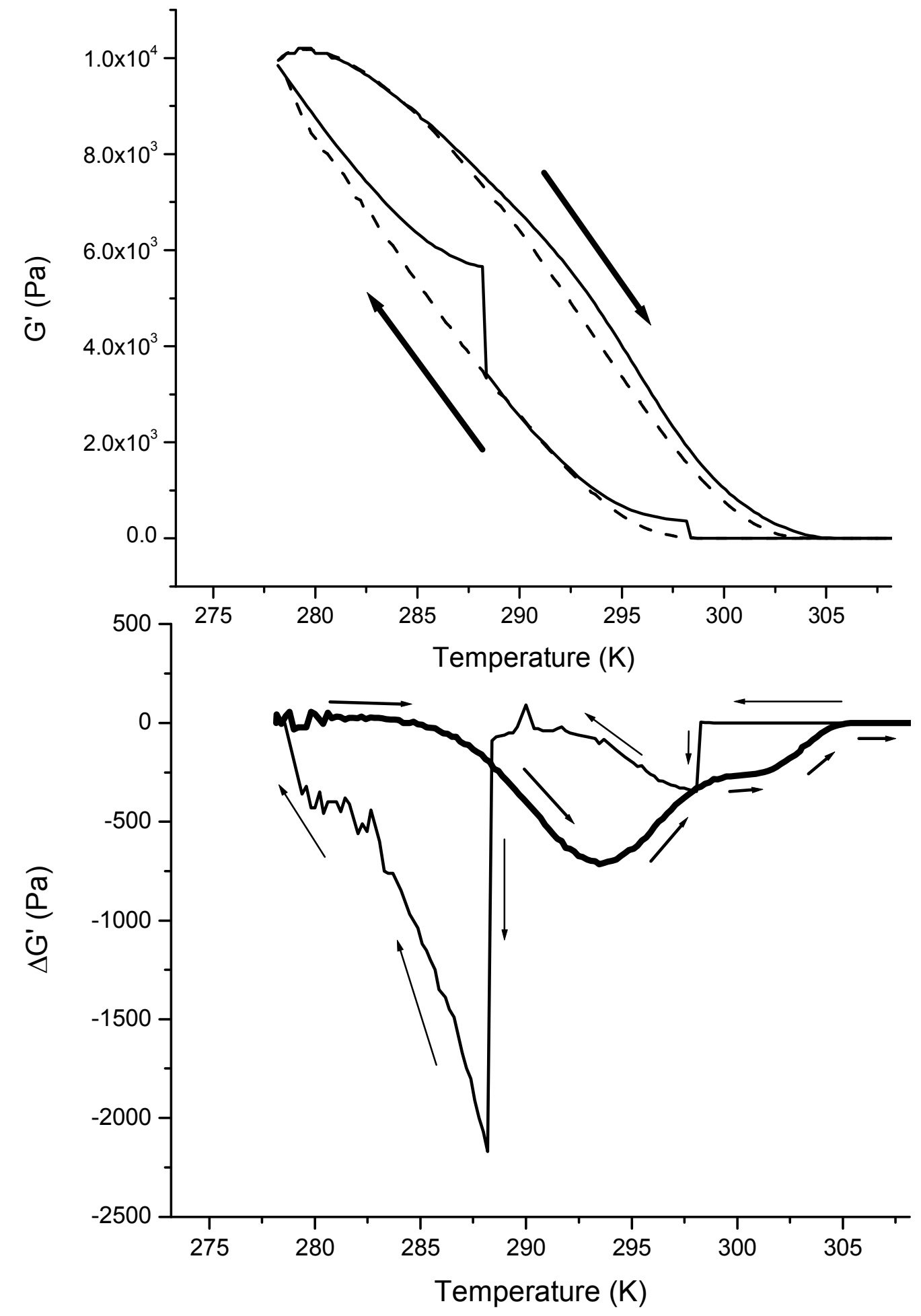

Figure 3.12 : Memory effect in the aging of a gelatine gel (measurement of the elastic modulus *G'*, from [65]). Two stops of 2 hours were made at 25 and 15°C during cooling. The upper part shows *G'* (solid line: with stops, dashed line: without stops) as a function of temperature. During the stops, *G'* increases slowly (aging, the gelatine gel stiffens). Upon re-heating, a wide-spread excess of *G'* is seen when compared with the curve obtained without stops. But, in the lower part of the figure which shows the difference plot, the memory of both stops is clearly revealed on re-heating.

Thus, it appears that aging effects in glasses in general can be considered as showing both "T-cumulative" and "rejuvenation and memory" contributions. The specificity of spin glasses might then be their ability to show sharp memory effects. However, the next section shows that the sharpness of these memory effects may be different in different spin glasses, and the further investigation of memory effects in *structural* glasses may bring other surprises.

# 4. Characteristic length scales for aging

As aging goes on, the dynamical response of the spin glass becomes slower. We have seen (Fig.2.3) that the time derivative of the magnetization relaxation after a field change gives access to an effective distribution of relaxation times, which shows a wide peak centred in the $log\ t = log\ t_w$ region [22]. For longer $t_w$'s, this distribution shifts towards the longer time region. We have no direct access to the spin configurations which correspond to these longer and longer response times, but it is reasonable to assume that longer response times are associated with flipping a larger number of correlated spins. This is the point of view that we have adopted above in this paper, discussing the multiple memory experiments (Fig.3.5) in terms of a hierarchy of embedded dynamical length scales selected by temperature (Fig.3.8) [53]. No simple symmetry allows an easy observation of these dynamical correlation lengths, but, considering that such characteristic dynamical lengths are underlying the aging phenomena, we have designed experiments which bring rather strong constraints on their properties. These experiments can be grouped in two classes: field variation and temperature variation experiments.

### 4.1 Length scales from field variation experiments

The idea of these experiments, based on [28], has been developed by R. Orbach and his group (UCLA and Riverside) [66]. It starts from the observation that the magnetization relaxation following a field change (as well in TRM as in ZFC procedure) becomes faster when a higher field amplitude is used, going beyond the linear response regime. An example is shown in Fig.4.1.

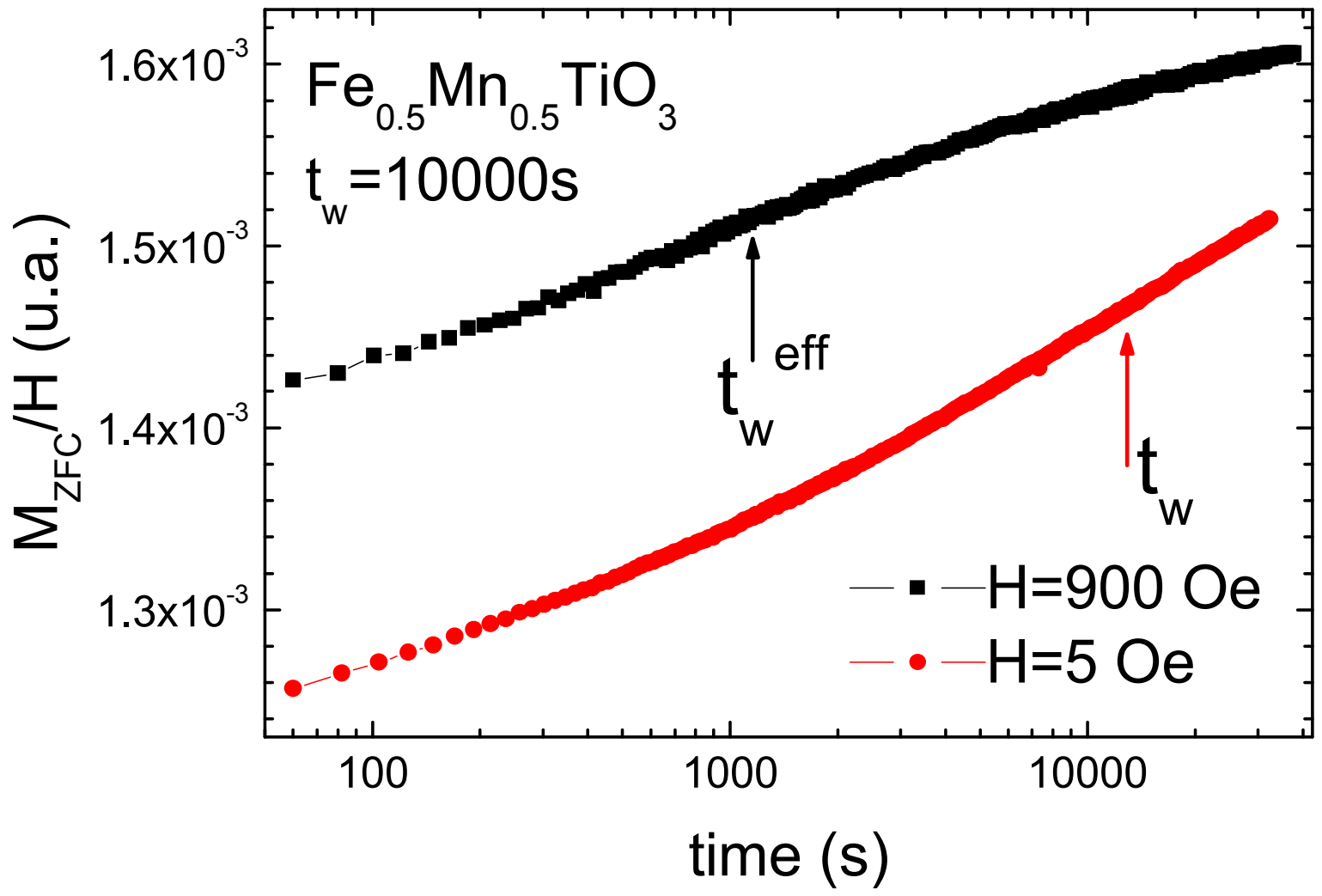

Figure 4.1 : ZFC relaxations of the $Fe_{0.5}Mn_{0.5}TiO_3$ Ising sample, for $t_w$=10000s and 2 different values of the field $H$ (data from [67]). The lower curve, taken with a low field $H$=5 Oe, shows an inflection point in the $t_w$ region. In the upper curve, taken with a much higher field $H$=900 Oe, the inflection point is found at a shorter time $t_w^{eff}$~1000s.

The lower curve in Fig.4.1 shows the ZFC relaxation obtained after applying a (small) 5 Oe field after $t_w=10000$s. Its inflection point is located as usual around $t \sim t_w$. The upper curve is obtained with a much higher field of 900 Oe, applied after the same $t_w$. This relaxation is faster than the first one in two respects: the initial rise up of the magnetization is higher, and the inflection point is found at shorter times, indicating that the distribution of the relaxation times now peaks at $t_w^{eff} \sim 1000$s, one order of magnitude smaller than $t_w = 10000$s. We propose to characterize the relaxation curves by their inflection point $t_i$ (time at which the relaxation rate is maximum), defining a typical free-energy barrier $U$ which can be overcome by thermal activation at temperature $T$ after a time $t_i$ with an attempt time $\tau_0$ ($\tau_0 \sim 10^{-12}$s is a paramagnetic fluctuation time):

$$U=k_BT\,Ln(t_i/\tau_0) \quad . \qquad (7)$$

In the case of the low field experiment with low field $H_0$, $t_i \cong t_w$, which defines a barrier $\Delta$ as

$$\Delta(H_0) = k_BT\,Ln(t_w/\tau_0) \ . \qquad (8)$$

In the experiment with a higher field $H$, the barrier $\Delta(H)=k_BT\,Ln(t_w^{eff}/\tau_0)$ is smaller since $t_w^{eff}<t_w$ . Assuming that, in a relaxation experiment performed after a given $t_w$, the spin correlations extend up to a typical number of spins $N_s(t_w)$, we propose to ascribe the free-energy reduction $\Delta(H_0)$- $\Delta(H)=E_Z(H)$ to the Zeeman energy of coupling of the magnetic field to the typical number of correlated spins $N_s(t_w)$ that must be flipped in the relaxation process [28,66]. In a low field experiment this Zeeman energy is negligible, and we have $t_i \cong t_w$, but for a higher field $H$ $E_Z(H)$ becomes significant, and we obtain it as the result of the measurement:

$$E_Z(H)= k_BT\,Ln(t_w/t_w^{eff}) \quad . \qquad (9)$$

The Zeeman energy is $E_Z=M.H$, $M$ being the magnetization of the $N_s$ spins. At this stage, we need to write explicitly the dependence of $M$ on $N_s$, which is not completely obvious for a disordered system. For a small number of spins $N_s$ in a random configuration, the magnetization is proportional to the typical fluctuation $N_s^{1/2}$, and is independent of the field: $E_Z= N_s^{1/2}\ \mu\ H$, where $\mu$ stands for the magnetic moment of 1 spin in the compound. On the other hand, at the *macroscopic* scale, the magnetization is an extensive quantity, proportional to the number of spins, and (to first order) proportional to the field via the susceptibility $\chi$ of 1 spin: $E_Z= N_s\,\chi\,H^2$.

It is likely that the general dependence of $E_Z$ on $N_s$ is a crossover shape from $H$ to $H^2$ dependence, but this would mean too many free parameters to interpret the results. In principle, the experiment should tell us which one is the dominant regime in the conditions of the measurement, since we can measure $E_z(H) \propto Ln\ t_w^{eff}(H)$ for various values of $H$, and conclude whether $E_z(H)$ has an $H$ or $H^2$ dependence. However, as in all fitting procedures, the result may depend on the range of fields explored, and the response is not completely unambiguous. Let us present now the experimental results that we obtained from various spin glass samples.

In an early series of experiments [66], we explored several spin glasses of different chemical nature: the insulating thiospinel $CdCr_{1.7}In_{0.3}S_4$, and the metallic alloys $Cu{:}Mn_{6\%}$ and $Ag{:}Mn_{2.6\%}$. With respect to spin anisotropy, these compounds are all Heisenberg-like [63]. For each sample, we measured ZFC relaxation curves for various amplitudes of the field $H$, at different temperatures $T$ and for various waiting times $t_w$. For fixed $t_w$ and $T$, the dependence of $Ln\ t_w^{eff}(H)$ versus $H^2$ was found to be significantly more linear than as a function of H, and we determined $N_s$ from the observed slope of $E_Z= N_s\ \chi\ H^2$ versus $H^2$[63]. The results are shown in Fig.4.2.

In this plot, which is presented as a function of the reduced variable $T/T_g Ln(t_w/\tau_0)$, the results from the 3 Heisenberg-like samples at 2 different temperatures do all fall on the same line. The number of correlated spins is, as expected, an increasing function of $t_w$, and the numbers reached in the experimental times are $\sim 10^4$-$10^6$, which means a range of 10-100 lattice units for the correlation length (assuming $L \sim N^{1/3}$).

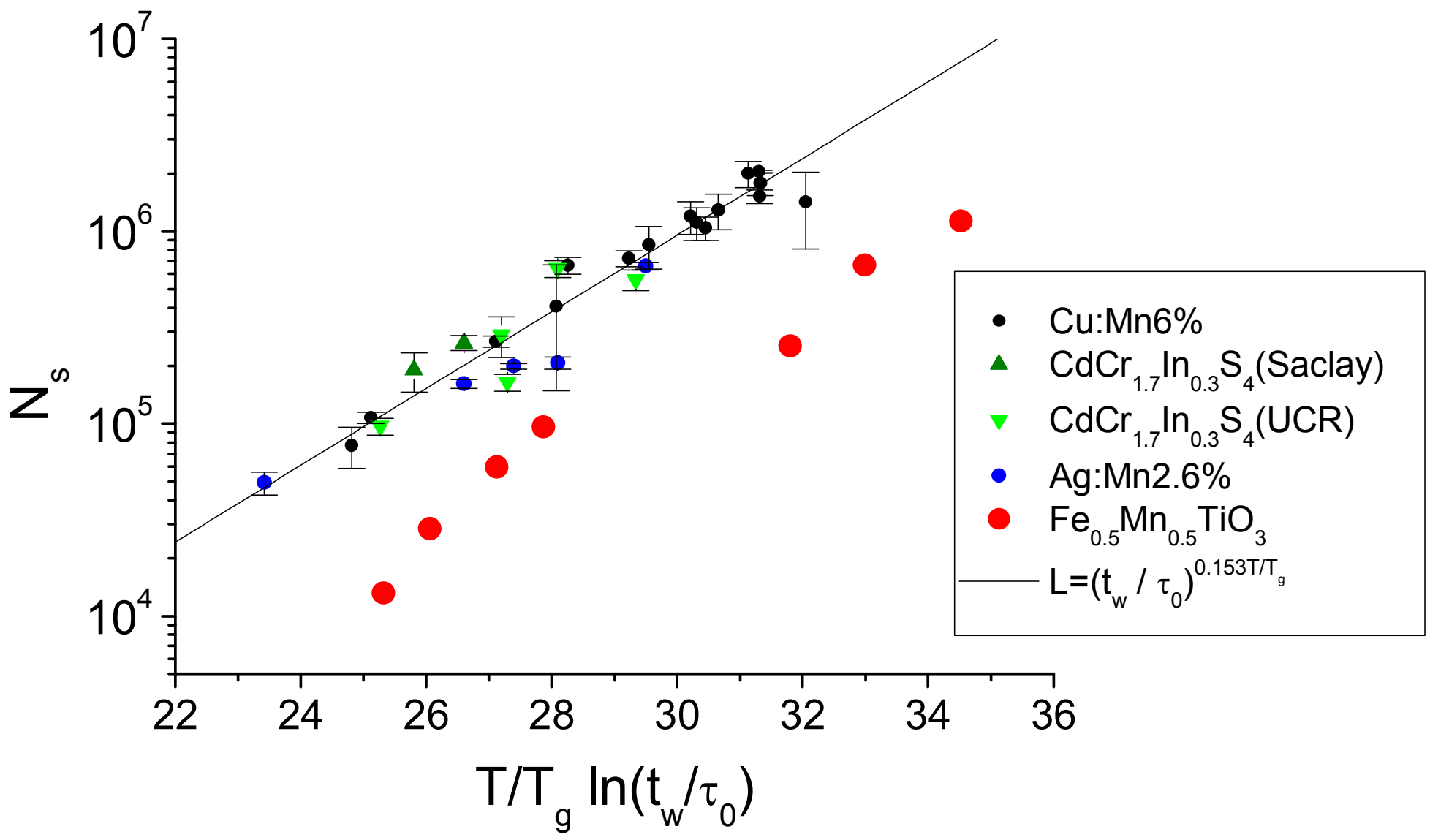

Figure 4.2 : Number of correlated spins extracted from field change experiments, as a function of the reduced variable $T/T_g Ln(t_w/\tau_0)$. The points with error bars correspond to Heisenberg-like spin glasses [66], they are well fitted by the straight line $N_S \sim (t_w/\tau_0)^{0.45T/Tg}$. The full circles lying below the others are from the $Fe_{0.5}Mn_{0.5}TiO_3$ Ising sample [67].

The 3 samples have a common (universal for Heisenberg-like?) behaviour, which is well fitted by a unique straight line. The solid line shown in the graph corresponds to the power law dependence $N_S=(t_w/\tau_0)^{0.45T/Tg}$. This was a rather big surprise because, soon after these experiments, numerical simulations of the Ising spin glass (Edwards-Anderson model) were performed in the aging regime by several groups, who could compute the four point correlation function and "directly obtain" an estimate of the correlation length $L(T,t_w)$ [13] . The numerical result, common to the different groups, is $L \simeq (t_w/\tau_0)^{0.15T/Tg}$ (recovering dynamic scaling $L \sim t^z$ of the equilibrium correlation length at $T_g$, $z=1/0.15 \cong 6$ being the usual dynamic exponent). This is the same result as in the experiments (if $N \propto L^3$). At this stage, the difference between Heisenberg-like (in experiments) and Ising (simulations) spins was not really discussed, and what was emphasized was the striking similarity between the simulations of the Edwards-Anderson model [13], performed up to $t_w/\tau_0 \sim 10^5$, and the experiments [66], which are performed 10 orders further in time in the $t_w/\tau_0 \sim 10^{12-17}$ regime.

This comparison motivated a second series of experiments [67], in which the properties of the strongly anisotropic system $Fe_{0.5}Mn_{0.5}TiO_3$ [68], considered a representative example of an Ising spin glass, were investigated using the same technique. We show in Fig.4.3 the measured $Ln\ t_w^{eff}$ as a function of $H^2$ and also $H$. In this Ising case, this is the linear behaviour of $Ln\ t_w^{eff}$ as a function of $H$ rather than $H^2$ which is favoured. Therefore we decided to analyse the Ising results in terms of $E_Z= N_s^{1/2}\ \mu\ H$. Checking afterwards the results of an analysis using $E_Z= N_s\ \chi\ H^2$ (only possible in the small field range), we found that it does not yield very different conclusions anyway. The results from the Ising sample are plotted in the same graph as those from the other samples in Fig.4.2 [67]. They lie – by almost a factor of 10 – lower than the others: in the Ising sample, after a given $t_w$, the number of correlated spins is smaller than in the Heisenberg-like samples. But, at fixed temperature, the $t_w$ dependence of $N_s$ is faster in the Ising case. The overall conclusion of this comparison is that the same simple power law, of the type $N \sim t_w^{aT/Tg}$, is not able to reproduce both sets of results from Heisenberg and Ising samples.

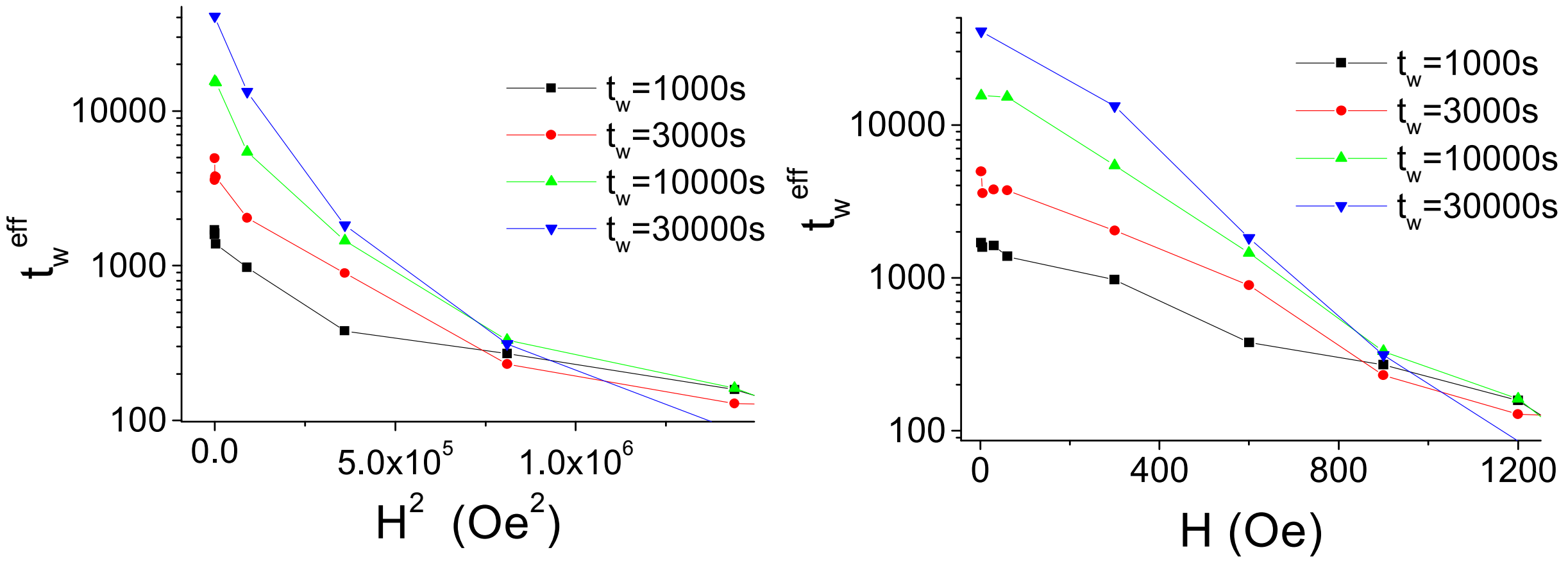

Figure 4.3 : Effective waiting times (log scale) obtained from the field change experiments on the $Fe_{0.5}Mn_{0.5}TiO_3$ Ising spin glass, as a function of $H^2$ (left) and $H$ (right) [67]. Four values of the waiting time $t_w$ have been explored.

The progress in computer simulations has finally allowed the numerical study of the Heisenberg spin glass, which is still more greedy in computer time. In [15], the comparison of the numerical results is presented using the same variables as in Fig.4.2. The correlation length $L(T,t_w)$ is found smaller in the Ising case, as in experiments. The extrapolation of the numerical data to the time regime of the experiments is rather hazardous, but it is possible that the long time slope of $L(T,t_w)$ versus $t_w$ at low $T$ becomes weaker for Heisenberg than for Ising (however, this is not the case in the numerical time range).

## 4.2 Length scales from temperature variation experiments

The temperature variation experiments bring a lot of information on the time-temperature relation in aging phenomena. In a negative temperature cycle experiment [44,19], the spin glass is aged during $t_1$ at $T$, then during $t_2$ at $T$-$\Delta T$ and finally during $t_3$ at $T$. A specific state of aging is established by this temperature history. If after this a field change is applied (like in the TRM procedure), the relaxation curve that is obtained reflects the properties of the state that has been prepared. For small $\Delta T$ values, it is possible to obtain the same relaxation curve after aging at constant temperature $T$ during a total waiting time $t_1+t_2^{eff}+t_3$, in such a way that the effect of waiting $t_2$ at $T$-$\Delta T$ is the same as waiting $t_2^{eff}$ at $T$. The identity of the relaxation curves tells us that the same state of aging has been established in both histories, at least for the aging processes whose time scales are probed in a *dc* relaxation experiment (~$10^0$ to $10^5$ s) [19]. Now, the idea of is to consider that this same aging state corresponds to the same dynamical length $L$ up to which correlations are established. Hence, from a couple of experiments as described, we constrain the time and temperature dependence of $L(t,T)$:

$$L(t_2, T\text{-}\Delta T)=L(t_2^{eff},T) \quad . \qquad (10)$$

We have performed TRM experiments with negative temperature cyclings on a series of representative spin glass samples, in order to better understand the differences between Ising and Heisenberg systems [67]. For this purpose, we have used a series of spin glasses which have also been studied in Orsay by torque measurements [63]. The torque measurements allowed sorting these spin glasses by their measured spin anisotropy (random anisotropy arising from Dzyaloshinsky-Moriya interactions). D. Petit and I. Campbell found [63] that the critical exponents at the spin glass transition present a systematic dependence on the spin anisotropy, ranging from Edwards-Anderson type exponents for the Ising example to chiral ordering exponents [69] in the most isotropic case. These samples are, $K_r$ being the relative anisotropy constant $K_r=(K/T_g)/(K/T_g)_{AgMn}$, normalized to the AgMn value [63]:

(1) $Fe_{0.5}Mn_{0.5}TiO_3$, $T_g$=20.7K, strongly anisotropic single crystal [68] (no $K_r$ estimate, but large)

(2) $(Fe_{0.1}Ni_{0.9})P_{16}B_6Al_3$, amorphous alloy with $T_g$=13.4K and $K_r$=16.5

(3) Au:$Fe_{8\%}$, diluted magnetic alloy with $T_g$=23.9K and $K_r$=8.25

(4) $CdCr_{1.7}In_{0.3}S_4$, insulating thiospinel with $T_g$=16.7K and $K_r$=5.0

(5) Ag:$Mn_{2.7\%}$, diluted magnetic alloy with $T_g$=10.4K and, by construction, $K_r$=1 (in the particular case of Ag:$Mn_{2.7\%}$, we use former data from [70]).

The experimental procedure is sketched in Fig.4.4.a, and a set of results with the thiospinel sample (#4, Heisenberg-like) is presented in Fig.4.4.b. In Fig.4.4.b, relaxation curves obtained after temperature cycling of amplitude $\Delta T$ are compared with those obtained after isothermal aging at $T=12K=0.7T_g$ during $t_w=t_1+t_3=1000$s (bottom solid curve) and $t_w=t_1+t_2+t_3=10000$s (top solid curve). If we look for instance at the curve obtained after temperature cycling $\Delta T=0.5K$ (full circles), we see that it almost lies on the isothermal $t_w=t_1+t_3=1000$s reference, far below the $t_w=t_1+t_2+t_3=10000$s reference. That is, in this case we have $t_2^{eff}\sim 0$, which means that aging during $t_2=9000$s at $T-\Delta T$ is almost of no influence on aging at $T$, even though $\Delta T$ is only of 0.5K: this is the "temperature microscope effect" that was invoked above to explain the possibility of multiple memories (Fig.3.5, with the same thiospinel sample). The comparison with the Ising sample is rather interesting (Fig.4.4.c).

For the Ising spin glass, $T_g$ is slightly different, but the temperatures are the same in units of $T_g$. The relaxations are performed at $T=15K=0.7T_g$, and we can look at the curve resulting from a temperature cycle with $\Delta T=0.6K=0.03T_g$ with solid squares (same fraction of $T_g$ as for the solid circles for the Heisenberg case in Fig.4.4.b). This curve lies in the middle region between the $t_w=t_1+t_3=1000$s and $t_w=t_1+t_2+t_3=10000$s references, which means that there is a significant effect of aging at $T-\Delta T$ on aging at $T$: in the Ising case, the $T$-microscope effect with temperature is not so strong as it is in a Heisenberg spin glass.

This visual appreciation of the curves can be expressed in quantitative terms. Using the scaling procedure described in Section 2, we can ascribe an effective waiting time $t_1+t_2^{eff}+t_3$ to each of the temperature cycled curves, adjusting precisely the value of $t_2^{eff}$ which allows the superposition of each of the $T$-cycled curves with a set of isothermally aged references. The result of each temperature cycling experiment is a value of $t_{eff}$ for a given $\Delta T$. In Figure 4.5, we present in the same graph the results obtained from the 5 samples for $T=0.85T_g$ (similar results have been obtained for $T=0.7T_g$) [67].

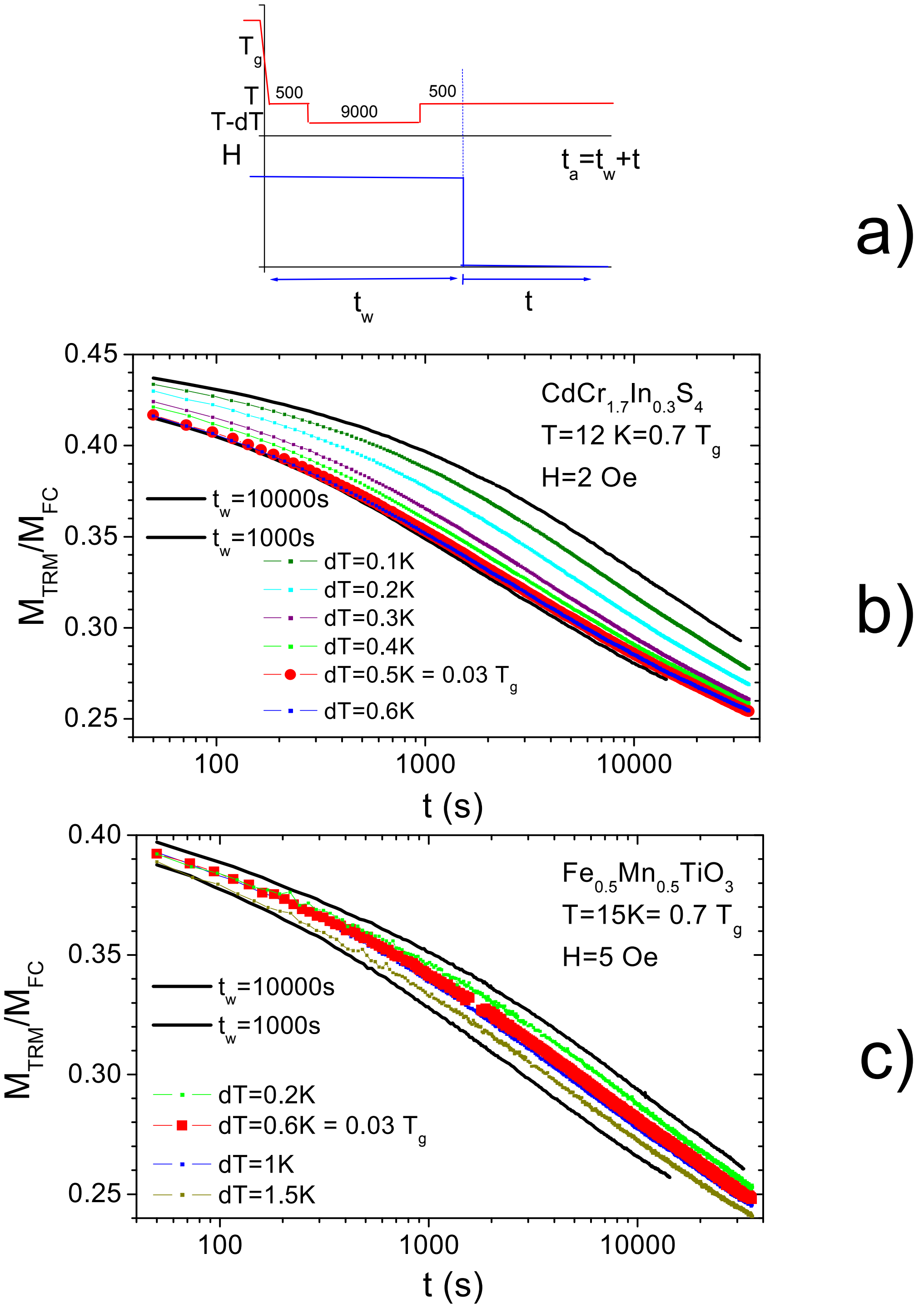

Figure 4.4 : TRM experiments with a negative temperature cycle from $T$ to $T\text{-}dT$ during the waiting time [67]. The extreme solid lines are reference curves, obtained after isothermal aging during $t_w$=1000 and 10000s. The thick full circles are obtained after a negative temperature cycling with $dT=0.03T_g$. Results from other $dT$ values are also presented. Top part: $CdCr_{1.7}In_{0.3}$ thiospinel (Heisenberg-like) spin glass. Bottom part: $Fe_{0.5}Mn_{0.5}TiO_3$ Ising spin glass.

Of course $t_{eff}$ is a decreasing function of $\Delta T$: for larger values of $\Delta T$, the contribution of aging at $T\text{-}\Delta T$ to aging at $T$ becomes weaker. Remarkably, we find that the slope of $t_2^{eff}(\Delta T)$ varies systematically with the spin anisotropy of the sample. The slope is weaker for the Ising sample than for the thiospinel (Heisenberg-like) sample #4, as expected from the trend observed in Fig.4.4bc, but the effect is systematic over the 5 samples studied. Going from Ising to Heisenberg situation, the weaker the spin anisotropy, the steeper the decrease of $t_2^{eff}(\Delta T)$, which means a stronger and stronger $T$-microscope effect.

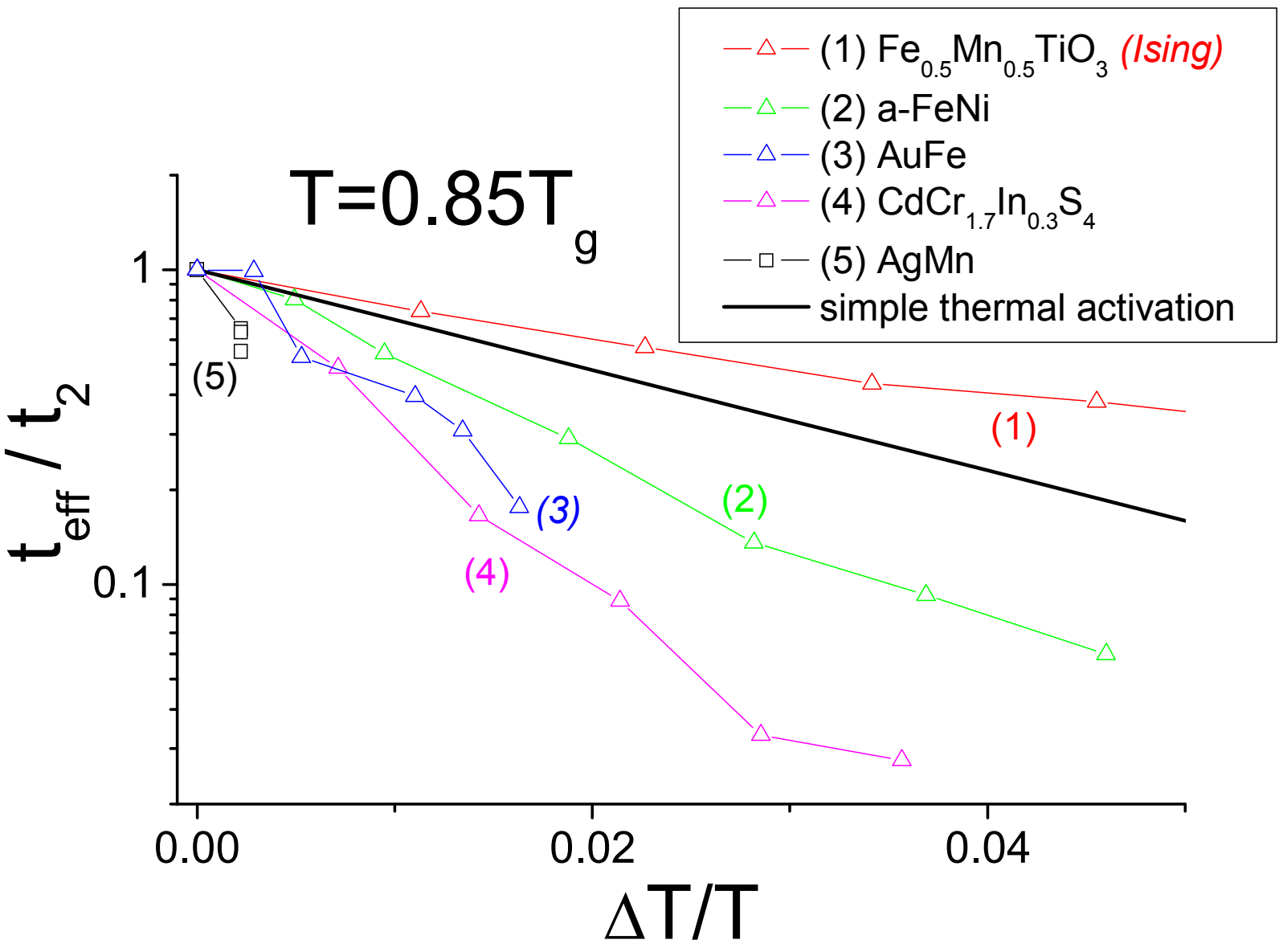

Figure 4.5 : Effective waiting times deduced from the temperature cycle experiments performed around $T=0.85T_g$, for the 5 samples investigated (ranked by decreasing anisotropy from #1 to #5) [67]. The straight line stands for usual thermal slowing down (constant energy barriers) with $\tau_0=10^{-12}$s.

We can compare the steepness of this decrease with usual thermal slowing down. A free-energy barrier $U(T-\Delta T)$ can be defined corresponding to the aging process during $t_2$ at $T-\Delta T$, and for usual thermal slowing down this barrier is the same as $U(T)$ corresponding to aging during $t_2^{eff}$ at $T$, which reads

$$U(T-\Delta T)=k_B(T-\Delta T)\, Ln(t_2/\tau_0), \quad U(T)= k_B T\, Ln(t_2^{eff}/\tau_0), \quad U(T)= U(T-\Delta T)\ . \ (11)$$

From Eq.(11), we obtain

$$Ln\ (t_2^{eff}/t_2)=-\ \Delta T/T\ Ln\ (t_2/\tau_0), \ (12)$$

which is a straight line of slope $-Ln\ (t_2/\tau_0)$ in the log-log plot of $t_2^{eff}/t_2$ versus $\Delta T/T$ in Fig.4.5 (for $\tau_0=10^{-12}$s, solid line in the figure). For samples #2-3-4-5, the slowing down is *stronger* than for usual thermal activation, a behaviour that was already observed in early experiments [70], and has been interpreted as the signature of a “super-activated” behaviour: the free-energy barriers $U$ increase as the temperature decreases, i.e. $U(T-\Delta T)>U(T)$ . These results cannot be ascribed to a decrease of $\tau_0$, which would then take unphysical small values (for the thisopinel sample #4, one would have $\tau_0=3.10^{-27}$s at $0.7T_g$, and even $\tau_0=6.10^{-48}$s at $0.85T_g$). On the other hand, an increase of $U(T)$ for decreasing $T$ is indeed what is expected from the hierarchical picture [44] sketched in Fig.3.6; as the temperature is lowered, free-energy barriers grow up, subdividing the valleys into new sub-valleys. Early temperature-cycling experiments on Heisenberg-like spin glasses [70] were already analyzed in terms of a barrier growth towards low temperatures, but the conclusions were somewhat different, since the rapid barrier growth was interpreted as an indication of divergences at all temperatures below $T_g$.

The behaviour of the Ising sample is rather surprising; the thermal slowing down is less steep than expected from usual thermal activation, and corresponds to an inverse temperature dependence of effective barriers, of the type $U(T-$ $\Delta T) < U(T)$, which seems quite unlikely. A way to understand this result is to consider that the hypothesis of a paramagnetic attempt time $\tau_0 = 10^{-12}$s is not valid in this case. The weak slope of the Ising results in Fig.4.5 means a smaller value of Ln $(t_2/\tau_0)$, implying a longer value for $\tau_0$, $\tau_0 \sim 2.10^{-7}$s. This renormalization of the microscopic attempt time can be due to critical fluctuations of the type encountered in the vicinity of $T_g$, which would have a stronger influence in the Ising case. Following this idea, we propose a common quantitative analysis of the 5 samples in the next section.

At this stage, an important remark should still be done. If aging corresponds to establishing correlations up to a typical length $L(t_w,T)$, our results in Fig.4.5 have clear-cut consequences on the possible time and temperature dependence of $L$, since in these experiments the same stage of aging (and hence the same $L$) can be obtained in 2 different temperature histories. If the $L$ dependence is of a simple power law type $L \sim (t_w/\tau_0)^{aT/Tg}$, as suggested earlier from the first (Heisenberg-like) experiments [66] and from the Ising simulations [13], then we should have

$$(t_2^{eff}/\tau_0)^{aT/Tg} = (t_2^{eff}/\tau_0)^{a(T-\Delta T)/Tg}, \qquad (13)$$

which is identical to $U(T) = U(T-\Delta T)$ in Eq.(11). In other words, a power law behaviour of $L$ would entail that, in a graph like Fig.4.5, all results from all samples lie on straight lines of slopes determined by the value of $\tau_0$. For the Ising sample this is not excluded, but for the Heisenberg-like spin glasses $\tau_0$ would then reach unphysical small values, smaller and smaller when approaching $T_g$. The conclusion from our temperature cycle experiments is that $L(t_w,T) \sim (t_w/\tau_0)^{aT/Tg}$ is cannot account for all results, and that one has to go beyond a power law behaviour for $L$, as already concluded from the field variation experiments described above (Fig.4.2).

Recent numerical simulations of Ising and XY spin glasses [71], using a new method for determining $L$, obtain results which are compatible with a power law behaviour of $L$ for both classes. However, in another set of simulations [16] following [15], Berthier and Young compare Heisenberg Ising spin glasses using the same procedure as in our temperature cycle experiments, that is, comparing the $t_{eff}(\Delta T)$ behaviours in both cases. The comparison with the experiments is rather puzzling. In [16], the $t_{eff}(\Delta T)$ line for the Ising case lies slightly above the line corresponding to simple thermal activation with constant barriers, as is the case in the experiments. But, at variance with the experiments, the numerical results for the Heisenberg case lie above the Ising ones. The authors [16] emphasize that this may be related to the difference in time scales. Therefore, they have explored the influence of increasing the time $t_2$ spent at $T-\Delta T$, and they do find that $t_{eff}(\Delta T)$ becomes steeper for increasing $t_2$, an effect which is much stronger in the Heisenberg than in the Ising case. The time scales explored experimentally and numerically remain very far from each other, but it is not completely excluded that, in the distant limit of experimental times, the numerical $t_{eff}(\Delta T)$ line for Heisenberg becomes lower than for the Ising case, in the same way as in the experiments.

## 4.3 The dynamical correlation length from both temperature and field variation experiments

In temperature variation experiments, a super-activated behaviour is observed for Heisenberg-like spin glasses, and the Ising results point towards a renormalization of the microscopic attempt time $\tau_0$ to a longer time scale $\tau_0'$. We propose to express $\tau_0'$ as a fluctuation time related to the correlation length with a usual dynamic scaling relation

$$\tau_0' = \tau_0 L^z , \qquad (14)$$

$z$ being the dynamic critical exponent which is measured above $T_g$, assuming that dynamic critical scaling may hold below $T_g$ in the same way as above $T_g$. To express the dependence on $L$ of the barrier $\Delta$ which must be crossed for flipping an ensemble of spins of size $L$, we follow the idea developed in the context of the droplet model for spin glasses [56] and write

$$\Delta(L,T) = \Upsilon(T)\, L^{\psi} \qquad (15)$$

($L$ being dimensionless, in units of lattice spacing), where the "stiffness" energy $\Upsilon(T)$ of the barrier

$$\Upsilon(T) = \Upsilon_0(1-T/T_g)^{\psi\nu} \qquad (16)$$

vanishes at $T_g$ with the same critical exponent $\nu$ which governs the divergence of the equilibrium correlation length $\xi \sim (T/T_g-1)^{-\nu}$ above $T_g$. In other words, we assume that (like in pinned ferromagnets) $\xi$ behaves in the same way below and above $T_g$, and that the barrier related to objects of size $L$ is

$$\Delta(L,T) = \Upsilon_0[L/\xi(T)]^{\psi} . \qquad (17)$$

Thermal activation over the barrier time $\Delta(L,T)$ yields the time $t$ needed for a rearrangement of spins at scale $L$ as $t=\tau_0' \exp [\Delta(L,T)]$, which reads explicitly [53]

$$t = \tau_0 L^z \exp\{\Upsilon_0(1-T/T_g)^{\psi\nu} L^{\psi}\} . \qquad (18)$$

This is a crossover expression between a purely critical regime $t = \tau_0 L^z$, obtained in the limit $L << \xi(T)$, and a superactivated regime in which the barriers grow as $(1-T/T_g)^{\psi\nu}$ when the temperature is decreased. It is clearly different from the power law $L \sim (t/\tau_0)^{aT/Tg}$ that was considered earlier, however Eq.(18) can also be written $t \sim L^{Zeff(T)}$ by defining

$$z_{eff}(T) = d \log t / d \log L = z + (\psi \Upsilon(T) L^{\psi})/k_B T , \qquad (19)$$

where $z_{eff}(T)$ is now an effective temperature (and length) dependent exponent, which is equal to the dynamic exponent $z$ at $T_g$ [53].

We have fitted the $t(L,T)$ expression (Eq.(18)) to both our *field* and *temperature* variation experiments, using the data of the 3 samples for which both kinds of measurements have been performed: Ising sample (#1), thiospinel (#3) and Ag:$Mn_{2.7\%}$ (#5) [67]. We have fixed $z\nu$ from published dynamic critical scaling data. We also fixed, to improve the global fit of all data, a geometrical factor $\alpha=2$ in the relation $N = \alpha L^3$ between the length $L$ and the number of spins $N$ . Apart from $\alpha$, which is the same for all samples, there are only 2 free parameters per sample in the adjustment of the whole set of data: $\Upsilon_0$ and $\psi$. A unique set of parameters is able to account for all the properties of each of the 3 samples (see Table 4.1). The fits are presented in Fig.4.6 and 4.7 for both sets of results.

Fig.4.6 shows the fit to the $N_s(t_w,T)$ results obtained from the relaxation experiments with various field amplitudes (Sect.4.1). In this representation, the simple power law behaviour of $L$ ($\sim N^{1/3}$) was represented by a straight line, but for the more complex crossover behaviour of Eq.(18), the time/temperature reduced variable $(T/T_g)ln(t_w/\tau_0)$ in the abscissa is no more relevant. Therefore, we have presented the results of the fit as curve segments, each segment representing, for one sample, the variation of $N_s$ as a function of $t_w$ at fixed temperature.

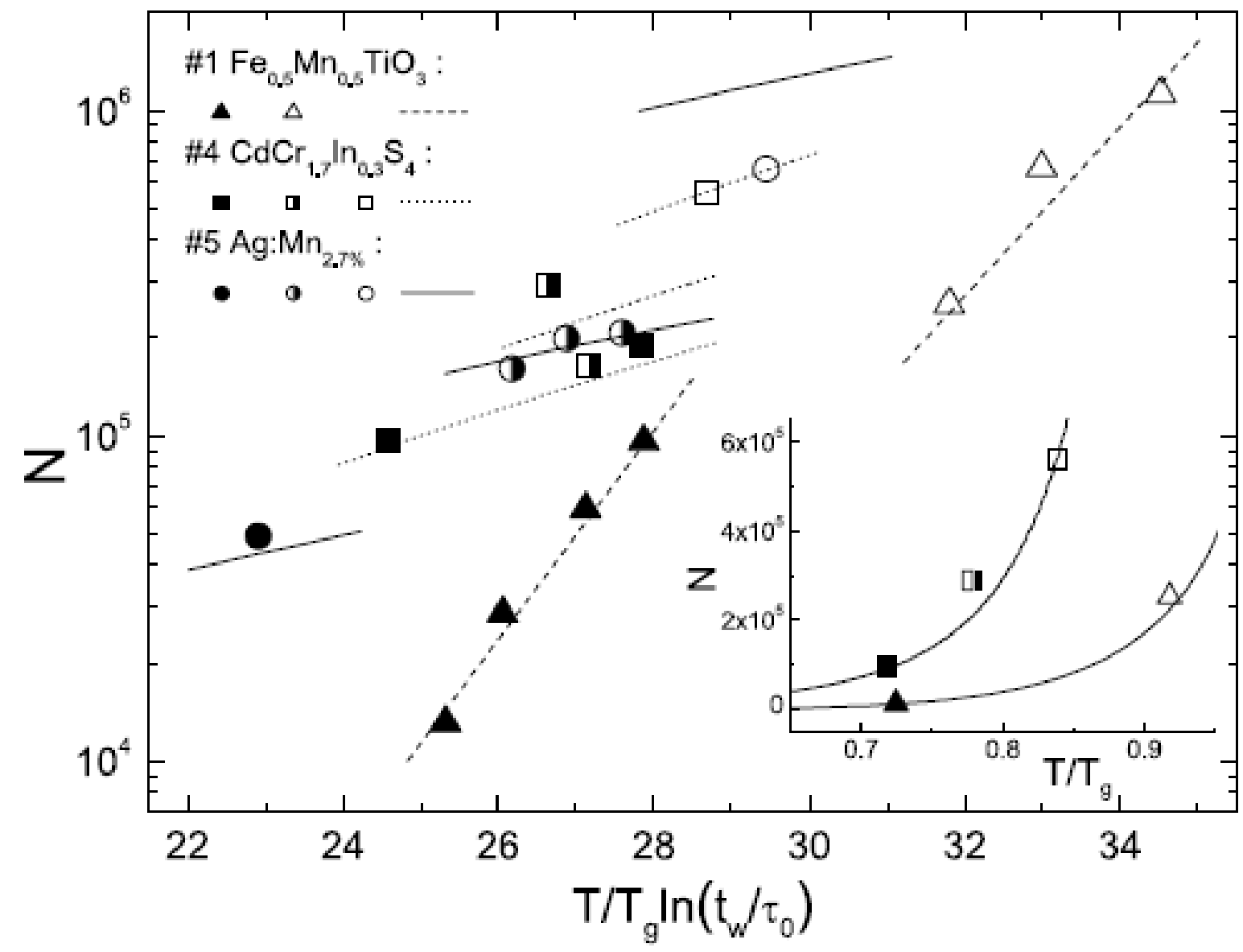

Figure 4.6 : Number of correlated spins from field change experiments (same data as in Fig.4.2), with the results of the *common* fit to both *field change* and *temperature variation* experiments [67]. Each curve segment is obtained at fixed temperature as a function of $t_w$. The inset shows the number of correlated spins $N$ as a function of temperature after $t_w$=1000s for samples #1 and #4, emphasizing their different behaviours.

Figure 4.7 shows the fits to the $t_{eff}(\Delta T)$ results from temperature cycle experiments (performed around $T=0.7$ and $0.85\ T_g$).

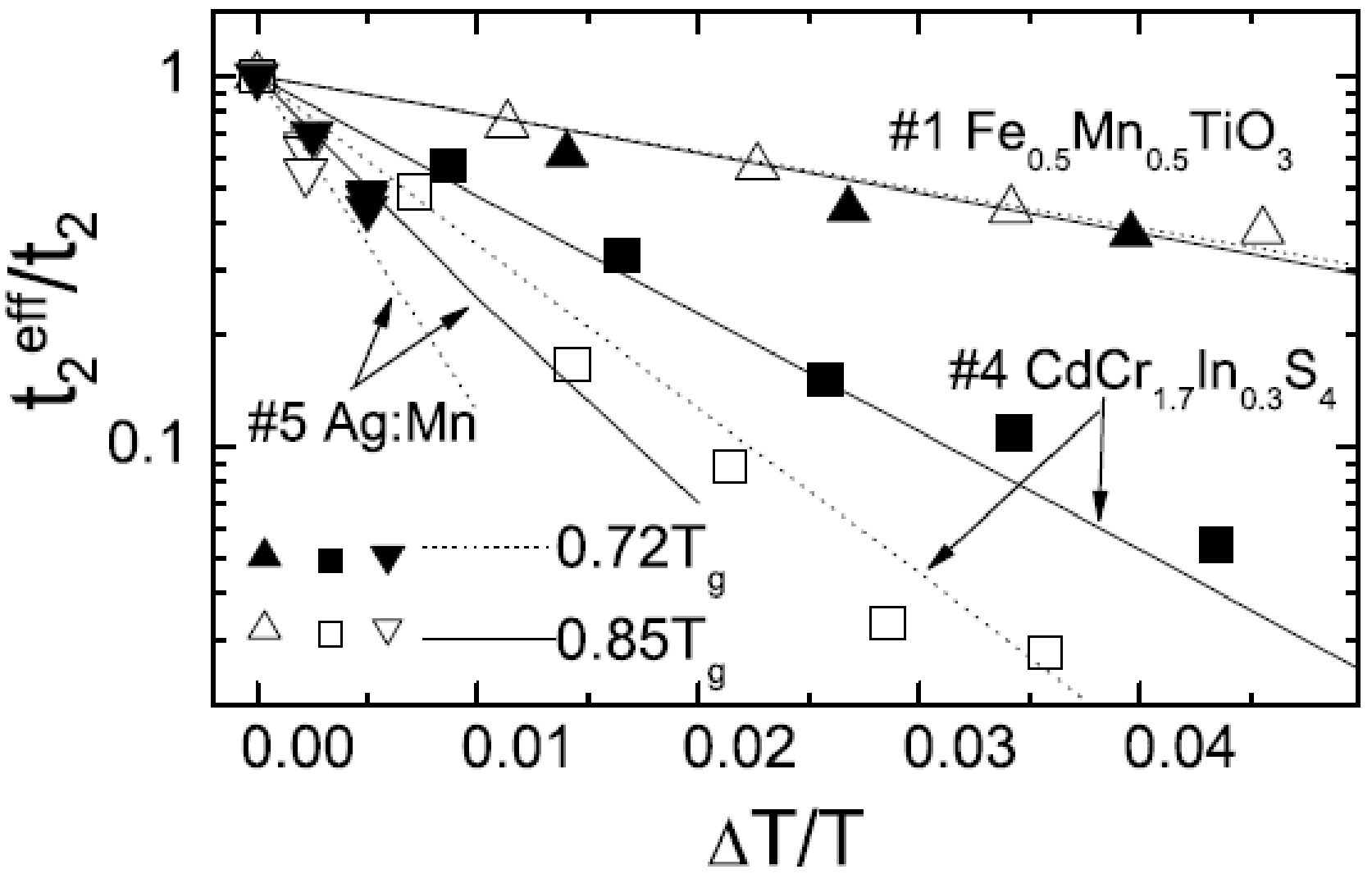

Figure 4.7: Effective waiting times deduced from the temperature cycle experiments (like in Fig.4.5), for 3 samples and 2 temperatures, together with the results (lines) of the common fit to both field change and temperature variation experiments [67].

In each figure 4.6 and 4.7 taken separately the quality of the fits is not excellent for all points, but it is important to remember that the results in both figures are fitted to Eq.(18) with a unique set of 2 parameters, and that the number of data points per sample is ~15 (see Table 4.1).

| | $\Upsilon_0$ | $\psi$ | Data points |
|---|---|---|---|
| $Fe_{0.5}Mn_{0.5}TiO_3$ (#1) | 14.5 | 0.03 | 16 |
| $CdCr_{1.7}In_{0.3}S_4$ (#4) | 1.2 | 1.1 | 17 |
| $Ag{:}Mn_{2.7\%}$ (#5) | 0.7 | 1.55 | 13 |

Table 4.1: Free parameters used in Eq.(18) to fit the temperature cycle and field variation experiments [67]. The "data points" column indicates the total number of data points that are fitted for each sample.

Actually, the parameters are not defined with a great quantitative accuracy, since their effects on the fit are strongly correlated. However, a consistent qualitative picture emerges. The main tendency is an *increasing* value of the barrier stiffness parameter $\Upsilon_0$ and a *decreasing* barrier exponent $\psi$ for *increasing* values of the anisotropy. This behaviour of the exponent is similar to that found in the analysis of previous *ac* temperature cycling experiments [72]. It contrasts with that derived from the time/frequency scaling of $\chi$" relaxations proposed in [73], which however is based on a less constrained analysis.

The extracted coherence length is noticeably smaller in the Ising sample (large $\Upsilon_0$) but grows faster with time (small $\psi$). At present, it is not clear how the strong single spin anisotropy in the Ising sample gives rise to both a high value of energy barriers and a very small value of the barrier exponent. Within a droplet description, $\psi \cong 0$ would imply that the droplet energy exponent $\theta$ is also zero, in agreement with recent numerical works on excitations in Ising spin glasses [74]. The case $\psi = 0$ also corresponds to barriers growing as the logarithm of the domain size. This behaviour has been argued by Rieger to hold in many disordered systems [75], including spin-glasses. In this case, the "effective exponent" $z_{eff}$ defined above becomes a true, temperature dependent, dynamical exponent.

Beyond the detailed values of the fitting parameters, the overall difference between an Heisenberg-like (thiospinel, #4) and the Ising samples is emphasized in the inset of Fig.4.6, which displays, at fixed $t_w=1000$s, the temperature dependence of $N$ for samples #1and 4. $N$ is larger for the Heisenberg-like sample, which leaves more space for building independent embedded active length scales at different temperatures, and also the temperature variation of $N$ is faster in the Heisenberg case, which signs up a faster separation of the active length scales with temperature (stronger temperature microscope effect). This should correspond to an increased sharpness of the memory dips in an experiment like that of Fig.3.5. This is indeed what has been observed in the very first experiments of comparison between an Ising and an Heisenberg spin glass [72]. Fig.4.8 shows the results of this *ac* memory dip experiment on the $Fe_{0.5}Mn_{0.5}TiO_3$ Ising spin glass, in which it was already visible (in comparison with Fig.3.5) that the memory effects are more spread out in temperature, in a way that we now understand in terms of a weaker $T$-microscope effect in the Ising case.

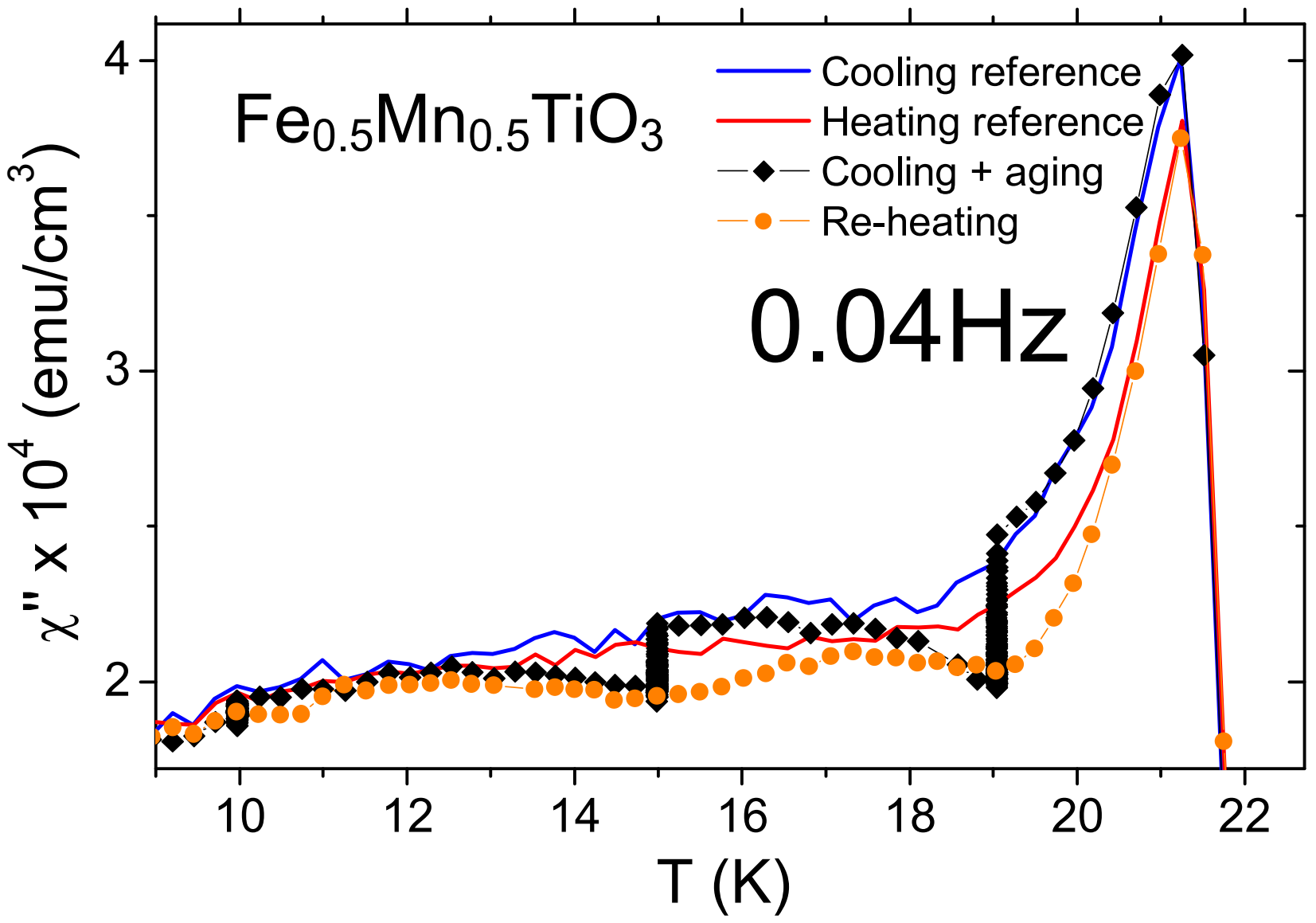

Figure 4.8 : Multiple rejuvenation and memory experiment with the $Fe_{0.5}Mn_{0.5}TiO_3$ Ising sample, from [72] (same as Fig.3.5, which is for the thiospinel Heisenberg-like sample). The solid lines show a reference behaviour for continuous cooling and reheating at 0.001 K/s (the reheating curve is slightly lower than the cooling curve). Diamonds: cooling with stops at 19, 15, and 10 K, during the stops $\chi''$ relaxes due to aging, and when cooling resumes $\chi''$ merges with the reference curve (rejuvenation). Circles: when reheating after cooling with stops for aging, the memory of aging is retrieved. The memory dips are not so sharply peaked in temperature as in the thiospinel (Heisenberg-like) sample (Fig.3.5).

## 4.4 Separation of time and length scales with temperature: how much?

In a spin glass, as the temperature is decreased, some aging processes become frozen (memory), while new ones are activated (rejuvenation). By "frozen", we mean that the time scale of a given relaxation process has become extremely large with respect to the experimental time window. In this sense, it is clear that there is a "separation of time scales" as a function of temperature in the spin glass, but it is interesting to see more precisely how far this time separation maps onto a "separation of length scales", as discussed by Berthier and Young in [16], of which we extract a characteristic figure as our last Figure 4.9.

In Fig.4.9, the authors have plotted the time variation of $L(T,t)$ using our parameterization (Eq.(18)) for an Heisenberg-like spin glass. This figure gives a precise idea of the length scales which are play in typical (Heisenberg) experiments and numerical simulations. In an experiment with aging during *10000*s at $T_1=0.825T_g$, the active length scale grows up to ~25 lattice units. The time separation with temperature is brutal, since $3.10^{21}$ years would be needed to obtain $L=25$ at $T_2=0.7T_g$. However, the active length that is reached after *10000*s at $T_2$ (starting from zero) is not that different, of the order of 15 : the "separation of *length* scales" from $T_1$ to $T_2$ takes place between 25 and 15, which is not spectacular, but enough to produce rejuvenation and memory effects, thanks to the fast separation of *time* scales.

Of course it is also very interesting to compare with the length scales that are reached at the time scale of the simulations, ~$10^5$ Monte Carlo steps. They are $L(T_1)$~6 and $L(T_2)$~4.5. This is not a powerful microscope in this case. Yet, due to the fast separation of the corresponding time scales rejuvenation and memory effects exist at the time scale of the simulations, and are now seen in the Heisenberg spin glass at d=3. They have not been found in the simulations of the Ising spin glass, probably because of a still weaker temperature microscope effect.

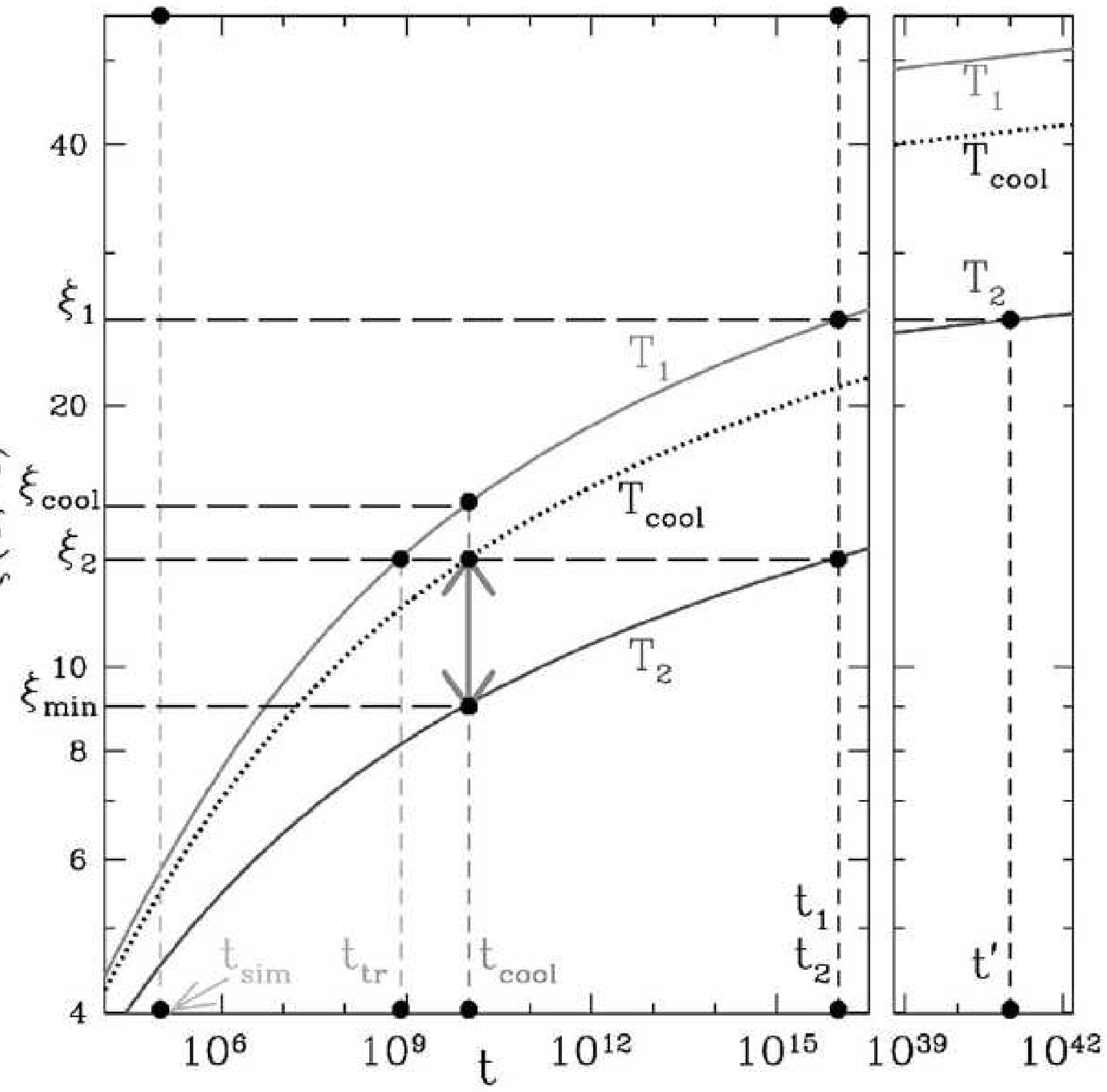

Figure 4.9 : From [16], growth of the dynamical correlation length as a function of time (in units of the elementary time $\tau_0$, $\tau_0 \sim 10^{-12}$s for the experiments and $\tau_0$=1 for the simulations), as obtained from the parameterization of our experimental results [67] in a Heisenberg-like case. See text (Section 4.4) and [16] for details.

# 5. Conclusions

In this rather general paper extracted from a summer school course, we have tried to review the most important features of the slow, out-of-equilibrium, dynamics of spin glasses. Perhaps the reader will have been convinced that, as stated in [53]:

"*Although spin glasses are totally useless pieces of material, they constitute an exceptionally convenient laboratory frame for theoretical and experimental investigations. ... There are at least two reasons for this: (a) the theoretical models are conceptually simpler (although still highly nontrivial), and (b) the use of very sensitive magnetic detectors allows one to probe in detail the* ac *and* dc *spin dynamics of these systems down to very small external fields. The corresponding mechanical measurements in other glassy systems are much more difficult to control, although some recent progress has been made*".

The waiting time dependence of the dynamical response (aging effect) is indeed a widely spread phenomenon observed in very different physical systems like polymer and structural glasses [20,36,42,43,64], disordered dielectrics [76,77], colloids and gels [32,33,34,35,65], foams, friction contacts [78], etc… Scaling laws of aging have been established in the rheology of glassy polymers [20], which precisely apply to the case of spin glasses [19]. In common with many different physical situations is also the subaging phenomenon, slight but systematic departure from pure $t/t_w$ scaling, of which we do not still know whether it is intrinsic or related to experimental artefacts (finite size effects [30,31], too slow cooling rates compared with microscopic times… [24]).

The response measurements can now be completed by direct measurements of the spontaneous fluctuations. The experiments of Ocio and Hérisson [29] could, for the first time, reveal the crossover to a modified fluctuation-dissipation relation when entering the strongly aging time regime of a spin glass. Further such experiments in spin glasses are needed, and an experimental way of normalizing the autocorrelation function has still to be found. In polymers and colloids, very interesting fluctuation dissipation studies could be performed these last years, which raise many new questions, among which the nature of the relationship between mechanical and dielectric properties of disordered systems [79].

The rejuvenation and memory experiments in spin glasses show that the effect on aging of the temperature history is highly non-trivial. The hierarchical structure of the numerous metastable states, proposed in the past [44,19], remains an efficient guideline to account for all details of the experiments, as discussed in various developments of Random Energy Models [50,47]. This "phase space" hierarchy can now be transcribed into a "real space" hierarchy of embedded length scales [53]. The basic ingredient is a strong separation of the time scales that govern the dynamics of the system on different length scales. Changing the temperature changes the length scale at which the system is observed, thereby allowing the coexistence of rejuvenation (that concerns short length scales) and memory (stored in long length scales). The relevance of "temperature-chaos" [61,56] for the occurrence of rejuvenation is still under debate [55]. In principle, rejuvenation may simply stem from the thermal variation of the equilibrium population rates of the metastable states, in the absence of any chaos effect [53], and in numerical simulations rejuvenation can indeed be observed without chaos [26]. However, it may well be that the experiments be influenced by chaos effects occurring at much larger length scales than can be directly explored [55].

A scenario of embedded active length scales is certainly at play in disordered ferromagnets, in which slow dynamics corresponds to hierarchical reconformations of elastic walls in a random pinning disorder [60,6]. The possible extension of this wall reconformation scenario to spin glasses raises some puzzling questions such as the nature of domains and walls in a spin glass.

The aging length scales can be captured in experiments which determine the dynamical correlation length that is growing during aging [67]. Several different sets of experiments can now be understood in terms of a unique form for the time and temperature dependence of the correlation length, which is a crossover between a critical regime and a super-activated regime, with energy barriers vanishing at $T_g$ [53]. From the study of five representative spin glass examples, we have found a clear trend to a stronger separation of active length scales with temperature when going from the Ising of the Heisenberg case (corresponding to sharper memory effects) [67]. The origin of this systematic dependence on spin anisotropy remains mysterious. Having again in mind the comparison with ferromagnets, a clue may be that less anisotropy should make broader walls, hence providing the spin glass with larger dynamical regions [80]. The comparison of Ising and Heisenberg spin glasses is intensively investigated in numerical simulations, which are now able to attack the time-consuming computation of Heisenberg spin dynamics. But the gap between numerical and experimental time scales remains immense [15,16,71].

The concept of a slowly growing and strongly temperature dependent dynamical correlation length allows understanding on the same basis the rejuvenation and memory effects and the cooling rate effects. It is now likely that this scenario of aging as a combination of "temperature specific" (rejuvenation and memory) and "temperature cumulative" processes [21], characterized in spin glasses, is also relevant for polymer and structural glasses, which were previously thought as dominated by cooling rate effects. Memory effects have now been observed in some polymers and gels [64,65]. It will be very interesting to see how far future experiments on various types of glasses may confirm the validity of a unique scenario for disordered systems which are made of such different building blocks.